\documentclass[12pt,a4paper]{article}    
\usepackage{jheppub}
\makeatletter
\usepackage{fontenc}
\usepackage[T1]{fontenc}
\usepackage[utf8]{inputenc}
\usepackage[svgnames,table]{xcolor}
\usepackage{booktabs,caption}
\usepackage{float}
\usepackage{units}
\usepackage{amsmath}
\usepackage{amssymb}
\usepackage{graphicx}
\usepackage{color}
\usepackage{esint}
\usepackage{youngtab}
\newcommand{\pd}[2]{\frac{\partial #1}{\partial #2}}

\usepackage{nicematrix}
\usepackage{bbold}

\renewcommand{\arraystretch}{1.2}

\usepackage{wrapfig}
\usepackage{slashed}
\usepackage{epsfig}
\usepackage{tensor}

\hypersetup{%
	pdftitle   = {Non-relativistic expansion of open strings and D-branes},
	pdfkeywords = {non-relativistic string theory, Newton-Cartan geometry, D-branes, Galilean electrodynamics},
	pdfauthor  = {Jelle Hartong, Emil Have},
	pdfcreator = {\LaTeX\ with package \flqq hyperref\frqq}}

\def\be{\begin{eqnarray}}
	\def\ee{\end{eqnarray}}
\newcommand{\nn}{\nonumber}

\def\Dslash{\,\,{\raise.15ex\hbox{/}\mkern-12mu D}}
\def\Dbarslash{\,\,{\raise.15ex\hbox{/}\mkern-12mu {\bar D}}}
\def\delslash{\,\,{\raise.15ex\hbox{/}\mkern-9mu \partial}}
\def\delbarslash{\,\,{\raise.15ex\hbox{/}\mkern-9mu {\bar\partial}}}
\def\pslash{\,\,{\raise.15ex\hbox{/}\mkern-9mu p}}
\def\calDslash{\,\,{\raise.15ex\hbox{/}\mkern-12mu {\cal D}}}

\newcommand{\D}{{\partial}}

\newcommand{\RR}{{\mathbb R}}

\newcommand{\zTeff}{T_{\text{NR}}}
\newcommand{\zaeff}{\alpha_{\text{NR}}'}
\newcommand{\zReff}{R_{\text{NR}}}

\def\lae{\mathrel{\mathop{\smash{\lower .5 ex \hbox{$\stackrel<\sim$}}}}}
\def\lae{\mathrel{\mathop{\smash{\lower .5 ex \hbox{$\stackrel>\sim$}}}}}

\usepackage[useregional]{datetime2}

\usepackage{float}
\usepackage{tikz}
\usepackage{pgfplots}
\pgfplotsset{compat=1.16}
\usetikzlibrary{calc,arrows,cd}
\usetikzlibrary{shapes,decorations}
\usepackage{makecell}
\pgfdeclarelayer{nodelayer}
\pgfdeclarelayer{edgelayer}
\pgfsetlayers{edgelayer,nodelayer,main}
\tikzstyle{ghost}=[fill={rgb,255: red,140; green,76; blue,150}, draw=black, shape=circle,scale=0.6]
\tikzstyle{real_ghost}=[fill=none, draw=none, shape=circle]
\tikzstyle{none}=[fill=none, draw=none, shape=circle]
\tikzstyle{Red_Circle}=[fill=none, draw=red, shape=circle]

\tikzstyle{BlackLine}=[-, draw=black, fill=white]
\tikzstyle{Arrow}=[<-, thick]
\tikzstyle{thin_black_line}=[-, fill=blue]
\tikzstyle{thin_black_line_red}=[-, fill=red]
\tikzstyle{thin_black_line_semi_purple}=[-, fill={rgb,255: red,1; green,1; blue,255}]
\tikzstyle{thin_black_line_purple}=[-, fill={rgb,255: red,200; green,1; blue,255}]
\tikzstyle{thin_black_line_turquoise}=[-, fill={rgb,255: red,1; green,200; blue,200}]
\tikzstyle{thin_black_line_gray}=[-, fill=gray]
\tikzstyle{thin_black_line_null}=[-, fill=orange]
\tikzstyle{Double_Arrow}=[<->]
\tikzstyle{Dashed_arrow}=[->, dashed, draw=red]
\tikzstyle{Dashed_arrow_gray}=[->, dashed, draw=gray]
\tikzstyle{BlackLine_dash}=[-, draw=red, fill=white, dashed]
\tikzstyle{Arrow_order}=[<-, draw=red]

\title{\bf 
	Non-relativistic expansion of open strings and D-branes
}

\author[1]{Jelle Hartong,}
\author[2]{Emil Have}

\affiliation[1]{School of Mathematics and Maxwell Institute for Mathematical Sciences,\\
	University of Edinburgh, Peter Guthrie Tait Road, Edinburgh EH9 3FD, UK}
\affiliation[2]{Niels Bohr International Academy, Niels Bohr Institute,\\ University of Copenhagen, Blegdamsvej 17, DK-2100 Copenhagen Ø, Denmark}

\emailAdd{j.hartong@ed.ac.uk}
\emailAdd{emil.have@nbi.ku.dk}

\abstract{We expand the relativistic open bosonic string in powers of $1/c^2$ where $c$ is the speed of light. We perform this expansion to next-to-leading order in $1/c^2$ and relate our results to known descriptions of non-relativistic open strings obtained by taking limits. Just as for closed strings the non-relativistic expansion is well-defined if the open string winds a circle in the target space. This direction must satisfy Dirichlet boundary conditions. It is shown that the endpoints of the open string behave as Bargmann particles in the non-relativistic regime. These open strings end on nrD$p$-branes with $p\le 24$. When these nrD$p$-branes do not fluctuate they correspond to $(p+1)$-dimensional Newton--Cartan submanifolds of the target space. When we include fluctuations and worldvolume gauge fields their dynamics is described by a non-relativistic version of the DBI action whose form we derive from symmetry considerations. The worldvolume gauge field and scalar field of a nrD$24$-brane make up the field content of Galilean electrodynamics (GED), and the effective theory on the nrD$24$-brane is precisely a non-linear version of GED. We generalise these results to actions for any nrD$p$-brane by demanding that they have the same target space gauge symmetries that the non-relativistic open and closed string actions have. Finally, we show that the nrD$p$-brane action is transverse T-duality covariant. Our results agree with the findings of Gomis, Yan and Yu in \cite{Gomis:2020fui}.}

\usepackage{todonotes}
\usepackage{mathrsfs}
\usepackage{amscd}
\newcommand{\diff}{\text{d}}

\usepackage{amsthm}

\theoremstyle{remark}

\usepackage{tcolorbox}
\usepackage{cancel}
\usepackage{xcolor}
\usepackage{manfnt}

\usepackage{tikz}

\DeclareFontFamily{U}{skulls}{}
\DeclareFontShape{U}{skulls}{m}{n}{ <-> skull }{}

\renewcommand{\i}{\text{i}}

\begin{document}
	\pagestyle{plain} \setcounter{page}{1}
	\newcounter{bean}
	\baselineskip16pt \setcounter{section}{0}
	\maketitle
	\flushbottom
	\section{Introduction}
	\label{sec:intro}

	Non-relativistic string theories (NRSTs) in their modern form trace their origins to~\cite{Gomis:2000bd,Danielsson:2000gi,Danielsson:2000mu}, where strings with non-relativistic dispersion relations where studied. 
	Following on from these pioneering works on what has since come to be known as the Gomis--Ooguri string, various authors have generalised the non-relativistic string theory to curved target spacetimes~\cite{Harmark:2017rpg,Kluson:2018egd,Bergshoeff:2018yvt,Harmark:2018cdl,Bergshoeff:2018vfn,Harmark:2019upf,Bergshoeff:2021bmc,Yan:2021lbe,Hartong:2021ekg,Bidussi:2021ujm,Hartong:2022dsx,Bidussi:2023rfs}. This was possible due to improvements in our understanding of non-Lorentzian geometries such as Newton--Cartan geometries and their stringy generalisations. There is by now a large class of non-Lorentzian string theories that have been constructed and studied where the non-Lorentzian geometry appears either in the target space or on the worldsheet (in some Polyakov formulation) or both. The study of these non-Lorentzian string theories has developed into a field in its own right. Examples are tensionless strings \cite{Isberg:1993av,Bagchi:2020fpr},  strings probing a string Carroll target space 
	\cite{Harksen:2024bnh,Bagchi:2023cfp}, obtained as an expansion around $c=0$ of relativistic strings, which have been found to provide a description of strings near black hole horizons in~\cite{Bagchi:2023cfp}, and limits of relativistic strings which belong to a duality web relating decoupling limits of string and M-theory on various non-Lorentzian backgrounds~\cite{Blair:2023noj,Gomis:2023eav}. 
	These non-Lorentzian string theories provide promising avenues for new versions of holographic dualities, to probe relativistic string/M-theory by taking new limits, and to study non-Lorentzian theories of quantum gravity. In particular, certain non-Lorentzian strings with a Galilean worldsheet are conjectured to be dual to spin matrix theory, which arises as a near-BPS limit of $\mathcal{N} = 4$ super Yang--Mills theory~\cite{Harmark:2014mpa,Harmark:2017rpg,Harmark:2018cdl,Harmark:2019upf,Harmark:2020vll,Bidussi:2023rfs}. Supersymmetric generalisations of NRST have been developed in~\cite{Gomis:2005pg,Kim:2007pc,Blair:2019qwi,Bergshoeff:2022pzk}, while connections to double field theory were established in~\cite{Ko:2015rha,Berman:2019izh,Cho:2019ofr,Blair:2020ops,Park:2020ixf,Gallegos:2020egk,Blair:2020gng}, and aspects of integrability as well as other types of NR strings can be found in \cite{Gomis:2005pg,Roychowdhury:2019sfo,Fontanella:2021hcb,Fontanella:2021btt,Roychowdhury:2021wte,Fontanella:2022fjd,Fontanella:2022pbm,Fontanella:2024rvn,deLeeuw:2024uaq}. For more details and a recent review see~\cite{Oling:2022fft}.

	Much of the focus of various non-Lorentzian string constructions has been with closed strings. In this paper we consider non-relativistic open strings. In fact this work is a continuation of \cite{Hartong:2021ekg,Hartong:2022dsx} by the authors, where non-relativistic expansions of closed bosonic strings were studied. Here we apply the same techniques to study the non-relativistic expansion of open strings. This naturally leads to the notion of a non-relativistic D-brane and we study their worldvolume theories. Previous work on non-relativistic limits of open strings and branes includes~\cite{Gomis:2004pw,Brugues:2004an,Gomis:2005bj,Brugues:2006yd,Mazzucato:2008tr,Kluson:2019avy,Roychowdhury:2019qmp,Kluson:2020kyp,Gomis:2020fui,Gomis:2020izd,Kluson:2020aoq,Ebert:2021mfu,Guijosa:2023qym,Lambert:2024yjk,Lambert:2024uue}, while the technology of expanding theories in inverse powers of $c$ was developed in~\cite{VandenBleeken:2017rij,Hansen:2018ofj,VandenBleeken:2019gqa,Hansen:2019vqf,Hansen:2020pqs,Ergen:2020yop}.

	Performing an expansion around $c=\infty$ in powers of\footnote{For simplicity, we restrict ourselves to even powers of $1/c$. We expect odd powers of $1/c$ to only matter beyond what we refer to as the NLO theory.} $1/c^2$ rather than taking a strict limit comes with several advantages. First of all, it allows us to go to any order in $1/c^2$ we may desire, leading to a tower of theories that include more and more relativistic corrections. More technically, the $c\rightarrow \infty$ limit of string theory is usually divergent and requires a so-called ``critical'' $B$-field that is fine-tuned in such a way that the divergence is cancelled. When doing expansions, there is no need for such a critical $B$-field. Some results do simplify when the $B$-field is critical at leading order (LO) and so we will sometimes make this assumption.

	In order to define a non-relativistic expansion we need the target spacetime to contain a circle that is wound by the string. This gives the string a rest energy with respect to which we can say that the centre-of-mass motion and the energy of the fluctuations are small, i.e., non-relativistic. This is true for both closed and open strings. Hence, for the case of an open string we need there to be a circle direction that is transverse to the D-brane the string is ending on. This compact direction that is wound by the string will be described by an adapted coordinate that we will denote by $v$. For the open strings studied in this work we assume that the embedding field for the $v$-direction obeys Dirichlet boundary conditions for both endpoints.\footnote{Choosing Neumann boundary conditions for the embedding field in the $v$-direction is either related to the NCOS (noncommutative open string) limit or to the DLCQ of a relativistic open string~\cite{Seiberg:2000ms,Gopakumar:2000na,Gomis:2020izd}.} Assuming all other directions satisfy NN boundary conditions, this leads to an open non-relativistic string defined on a non-relativistic D$24$-brane, or ``nrD$24$-brane''. 
	The dynamics of this brane is captured by a string $1/c^2$ expansion of the DBI action of a relativistic D24-brane. We name these objects non-relativistic D-branes or nrD-branes.

	From the perspective of a $1/c^2$ expansion, the requirement of Dirichlet boundary conditions in the $v$-direction is quite natural: while the endpoints of a relativistic open string travel at the speed of light for spacetime-filling D$25$-branes, this is no longer the case for general D$p$-branes with $p<25$, where the endpoints become massive. This makes them ideally suited for a non-relativistic expansion, and we shall demonstrate that the string endpoints on a nrD$24$-brane become non-relativistic (Bargmann) particles.\footnote{A non-relativistic open string moving in a flat target space and ending on a flat nrD24-brane possesses global symmetries that form the Bargmann algebra.}
	
	When we include worldvolume fields on the nrD24-brane (with a flat target spacetime for the open string), these fields transform in a certain way under the Bargmann transformations. For a nrD24-brane the worldvolume fields (a scalar and a connection 1-form) can be shown to transform exactly like the fields of Galilean electrodynamics (GED) \cite{LeBellac:1973,Santos:2004pq,GEDreview,Bagchi:2014ysa,Festuccia:2016caf}, where the scalar is related to transverse fluctuations of the nrD$24$-brane. The theory describing the nrD24-brane is then a non-linear version of Galilean electrodynamics.
	
	In order to study nrD$p$-branes with $p<24$ we study transverse T-duality. Here transverse means any direction that is not the $v$-direction. We study T-duality for closed strings in general background fields which leads to the Buscher rules for $1/c^2$ target space geometries. We then study T-duality for open strings in a flat target space ending on a nrD24-brane with GED fields on it and show that this leads to a NR string ending on a nrD23-brane, where the component of the GED gauge field in the direction along which the T-duality is performed turns into a fluctuation scalar for the brane in the T-dual (transverse) direction, perfectly mimicking the equivalent scenario in relativistic string theory.

	Finally, we derive the action describing general nrD$p$-branes by 1). expanding the relativistic DBI action in powers of $1/c^2$, and 2). using an intrinisally non-relativistic construction that is based on demanding the action to possess the same target space gauge symmetries as open and closed strings have. In both these approaches we assume that the relativistic theory has a $B$-field that is critical at LO in the $1/c^2$ expansion.

	\subsection*{Outline}
	
	This paper is structured as follows. In Section~\ref{sec:NR-open-strings}, we work out the string $1/c^2$ expansion for open strings moving in a flat target spacetime. We perform the expansion of the Nambu--Goto and Polyakov actions to NLO and show that the NLO theory agrees with previous constructions of non-relativistic open strings. We derive the gauge-fixed mode expansions and we compute the energy of the string. In Section~\ref{sec:Bargmann-particles}, we demonstrate that the endpoints of the NLO string, which we take to move on a nrD$24$-brane, behave as Bargmann particles.

	In Section~\ref{sec:NL-GED} we consider non-relativistic open strings in a flat target spacetime ending on a nrD24-brane with worldvolume fields turned on. We first discuss the $1/c^2$ expansion of an open string ending on a D24 with a scalar and a gauge field turned on and we work out how the NR versions of these worldvolume fields transform under the Bargmann transformations of Section~\ref{sec:Bargmann-particles}. This shows that these fields transform in the same way as the fields of Galilean electrodynamics (GED) which in turn suggests that the action for a nrD24-brane is a non-linear version of GED. In Section~\ref{sec:expansion-of-DBI}, we confirm this by performing the $1/c^2$ expansion of the DBI action for a D$24$-brane with a critical $B$-field, which leads to a non-linear GED action describing the dynamics of a nrD$24$-brane. By expanding in inverse powers of the string tension, we show how the GED action itself emerges from the non-linear theory.

	Section~\ref{sec:transverse-T-duality} is about transverse T-duality. We begin in Section~\ref{sec:T-duality-closed-strings} by considering transverse T-duality for closed NLO strings; first at the level of the mode expansions, and then in an arbitrary curved background using the Ro\v cek--Verlinde procedure which allows us to derive the Buscher rules for T-dual string Newton--Cartan backgrounds. In Section~\ref{sec:T-duality-open-strings} we repeat this for the non-relativistic open string, for a flat target space, but with nrD-brane worldvolume fields turned on.

	In Section~\ref{sec:nrDp-branes}, we consider general nrD$p$-branes for $p\leq 24$ in general backgrounds. We first derive the non-relativistic DBI action by expanding the relativistic DBI action in powers of $1/c^2$ in Section~\ref{sec:nrDBI-from-exp}, followed by a derivation based on target space gauge symmetries and T-duality covariance in Section~\ref{sec:DBI-from-T-duality-covariance}. In particular, we show that the two approaches give the same result. To set the stage for the discussion of the target space gauge symmetries we include a discussion of the gauge symmetries of the action of a string moving in a string Newton--Cartan background with a Kalb--Ramond $B$-field in Section~\ref{sec:syms}. We conclude with a discussion in Section~\ref{sec:discussion}.

	\subsection*{Notation and conventions}
	\begin{table}[!ht]
		\centering
		\rowcolors{2}{red!10}{white}
		\renewcommand{\arraystretch}{1.3}
		\begin{tabular}{>{$}l<{$}|>{$}l<{$}|>{$}r<{$}}\toprule
			\multicolumn{1}{c|}{index name}                                                                 & \multicolumn{1}{c|}{description}                                                                                        & \multicolumn{1}{c}{range} \\
			\midrule
			M,N,\dots                                    &  \text{spacetime indices}                                         & 0,\dots,d+1                    \\ 
			\mu,\nu,\dots   & \text{spacetime indices excluding $v$}  & 0,\dots,d                          \\ 
			i,j,\dots                                                      & \text{transverse spacetime indices}  & 1,\dots,d                          \\ 
			a,b,\dots  & \text{indices on NN directions} & 0,\dots,p                          \\I,J,\dots         & \text{indices on DD directions}                &  p+1,\dots,d+1 \\
			\bar M,\bar N,\dots   & \text{spacetime indices excluding $\theta$}  & 0, \dots, d-1, d+1                          \\
			\bar\mu,\bar\nu,\dots   & \text{spacetime indices excluding $\theta$ and $v$}  & 0,\dots,d-1                          \\
			\bar i,\bar j,\dots  & \text{transverse spacetime indices excluding $\theta$}  & 1,\dots,d-1\\
			\alpha,\beta,\dots                                    &  \text{worldsheet indices}                                         & 0,1                    \\ 
			\hat\alpha,\hat \beta,\dots  & \text{worldvolume indices on (nr)D$p$-branes }  & 0,\dots,p\\
			A,B,\dots & \text{longitudinal tangent space indices} & 0,1                   \\ 
			\bottomrule
		\end{tabular}  
		\caption{Overview of the index conventions adopted in this work. For most applications $d=24$.}
		\label{tab:indices}
	\end{table}
	
	As an aid to the reader, we lay out our conventions for the notation employed in this work. The string theories we consider live on $(d+2)$-dimensional target spaces where $d=24$ for the critical string. Components of tensors on these backgrounds carry spacetime indices $M,N,\dots$, where $M=0,\dots,d+1$. We will assume that the target spacetime admits a circle that in adapted coordinates is parametrised by $v$.
	The time coordinate will be $x^0=t$ and we often abuse notation by writing $t$ instead of $0$ as a value of the index $M$ (and similarly for other distinguished coordinates).
	The index $M$ then splits according to $M = (\mu,v) = (t,i,v)$, where $i = 1,\dots,d$ ranges over the $d$ transverse directions and where we identified $x^{d+1}=v$.
	Sometimes we shall find it useful to introduce vielbeine in the longitudinal directions, which carry a two-dimensional longitudinal tangent space index $A = 0,1$. When we work in flat space, we will simply use $A$ to mean $(t,v)$. 
	Finally, when we consider transverse T-dualities the target space includes an additional compact transverse direction which we call the $\theta$-direction. There we shall find it useful to split $M$ into $\bar M=(t,\bar i,v)$ and $\theta$ where $\bar i=1,\ldots,d-1$. We identify $x^{d}$ with $\theta$. We have summarised the properties of these indices as well as a few others in Table~\ref{tab:indices}.

	Throughout this work, we will frequently use lightcone coordinates both on the worldsheet in conformal gauge and in the longitudinal sector of the target space. On the worldsheet, which has coordinates $\sigma^\alpha$, we define the lightcone coordinates as $\sigma^\pm = \sigma^0\pm \sigma^1$. The two-dimensional Minkowski metric on the worldsheet will be denoted by $\eta_{\alpha\beta}$ which has the lightcone components
	\begin{align}
		\begin{aligned}
			\eta_{-+} &= \eta_{+-} = -\frac{1}{2}\,,\qquad \eta_{--} = \eta_{++} = 0\,,\\
			\eta^{-+} &= \eta^{+-} = -2\,,\qquad \eta^{--} = \eta^{++} = 0\,.
		\end{aligned}
	\end{align} 
	Similarly, the Levi-Civita symbols $\varepsilon_{\alpha\beta}$ and $\varepsilon^{\alpha\beta}$, which we take to satisfy $\varepsilon_{01} = +1$ and $\varepsilon^{01} = -1$, have lightcone components
	\begin{align} 
		\label{eq:levicivitalightcone}
		\begin{aligned}
			\varepsilon^{-+} &= -\varepsilon^{+-} = -2\,,\qquad \varepsilon^{++} = \varepsilon^{--} = 0\,,\\
			\varepsilon_{+-} &= -\varepsilon_{-+} = -\frac{1}{2}\,,\qquad \varepsilon_{++} = \varepsilon_{--} = 0\,.
		\end{aligned}
	\end{align}
	We emphasise that the convention for the Levi-Civita symbol differ from those employed in~\cite{Hartong:2021ekg,Hartong:2022dsx}, which will lead to differences in the signs of certain expressions.

	\section{Non-relativistic open strings from a \texorpdfstring{$1/c^2$}{inverse c} expansion}
	\label{sec:NR-open-strings}
	In this section, we first review the string $1/c^2$ expansion developed in~\cite{Hartong:2021ekg,Hartong:2022dsx} for closed bosonic strings. We then generalise the procedure to open strings in a flat target space with topology $\RR^{1,24}\times S^1_R$ where $R$ is the radius of the circle. We will derive the leading order (LO) and next-to leading order (NLO) open string actions and determine their energy. Finally, we will demonstrate that the endpoints of the NLO open string behave as Bargmann particles (i.e. standard non-relativistic particles).
	
	\subsection{The string \texorpdfstring{$1/c^2$}{inverse c} expansion \textit{\&} string Newton--Cartan geometry}
	\label{sec:string-expansion}
	
	Perturbative string theory comes with an intrinsic length scale: the string length $\ell_{\text{s}} = \sqrt{\alpha' \hbar /c}$. Hence, to define a dimensionless parameter in terms of which we can perform an expansion, we require an additional quantity with dimensions of length. This is achieved by compactifying one of the directions in the target space, which we will refer to as the $v$-direction. Together with the timelike direction, this compact direction will form part of a two-dimensional ``longitudinal'' Lorentzian subspace of the target space which is endowed with its own characteristic velocity $\tilde c$~\cite{Bidussi:2023rfs}. In a $26$-dimensional\footnote{The critical dimension of non-relativistic string theory is, just like for relativistic bosonic string theory, $26$ dimensions~\cite{Gomis:2000bd}.} flat target space with a compact direction, $\RR^{1,24}\times S^1_R$, we write the metric as
	\begin{equation}
		\label{eq:flat-coordintes}
		ds^2 = \eta_{MN} dx^M dx^N = c^2(-dt^2 + \tilde c^{-2}dv^2) + dx^i dx^i\,,
	\end{equation}
	where $M,N = 0,1,\dots,25$ are spacetime indices with $(t,x^i)$ for $i = 1,\dots,24$ coordinates on $\RR^{1,24}$. We denote $x^0=t$ and $x^{25}=v$ and often abuse notation by letting $M$ take the ``values'' $t$ and $v$ as opposed to 0 and 25. The components of $\eta_{MN}$ are 
	\begin{equation}
		\label{eq:flat-metric}
		\eta_{MN} = c^2(-\delta_M^t \delta_N^t + \tilde c^{-2}\delta^v_M \delta^v_N) + \delta^i_M \delta^i_N\,.
	\end{equation}
	The compact direction denoted by $v$ which is periodically identified according to
	\begin{equation}
		\label{eq:compact-v}
		v \sim v + 2\pi \zReff\,,
	\end{equation}
	where 
	\begin{equation}
		\zReff = R\tilde c/c\,,
	\end{equation}
	has dimensions of length and its radius $R_{\text{NR}}$ is assumed to be independent of $c$. This allows us to define a dimensionless parameter of order $c^{-2}$ via
	\begin{equation}
		\label{eq:dimless-param}
		\epsilon = \frac{\ell_s^2}{R^2} = \frac{\tilde c^2}{c^2}\frac{\ell^2_{s,\text{NR}}}{ \zReff^2} \,,
	\end{equation}
	where we introduced the string length of the non-relativistic string as 
	\begin{equation}\label{eq:lsNR}
		\ell^2_{s,\text{NR}} = \zaeff \hbar/\tilde c\,,
	\end{equation}
	and where we defined the, by assumption, $c$-independent combination
	\begin{equation}
		\zaeff = \frac{\tilde c}{c}\alpha'=\frac{1}{2\pi T_{\text{NR}}}=\frac{\tilde c}{c}\frac{1}{2\pi T}\,.
	\end{equation}
	In writing this, we defined the tension of the non-relativistic string as $T_{\text{NR}}=c T/\tilde c$. 
	
	It is the dimensionless parameter $\epsilon$ that we expand everything in terms of. In particular, this may either be considered as an expansion in $1/c^2$ or in $1/\zReff^2$; in other words, the expansion of string theory in terms of $\epsilon$ could equivalently be interpreted as an expansion around the decompactification limit $\zReff \rightarrow \infty$. The rationale behind this is that we need the circle to create a large rest energy with respect to which the energy of the string centre-of-mass momentum and fluctuations are small. The winding of the string along the $v$-direction is what leads to a large rest energy when the circle is large. 
	
	More generally, if we take the target space geometry to be $(d+2)$-dimensional, where $d$ is the number of transverse directions, the string $1/c^2$ expansion of a $(d+2)$-dimensional Lorentzian geometry as developed in~\cite{Hartong:2021ekg,Hartong:2022dsx} starts with splitting the metric $g_{MN}$ and its inverse $g^{MN}$ according to
	\begin{equation}
		\label{eq:PNR-decomp}
		\begin{split}
			g_{MN} &= c^2 \eta_{AB} T^A_M T^B_N + \Pi^\perp_{MN}\,,\qquad
			g^{MN} = c^{-2}\eta^{AB} T^M_A T^N_B + \Pi_\perp^{MN}\,,
		\end{split}
	\end{equation}
	where $A,B = 0,1$ are longitudinal tangent space indices, and where the signature of $\Pi_{MN}^\perp$ is $(0,0,1,\ldots,1)$. In writing these expressions we defined the two-dimensional longitudinal Minkowski metric 
	\begin{equation}
		\label{eq:longitudinal-mink}
		\eta_{AB} = \begin{pmatrix}
			-1 ~ &~0\\
			0~&~\tilde c^{-2}
		\end{pmatrix}\,,
	\end{equation}
	which includes the longitudinal lightspeed $\tilde c$. The variables that appear in the decomposition~\eqref{eq:PNR-decomp} satisfy the relations
	\begin{equation}
		T^M_A\Pi^\perp_{MN} = T_M^A\Pi_\perp^{MN} = 0\,,\qquad T^M_A T^B_M = \delta^B_A\,,\qquad \delta^M_N = \Pi^\perp_{NP}\Pi_\perp^{PM} + T^A_N T^M_A\,.
	\end{equation}
	We assume that the fields appearing in these expressions admit Taylor expansions in $1/c^2$, i.e., they admit expansions of the form
	\begin{equation}
		\label{eq:exp-of-PNR}
		T_M^A = \tau_M^A + c^{-2}m_M^A + \mathcal{O}(c^{-4})\,,\qquad \Pi^\perp_{MN} = H^\perp_{MN} + \mathcal{O}(c^{-2})\,,
	\end{equation}
	with similar expansions for $T^M_A$ and $\Pi_\perp^{MN}$. This means that the metric $g_{MN}$ acquires an expansion of the form
	\begin{equation}
		\label{eq:metric-exp}
		g_{MN} = c^2 \tau_{MN} + H_{MN} + \mathcal{O}(c^{-2})\,,
	\end{equation}
	where we defined the combinations
	\begin{equation}
		\label{eq:combinations}
		\tau_{MN} = \eta_{AB}\tau^A_M\tau^B_N\,,\qquad H_{MN} = H^\perp_{MN} + 2\eta_{AB}\tau_{(M}^A m_{N)}^B\,.
	\end{equation}
	We discuss the gauge symmetries of the $1/c^2$ expanded geometry (as seen by a string probe) in Section~\ref{sec:syms}. In terms of these structures, the flat space metric~\eqref{eq:flat-metric} corresponds to
	\begin{equation}
		\tau^A_M = \delta^A_M\,,\qquad H^\perp_{MN} = \delta^i_M\delta_N^i\,,\qquad m_M^A = 0\,.
	\end{equation}
	For more details about the string $1/c^2$ expansion and string Newton--Cartan geometry, we refer to~\cite{Hartong:2022dsx}.

	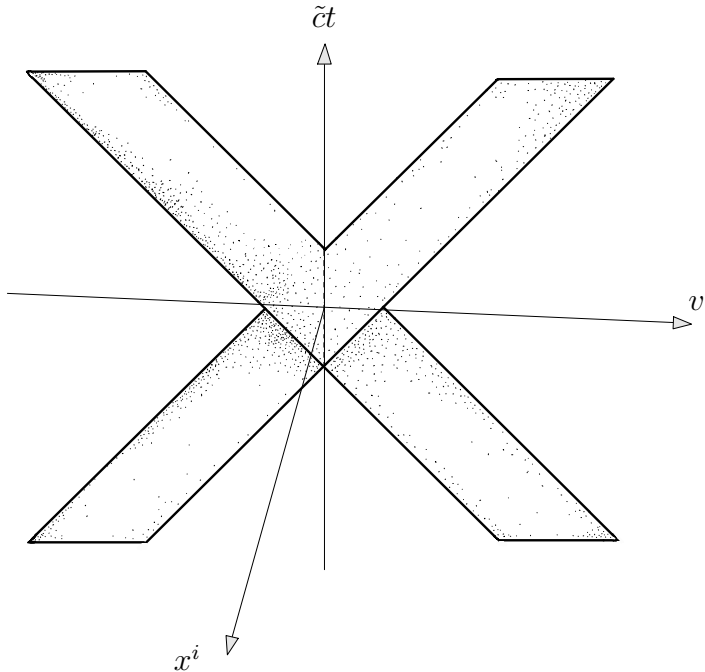
\begin{figure}[t!]
		\centering
		\includegraphics[width=0.7\textwidth]{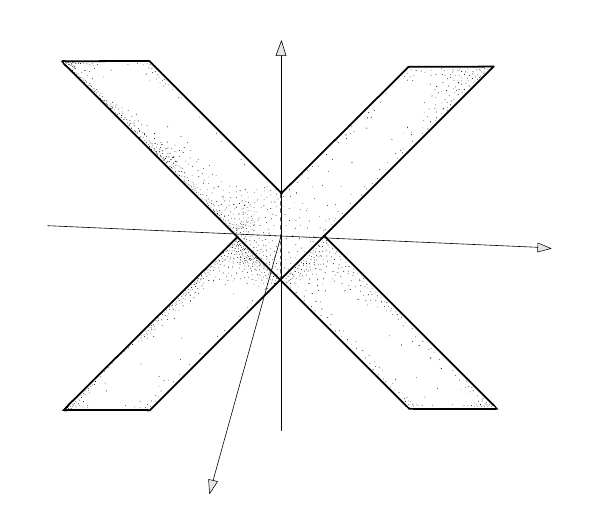}
		\begin{tikzpicture}[overlay]
			\begin{pgfonlayer}{nodelayer}
				\node [style=real_ghost] (0) at (-0.8, 5) {$v$};
				\node [style=real_ghost] (1) at (-7.5, 0.3) {$x^{i}$};
				\node [style=real_ghost] (2) at (-5.7, 8.8) {$\tilde ct$};
			\end{pgfonlayer}
		\end{tikzpicture}
		\caption{The local causal structure of SNC geometry. When taking the limit $c\to\infty$, the lightcone degenerates into two ``lightwedges'' which are hyperplanes defined by $v\pm \tilde c t = 0$.}
		\label{fig:SNC-causality}
	\end{figure}
	
	When the target space geometry only contains a metric (and no $B$-field) an important ingredient in the construction of these geometries is the condition that the longitudinal one-forms $\tau^A$ give rise to a codimension-2 foliation, i.e., that they satisfy the Frobenius condition
	\begin{equation}
		\label{eq:foliation-constraint}
		d\tau^A = \alpha^A{}_B\wedge \tau^B\,,
	\end{equation}
	where $\alpha^A{}_B$ are arbitrary one-forms. As demonstrated in~\cite{Hartong:2021ekg,Hartong:2022dsx}, this condition arises at LO in the string $1/c^2$ expansion of the vacuum Einstein equation. A stronger version of this ``foliation constraint'' was imposed in~\cite{Bergshoeff:2018vfn,Bergshoeff:2019ctr}, which sets $\alpha^A{}_B = \omega \varepsilon^A{_B}$, where $\omega$ is some one-form. When there is an NSNS $B$-field present this condition can be removed in favour of a so-called critical $B$-field (see further below). 
	
	Finally, let us comment on the local causal structure of string Newton--Cartan geometry (see also~\cite{Bergshoeff:2022fzb,Bidussi:2023rfs} for a discussion of the causal structure of string Newton--Cartan geometry). For a flat spacetime the lightcone at the origin is defined by the quadric 
	\begin{equation}
		\label{eq:quadric}
		0 = -t^2 + \tilde c^{-2}v^2 + c^{-2}x^{i}x^{i}\,.
	\end{equation}
	In the limit $c\to \infty$, the quadric~\eqref{eq:quadric} degenerates into two hyperplanes, or ``light- wedges'', defined by $ v \pm \tilde c t = 0$ as shown in Figure~\ref{fig:SNC-causality}. The lightcone direction along the $x^i$ flatten out. For $v=0$ we get a standard non-relativistic lightcone in the large-$c$ limit.

	The SNC geometry can viewed as consisting of a 2-dimensional Lorentzian base manifold over which a $d$-dimensional Riemannian space has been fibered. The metric on the leaves of this fibration is given by $H_{\perp MN}dx^M dx^N$ evaluated at a point on the base. The total space of this fibration is the SNC geometry. The Lorentzian base manifold will be assumed to have the topology of a cylinder as we need a longitudinal circle that is wrapped by the string to define the non-relativistic limit.

	\subsection{Expanding the open string action in flat target space}
	\label{sec:string-expansions-without-WV-fields}
	Let us now consider the $1/c^2$ expansion of relativistic open strings with target space $\RR^{1,24}\times S^1_R$. In this section, these differ only from closed strings as considered in~\cite{Hartong:2021ekg,Hartong:2022dsx} by virtue of having different boundary conditions, while other essential open-string features such as D-brane worldvolume gauge fields will be discussed in Section~\ref{sec:NL-GED}. 
	
	The open string is described by the Nambu--Goto (NG) action
	\begin{equation}
		\label{eq:Rel-NG}
		S_{\text{NG}} = -cT\int d^2\sigma\,\sqrt{-\det(\eta_{MN}\D_\alpha X^M \D_\beta X^N)}\,,
	\end{equation}
	where $\eta_{MN}$ is the Minkowski metric~\eqref{eq:flat-metric}, and where $\sigma^\alpha$ for $\alpha =0,1$ are dimensionless coordinates on the worldsheet. The embedding scalars are such that $\eta_{MN}\D_\alpha X^M \D_\beta X^N$ has the dimension of a length squared. The tension $T$ has dimension mass per unit length. The dynamics of the string is equivalently described by the Polyakov action, which in conformal gauge reads
	\begin{equation}
		\label{eq:Pol-action}
		S_{\text{P}} = -\frac{cT}{2} \int d^2\sigma~\eta^{\alpha\beta}\eta_{MN}\D_\alpha X^M\D_\beta X^N\,,
	\end{equation}
	where $\eta_{\alpha\beta} =\text{diag}(-1,+1)$ is the metric on the two-dimensional worldsheet. The equation of motion of the embedding fields $X^M$ is
	\begin{equation}
		\label{eq:rel-EOM}
		\text{EOM}_M= \eta_{MN}\eta^{\alpha\beta}\D_\alpha \D_\beta X^N = 0\,,
	\end{equation}
	while the Virasoro constraints read
	\begin{equation}
		\label{eq:rel-Vir}
		T_{\alpha\beta} = \eta_{MN} \D_\alpha X^M \D_\beta X^N - \frac{1}{2} \eta_{\alpha\beta}\eta^{\gamma\delta}\eta_{MN}\D_\gamma X^M \D_\delta X^N = 0\,.
	\end{equation}
	
	For open strings, the spatial coordinate on the worldsheet $\sigma^1$ conventionally takes values in the interval $\sigma^1\in[0,\pi]$, and the on-shell\footnote{In this context, on-shell means that we both impose the equations of motion, $\D_\alpha\D^\alpha X^M = 0$, and assume that the variation $\delta X^M$ vanishes at the temporal endpoints.} variation of the Polyakov action becomes
	\begin{equation}
		\delta S_{\text{P}}\big\vert_{\text{on-shell}} = -cT \left[\int d\sigma^0~\eta_{MN}X'^M\delta X^N\right]_{\sigma^1 = 0}^{\sigma^1=\pi}\,,
	\end{equation}
	where the prime denotes differentiation with respect to $\sigma^1$. Hence, boundary conditions must be imposed in such a way that
	\begin{equation}
		\label{eq:rel-BC}
		\eta_{MN}X'^M\delta X^N\big\vert_{\text{ends}}=0\,,
	\end{equation}
	which means that for any given component $X^\bullet$, we can impose either Neumann boundary conditions:
	\begin{equation}
		\label{eq:neumann-bcs}
		X'^\bullet\big\vert_{\text{ends}} = 0\,,
	\end{equation}
	or Dirichlet boundary conditions:
	\begin{equation}
		\label{eq:dirichlet-bcs}
		\delta X^\bullet\vert_{\text{ends}} = 0\Rightarrow X^\bullet = k^\bullet\,,
	\end{equation}
	where $k^\bullet$ is a constant. While it is possible to consider mixed boundary conditions, we will consider only scenarios where both ends satisfy either Neumann (i.e., NN) or Dirichlet (i.e., DD) boundary conditions.
	
	If we take NN boundary conditions for $X^{a}$ with $a = 0,\dots,p$ and DD boundary conditions for $X^{I}$ where $I = p+1,\dots,d+2$, this corresponds to the following at the string endpoints 
	\begin{align}
		\label{eq:BC2}
		\begin{split}
			\text{NN:}\quad X'^{a} &= 0\,,\qquad a = 0,\dots,p\,,\\
			\text{DD:}\quad X^{I} &= k^{I}\,, \qquad I =p+1,\dots,d+2\,,
		\end{split}
	\end{align}
	which fixes the two endpoints of the string to lie on the $p$-dimensional (not counting time) hyperplane defined by $X^{I} = k^{I}$: a D$p$-brane. In particular, this breaks the Lorentz group according to $\operatorname{SO}(d+1,1)\rightarrow \operatorname{SO}(p,1)\times \operatorname{SO}(d+1-p)$. Furthermore the translations perpendicular to the brane are broken by the Dirichlet boundary conditions. 
	
	A compact target space direction transverse to the brane satisfies Dirichlet boundary conditions, while a compact direction wrapped by the brane satisfies Neumann boundary conditions. T-duality interchanges the two. We will have more to say about the $1/c^2$ expansion of D$p$-branes in Sections~\ref{sec:expansion-of-DBI} and \ref{sec:DBI-from-T-duality-covariance}, while we study (transverse) T-duality in the context of $1/c^2$ expansions in Section~\ref{sec:transverse-T-duality}.

	The Polyakov action~\eqref{eq:Pol-action} in conformal gauge retains residual gauge symmetries consisting of diffeomorphisms of the form
	\begin{equation}
		\label{eq:residual-gauge-trafos}
		\delta_\xi X^M = \xi^\alpha\D_\alpha X^M\,,\qquad \xi(\sigma) = \xi^-(\sigma^-)\D_- + \xi^+(\sigma^+)\D_-\,,
	\end{equation}
	where we introduced worldsheet lightcone coordinates
	\begin{equation}
		\label{eq:WS-LC-coords}
		\sigma^\pm = \sigma^0\pm \sigma^1\,.    
	\end{equation}
	These residual gauge transformations must furthermore respect the boundary conditions. They will play an important role when writing down gauge-fixed mode expansions for the embedding fields below.

	As explained in~\cite{Hartong:2021ekg,Hartong:2022dsx}, one must expand both the target space metric, the worldsheet metric and the embedding scalars in powers of $1/c^2$. However, since we specialise to a flat target space with metric~\eqref{eq:flat-metric}, the associated $1/c^2$ expansion is simply given by that same equation. Furthermore, going to conformal gauge in the relativistic theory commutes with gauge-fixing the $1/c^2$ expansion of the theory order by order~\cite{Hartong:2021ekg,Hartong:2022dsx}. In practice, this means that we may $1/c^2$ expand the (partially) gauge-fixed Polyakov action~\eqref{eq:Pol-action} along with the equations of motion~\eqref{eq:rel-EOM} and the Virasoro constraints~\eqref{eq:rel-Vir} to obtain the corresponding theories of non-relativistic open strings on a flat target space to any given order in $1/c^2$. Expanding the embedding scalars according to
	\begin{equation}
		X^M = x^M + c^{-2}y^M + \mathcal{O}(c^{-4})\,,
	\end{equation}
	we find that the Polyakov action~\eqref{eq:Pol-action}, the equations of motion~\eqref{eq:rel-EOM} and the Virasoro constraints~\eqref{eq:rel-Vir} admit expansions in powers of $1/c^2$ of the form (to NLO)
	\begin{equation}
		\label{eq:expansions}
		\begin{split}
			S_{\text{P}} &= c^2 S_{\text{P-LO}} + S_{\text{P-NLO}} + \mathcal{O}(c^{-2})\,,\\
			\text{EOM}_M &= c^2\text{EOM}_M^{\text{LO}} + \text{EOM}_M^{\text{NLO}}+ \mathcal{O}(c^{-2})\,,\\
			T_{\alpha\beta} &= c^2T_{\alpha\beta}^{\text{LO}} + T_{\alpha\beta}^{\text{NLO}}+ \mathcal{O}(c^{-2})\,.
		\end{split}
	\end{equation}
	We will study the LO and NLO theories in more detail below.

	\subsubsection{The LO open string}
	\label{sec:LO-open-string}
	The LO Polyakov action as it appears in~\eqref{eq:expansions} is given by
	\begin{equation}
		\label{eq:LO-P-action}
		\begin{split}
			S_{\text{P-LO}} &= \frac{\tilde c\zTeff}{2}\int d^2\sigma\left(\eta^{\alpha\beta}\D_\alpha x^t \D_\beta x^t - \tilde c^{-2}\eta^{\alpha\beta}\D_\alpha x^v \D_\beta x^v\right) \\
			&= - \frac{\tilde c\zTeff}{2}\int d^2\sigma\,\eta^{\alpha\beta}\eta_{AB}\D_\alpha x^A \D_\beta x^B\,,
		\end{split}
	\end{equation}
	where the effective string tension
	\begin{equation}
		\zTeff = \frac{c}{\tilde c}T\,,
	\end{equation}
	is $c$-independent. Parenthetically, we remark that this is identical in form to the relativistic Polyakov action with a two-dimensional target space. The equations of motion may be found either by varying~\eqref{eq:LO-P-action} directly or from~\eqref{eq:expansions}. They are given by
	\begin{equation}
		\text{EOM}_A^{\text{LO}} = \eta_{AB}\D_-\D_+ x^B = 0\,,
	\end{equation}
	where we remind the reader that the target space $v$-direction is compact, cf.~Eq.~\eqref{eq:compact-v}. The LO Virasoro constraints appearing in~\eqref{eq:expansions} are 
	\begin{equation}
		\label{eq:LO-nonLC-Vir}
		T_{\alpha\beta}^{\text{LO}} = \eta_{AB} \D_\alpha x^A \D_\beta x^B - \frac{1}{2} \eta_{\alpha\beta}\eta^{\gamma\delta}\eta_{AB}\D_\gamma x^A \D_\delta x^B = 0\,.
	\end{equation}
	
	The well-posedness of the variational principle for the action~\eqref{eq:LO-P-action}, or equivalently, the LO term in the $1/c^2$ expansion of~\eqref{eq:rel-BC}, produces the condition that 
	\begin{equation}\label{eq:bdrycondition}
		\eta_{AB}x'^A\delta x^B\big\vert_{\text{ends}}=0\,.
	\end{equation}
	That means that we have the option to impose either Neumann \eqref{eq:neumann-bcs} or Dirichlet \eqref{eq:dirichlet-bcs} boundary conditions. For $x^t$, we impose Neumann boundary conditions, which at $\sigma^1 = 0$ implies that 
	\begin{equation}
		0 = x'^t(\sigma^0,0) = \D_+ x^t(\sigma^0,0) - \D_-x^t(\sigma^0,0)\,,
	\end{equation}
	and if we define $f^t(\sigma^+) = \D_+x^t$ and $g^t(\sigma^-) = \D_-x^t$, the equation above implies that $f^t(\sigma^0) = g^t(\sigma^0)$ for all values of $\sigma^0$, and since $\sigma^\pm =\sigma^0$ for $\sigma^1=0$, we conclude that $f^t$ and $g^t$ are the same function. At the other endpoint, we get the relation
	\begin{equation}
		f^t(\sigma^0+\pi) = g^t(\sigma^0-\pi)\,,
	\end{equation}
	which, using $f^t = g^t$, implies that $f^t(\sigma^0+\pi) = f^t(\sigma^0-\pi)$, or, equivalently, $f^t(\sigma^0) = f^t(\sigma^0 + 2\pi)$: the function $f^t$ is periodic with period $2\pi$. Hence, we get the following mode expansions for the lightcone derivatives
	\begin{equation}
		\D_- x^t = \frac{1}{\sqrt{4\pi\tilde c\zTeff}}\sum_{k\in\mathbb{Z}}\alpha^t_ke^{-ik\sigma^-}\,,\qquad \D_+ x^t = \frac{1}{\sqrt{4\pi \tilde c\zTeff}}\sum_{k\in\mathbb{Z}}\alpha^t_ke^{-ik\sigma^+}\,,
	\end{equation}
	which leads to the mode expansion
	\begin{equation}
		x^t = x^t_0 + \frac{1}{\sqrt{\pi \tilde c\zTeff}}\alpha_0^t\sigma^0 + \frac{i}{\sqrt{\pi\tilde c\zTeff}}\sum_{k\neq 0}\frac{1}{k}\alpha_k^te^{-ik\sigma^0}\cos(k\sigma^1)\,.
	\end{equation}
	As in~\cite{Gomis:2020fui}, we impose Dirichlet boundary conditions for the embedding field $x^v$ in the compact $v$-direction. This is because the endpoint of a string ending on a D$p$-brane with $p\le 24$ is a massive particle from the point of view of the worldvolume theory, whilst the endpoint of a string ending on a spacefilling D25-brane behaves as a massless particle. Since the $1/c^2$ expansion of a massive particle leads to standard non-relativistic particles we choose to work with D$p$-branes with $p\le 24$. Furthermore, we need the open string to wind a spatial circle to give it a rest energy with respect to which we define a non-relativistic regime. As we will see this circle has to be longitudinal to the string and thus transverse to the brane. 
	
	The dynamics of the D$p$-branes after we have taken the non-relativistic expansion is described by a $1/c^2$ expansion of the DBI action (see Section~\ref{sec:expansion-of-DBI}) and so we will denote the branes on which a non-relativistic string ends as a non-relativistic D$p$-brane which we abbreviate as nrD$p$-brane.

	For most of this paper we will consider open strings ending on D24-branes and once we understand their non-relativistic limit we will generalise to D$p$-branes with $p<24$ in Section~\ref{sec:nrDp-branes}. 
	
	DD boundary conditions in the $v$-direction imply that
	\begin{equation}
		x^v(\sigma^0,0) = k^v\qquad\text{and}\qquad x^v(\sigma^0,\pi) = k^v\qquad (\text{mod}~2\pi \zReff)\,,
	\end{equation}
	and hence $\dot x^v(\sigma^0,0) = \D_+ x^v(\sigma^0,0) + \D_-x^v(\sigma^+,0)$. Defining, as above, $f^v(\sigma^+) = \D_+x^v$ and $g^v(\sigma^-) = \D_- x^v$, we find that $f^v(\sigma^0) = -g^v(\sigma^0)$ at the endpoint with $\sigma^1 = 0$, and thus $f^v=-g^v$ as functions of their respective arguments. As above, the same argument at the other endpoint produces the result that $f^v$ is periodic with period $2\pi$, and hence the mode expansion of $x^v$ takes the form
	\begin{equation}
		x^v = k^v + 2w\zReff\sigma^1 - \frac{1}{\sqrt{\pi\tilde c\zTeff}}\sum_{k\neq 0}\frac{1}{k}\alpha_k^ve^{-ik\sigma^0}\sin(k \sigma^1)\,,
	\end{equation}
	where we have included a winding term, where $w\in \mathbb{N}$ is the winding number counting the number of times the open string winds the compact $v$-direction, and where the factor of two in the winding term comes from the fact that the range of $\sigma^1$ is $[0,\pi]$. As we will see the winding is related to the rest energy of the string, which will be required to be nonzero, so that we can take it to be positive.

	For the lightcone combinations $x^\pm$, 
	\begin{equation}
		\label{eq:lightcone-coords}
		x^\pm =  x^t \pm \tilde c^{-1} x^v\,,
	\end{equation}
	which have dimensions of time, we get the mode expansions
	\begin{align}
		\begin{split}
			x^\pm &= x_0^t\pm \tilde c^{-1}k^v + \left( \frac{1}{2}\sqrt{\frac{1}{\pi \tilde c\zTeff}}\alpha_0^t \pm \frac{w\zReff}{\tilde c} \right) \sigma^+ + \left( \frac{1}{2}\sqrt{\frac{1}{\pi\tilde c \zTeff}}\alpha_0^t \mp \frac{w\zReff}{\tilde c} \right) \sigma^-\\
			&\quad+ \frac{i}{\sqrt{4\pi \zTeff}}\sum_{k\neq 0}\frac{1}{k}\left[ \alpha^\pm_ke^{-ik\sigma^-} + \alpha^\mp_k e^{-ik\sigma^+} \right] \,, 
		\end{split}
	\end{align}
	where $\alpha_k^\pm = \alpha^t_k\pm \tilde c^{-1}\alpha^v_k$. We have yet to impose the LO Virasoro constraints \eqref{eq:LO-nonLC-Vir}. These are
	\begin{equation}
		\partial_+ x^+\partial_+ x^-=0\,,\qquad
		\partial_+ x^+\partial_+ x^-=0\,,
	\end{equation}
	to which there are four classes of solutions. Either one of the fields $x^+$ or $x^-$ is constant which implies that $\alpha_0^t=0=w$ or one of them depends on $\sigma^+$ and the other on $\sigma^-$. We will here only consider solutions with nonzero winding. We can then without loss of generality take $x^+$ ($x^-$) to be a function of $\sigma^+$ ($\sigma^-$). Hence, the LO Virasoro constraints can be written as 
	\begin{equation}
		\label{eq:LO-virasoro}
		\D_-x^+ = \D_+x^- = 0\,,
	\end{equation}
	which thus imply the equations of motion $\D_+\D_-x^A=0$. They also tell us that
	\begin{equation}
		\frac{1}{2}\sqrt{\frac{\tilde c}{\pi \zTeff}}\alpha_0^t = w\zReff
	\end{equation}
	and that $x^\pm$ has no $\sigma^\mp$-oscillations, implying that
	\begin{equation}
		\alpha^+_k = 0\qquad \forall k\neq 0\,,
	\end{equation}
	so that the mode expansions can be written as
	\begin{equation}
		x^\pm = \tilde c x_0^t\pm k^v + 2 w\zReff \sigma^\pm + \frac{i}{\sqrt{4\pi \zTeff}}\sum_{k\neq 0}\frac{1}{k} \alpha^-_ke^{-ik\sigma^\pm}\,.
	\end{equation}
	
	Just as for the closed string~\cite{Hartong:2021ekg,Hartong:2022dsx}, the residual gauge transformations~\eqref{eq:residual-gauge-trafos}, which are expanded according to
	\begin{equation}
		\label{eq:2ddiffeoexp}
		\xi^\alpha = \xi^\alpha_{(0)} + c^{-2}\xi^\alpha_{(2)} + \cdots\,,    
	\end{equation}
	allow us to set the remaining oscillations equal to zero, but the argument is not quite identical. The vanishing of the boundary term \eqref{eq:bdrycondition} for open strings under the residual gauge transformations implies that 
	\begin{equation}
		(x'^+\delta x^- + x'^-\delta x^+)\big\vert_{\text{ends}} = 0\,,
	\end{equation}
	and using the Virasoro constraints and $\delta_{\xi_{(0)}}x^\pm = \xi^\pm_{(0)}\D_\pm x^\pm$, this becomes 
	\begin{equation}
		\D_-x^-\D_+x^+(\xi_{(0)}^- - \xi^+_{(0)})\big\vert_{\text{ends}} = 0\,.
	\end{equation}
	At the endpoint $\sigma^1 = 0$, we thus find that $\xi^\pm_{(0)}(\sigma^\pm)$ are the same function (call it $\xi_{(0)}$) of their respective arguments, while at the endpoint $\sigma^1 = \pi$, we find that $\xi_{(0)}$ is $2\pi$-periodic. Hence, we may fix the residual gauge invariance at LO by setting 
	\begin{equation}
		\alpha^-_k = 0\qquad \forall k\neq 0\,,
	\end{equation}
	leading to the fully gauge-fixed mode expansions
	\begin{equation}
		\label{eq:fully-gauge-fixed-LO}
		x^\pm =  x_0^t\pm \tilde c^{-1}k^v + 2w\tilde c^{-1} \zReff \sigma^\pm \,,
	\end{equation}
	at LO, which we may alternatively express as 
	\begin{equation}
		\label{eq:fully-gauge-fixed-LO-non-LC}
		x^t = x_0^t + 2w \tilde c^{-1}\zReff\sigma^0\,,\qquad x^v = k^v + 2w\zReff \sigma^1\,.    
	\end{equation}
	This means that the LO energy is given by
	\begin{equation}
		\label{eq:LO-energy}
		E_{\text{LO}} = -\int_0^\pi d\sigma^1\pd{\mathcal{L}_{\text{P-LO}}}{\dot x^t}\bigg\vert_{\text{on-shell}} =\tilde c\zTeff \int_0^\pi d\sigma^1\dot x^t = \frac{w\zReff}{\zaeff}\,,
	\end{equation}
	which is the same as the energy of the closed LO string \cite{Hartong:2022dsx}. This actually has the dimension of a mass which is due to the fact that we factored out $c^2$ in \eqref{eq:expansions}.
	
	\subsubsection{The NLO open string}
	\label{sec:NLO-open-string}
	The action describing the NLO theory of an {open string} on a flat target space in flat gauge as it appears in~\eqref{eq:expansions} is given by 
	\begin{equation}
		\label{eq:P-NLO-action-flat}
		\begin{split}
			S_{\text{P-NLO}} &= - \frac{\tilde c\zTeff}{2}\int d^2\sigma\left[ \eta^{\alpha\beta}\D_\alpha x^i\D_\beta x^i + 2\eta^{\alpha\beta}\eta_{AB}\D_\alpha y^A\D_\beta x^B \right]\\
			&=\int d^2\sigma\left[ - \frac{\tilde c\zTeff}{2}\eta^{\alpha\beta}\D_\alpha x^i\D_\beta x^i + y^A\frac{\delta S_{\text{P-LO}}}{\delta x^A} \right] - \frac{\zTeff}{\tilde c}\int d\sigma^0\big[ y^v \D_1 x^v \big]_{\sigma^1=0}^{\sigma^1 = \pi}\,,
		\end{split}
	\end{equation}
	
	where we used that $x^t$ is a NN direction and so satisfies $x'^t = 0$ at the endpoints and where we dropped a total time derivative. In here $y^A$ denotes $y^t$ and $y^v$. The NLO equations of motion are
	\begin{equation}
		\eta^{\alpha\beta}\D_\alpha\D_\beta x^i = 0\,,\qquad \eta^{\alpha\beta}\D_\alpha\D_\beta x^A = 0\,,\qquad \eta^{\alpha\beta}\D_\alpha\D_\beta y^A = 0\,.
	\end{equation}
	In order that the on-shell variations vanish we must impose boundary conditions such that
	\begin{align}\label{eq:bdrytermsNLO}
		\begin{split}
			\eta_{AB}x'^A\delta y^B \big\vert_{\text{ends}}&=0\,,\\
			\eta_{AB}y'^A\delta x^B \big\vert_{\text{ends}}&=0\,,\\
			\delta_{ij}x'^i\delta x^j \big\vert_{\text{ends}}&=0\,.
		\end{split}
	\end{align}
	These follow either directly from varying~\eqref{eq:P-NLO-action-flat} or from expanding~\eqref{eq:rel-BC} to NLO. From these, we learn that $x^A$ and $y^A$ must satisfy the same boundary conditions, and so $y^v$ satisfies DD boundary conditions, while $y^t$ satisfies NN boundary conditions. The transverse scalars $x^i$ must satisfy either NN or DD boundary conditions, depending on the dimensionality of the nrD$p$-brane. As stated previously we will mostly work with a single nrD$24$-brane so that all the $x^i$ obey NN boundary conditions. We will assume that the location of the D$24$-brane does not depend on $c$ so that $y^v$ vanishes at the endpoints.

	Exactly as above, the mode expansions for the NLO fields $y^A$ and $x^i$ implied by their respective boundary conditions are
	\begin{align}
		\begin{split}
			y^t &= y^t_0 - \frac{1}{\pi \tilde c \zTeff}p_{(0)t}\sigma^0 + \frac{i}{\sqrt{\pi\tilde c\zTeff}}\sum_{k\neq 0}\frac{1}{k}\beta_k^te^{-ik\sigma^0}\cos(k\sigma^1)\,,\\
			y^v &=   - \frac{1}{\sqrt{\pi\tilde c\zTeff}}\sum_{k\neq 0}\frac{1}{k}\beta_k^ve^{-ik\sigma^0}\sin(k \sigma^1)\,,\\
			x^i &= x^i_0 + \frac{1}{\pi\tilde c \zTeff}p_{(0)i}\sigma^0 + \frac{i}{\sqrt{\pi\tilde c\zTeff}}\sum_{k\neq 0}\frac{1}{k}\alpha_k^i e^{-ik\sigma^0}\cos(k\sigma^1)\,,
		\end{split}
	\end{align}
	where, as in \cite{Hartong:2021ekg}, we do not assign winding to subleading fields. The new Virasoro constraints at NLO obtained by expanding~\eqref{eq:rel-Vir} as in~\eqref{eq:expansions} are
	\begin{equation}
		\label{eq:NLO-Virasoro}
		\D_+ y^- = \frac{\tilde c}{2w\zReff} \D_+ x^i \D_+ x^i\,,\qquad \D_- y^+ = \frac{\tilde c}{2w\zReff} \D_- x^i \D_- x^i\,,
	\end{equation}
	where
	\begin{equation}
		\D_\pm x^i = \tilde c^{-1}\zaeff p_{(0)i} + \frac{1}{2\sqrt{\pi\tilde c \zTeff}}\sum_{k\neq 0}\alpha^i_k e^{-ik\sigma^\pm}\,,
	\end{equation}
	and where the lightcone combinations $y^\pm$ are as in~\eqref{eq:lightcone-coords}.

	We have yet to fix the residual gauge tranformations at NLO, which (using the gauge fixed expressions for $x^\pm$ in~\eqref{eq:fully-gauge-fixed-LO}) act on $y^A$ as
	\begin{equation}
		\label{eq:NLO-residual-trafos}
		\delta_{\xi_{(2)}}y^- = 2w\tilde c^{-1}\zReff \xi^-_{(2)}(\sigma^-)\,,\qquad \delta_{\xi_{(2)}}y^+ = 2w\tilde c^{-1}\zReff \xi^+_{(2)}(\sigma^+)\,,
	\end{equation}
	where we used the expansion~\eqref{eq:2ddiffeoexp} of the residual gauge transformations~\eqref{eq:residual-gauge-trafos}. For $y^\pm$ we can write
	\begin{equation}
		\label{eq:NLO-mode-expansions}
		y^\pm = \tilde c^{-1}y^t_0 -  \frac{1}{2\pi\tilde c\zTeff}p_{(0)t}(\sigma^+ + \sigma^-) + \frac{i}{\sqrt{4\pi\tilde c \zTeff}}\sum_{k\neq 0}\frac{1}{k}\left( \beta^\pm_ke^{-ik\sigma^-} + \beta^\mp_k e^{-ik\sigma^+} \right) \,, 
	\end{equation}
	with $\beta_k^\pm = \beta^t_k\pm \tilde c^{-1}\beta^v_k$. As above, the boundary terms (the left hand side of \eqref{eq:bdrytermsNLO}) must vanish under the NLO residual gauge transformations, which in particular means that
	\begin{equation}
		\label{eq:NLO-residual-bdary}
		(x'^+\delta y^- + x'^-\delta y^+)\big\vert_{\text{ends}} = 0\,.
	\end{equation}
	Using \eqref{eq:fully-gauge-fixed-LO} and the transformations \eqref{eq:NLO-residual-trafos} this gives 
	\begin{equation}
		(\xi^-_{(2)}(\sigma^-)-\xi^+_{(2)}(\sigma^+))\big\vert_{\text{ends}} = 0\,,
	\end{equation}
	which, as above, tells us that $\xi^\pm_{(2)}(\sigma^\pm))$ are the same function (i.e., $\xi^+_{(2)}=\xi^-_{(2)}=\xi_{(2)}$) and that this function is $2\pi$-periodic. The mode expansions \eqref{eq:NLO-mode-expansions} thus tell us that we can use $\xi_{(2)}$ to set 
	\begin{equation}
		\beta^-_k = 0 \qquad \forall k\neq 0\,,
	\end{equation}
	leaving only $\beta^+$ and yielding the fully gauge-fixed mode expansions for the longitudinal $y^A$ fields,
	\begin{equation}
		\label{eq:gauge-fixed-mode-expansions-NLO}
		y^\pm = \tilde c^{-1}y^t_0 - \frac{1}{2\pi\tilde c \zTeff}p_{(0)t}(\sigma^+ + \sigma^-) +  \frac{i}{\sqrt{4\pi\tilde c\zTeff}}\sum_{k\neq 0}\frac{1}{k}\beta_k^+ e^{-ik\sigma^\mp}\,.
	\end{equation}
	
	The NLO Virasoro constraints \eqref{eq:NLO-Virasoro} as well as \eqref{eq:gauge-fixed-mode-expansions-NLO} imply that the (classical) NLO energy takes the form
	\begin{equation}
		E_{\text{NLO}} = -\int_0^\pi\diff\sigma^1\pd{\mathcal{L}_{\text{P-NLO}}}{\dot x^t}=-p_{(0)t}=\frac{\zaeff}{2w\zReff}(p_{(0)i})^2 + \frac{\tilde c N_{(0)}}{2w\zReff}\,,
	\end{equation}
	where 
	\begin{equation}
		N_{(0)} := \sum_{k=1}^\infty\alpha_{-k}^i\alpha_k^i\,,
	\end{equation}
	is the number operator which has the same dimension as $\hbar$.
	
	\subsection{NLO string endpoints as Bargmann particles}
	\label{sec:Bargmann-particles}
	It is well-known that the endpoints of relativistic strings ending on a D$25$-brane behave as massless particles. However, relativistic string endpoints ending on branes that are not spacetime-filling instead behave as \textit{massive} particles. To see this, we consider the Virasoro constraints~\eqref{eq:rel-Vir} 
	which take the form
	\begin{equation}
		\label{eq:vir-cons}
		c^2\eta_{AB}\dot X^A X'^B+X'^i\dot X^i = 0\,,\qquad c^2\eta_{AB}(\dot X^A \dot X^B + X'^A X'^B)+\dot X^i\dot X^i + X'^i X'^i = 0\,.
	\end{equation}
	We will evaluate these at the endpoints with DD boundary conditions for $X^v$, and NN boundary conditions for $X^i$ and $X^t$. At the endpoints the first of these constraints is identically satisfied while the second one gives
	\begin{equation}
		\left(-c^2\dot X^t\dot X^t+\frac{c^2}{\tilde c^2}X'^v X'^v+\dot X^i\dot X^i\right)\Big\vert_{\text{ends}}=0\,.
	\end{equation}
	The speed of the endpoint of the open string ending on a D24 brane is given by
	\begin{equation}
		\frac{\dot X^i}{\dot X^t}\frac{\dot X^i}{\dot X^t}\Big\vert_{\text{ends}} = c^2\left(1-\frac{1}{\tilde c^2}\left(\frac{X'^v}{\dot X^t}\Big\vert_{\text{ends}}\right)^2\right)\,,
	\end{equation}
	implying that the endpoints move at a speed slower than $c$ and therefore describe massive particles. The left-hand side is greater than or equal to zero and zero only when $\dot X^i=0$. In that case the string does not oscillate and the endpoint is not moving along the D24-brane. We then have $\frac{X'^v}{\dot X^t}=\tilde c$. In all other cases the left-hand side is positive and $\frac{X'^v}{\dot X^t}<\tilde c$ implying that the endpoint moves at a speed less than $c$.

	A similar argument would show that the endpoint of a relativistic open string ending on a D$p$-brane with $p<24$ also behaves like a massive particle. By studying the global symmetries of said open strings we can see that these are representations of the $(p+1)$-dimensional Poincar\'e algebra preserved by the flat worldvolume of the D$p$-brane. It is well-known how to expand massive particles in powers of $1/c^2$ (see, e.g., \cite{Hansen:2020pqs}) and that the result leads to a non-relativistic point particle moving in a Newton--Cartan geometry. Furthermore, we need that the open string can wind a compact direction in order to give it a rest energy with respect to which we can take a non-relativistic approximation. This direction is longitudinal to the string and thus transverse to the D-brane. Hence, to summarise, we consider the $1/c^2$ expansion of a relativistic open string ending on a D$p$-brane with $p\le 24$ and with a compact circle transverse to the brane that is wound by the open string.

	As we shall now demonstrate, for a flat target spacetime, the endpoints of an open non-relativistic string ending a nrD24-brane as described by~\eqref{eq:P-NLO-action-flat} behave as non-relativistic particles that are in a representation of the Bargmann algebra in $24+1$ dimensions. 
	
	For the case of a non-relativistic closed string the target space symmetries of the NLO theory~\eqref{eq:P-NLO-action-flat} were worked out in \cite{Hartong:2022dsx} and these are
	\begin{align}
		\label{eq:string-bargmann-trafos}
		\begin{split}
			\delta x^A &= \lambda^A_{(0)B}x^B + k^A_{(0)}\,,\\
			\delta x^i &= \lambda^i_{(0)j}x^j+ \lambda^i_{(0)A}x^A + k^i_{(0)}\,,\\
			\delta y^A &= \lambda^A_{(0)B}y^B + \lambda^A_{(2)B}x^B + \lambda^A_{(0)i}x^i + k^A_{(2)}\,,
		\end{split}
	\end{align}
	where $\lambda^A_{(0)B}=\eta^{AC}\lambda_{(0)CB}$ with $\lambda_{(0)CB}=-\lambda_{(0)BC}$ parametrise infinitesimal longitudinal Lorentz transformations, $k^A_{(0)}$ parameterise longitudinal translations, $\lambda^i_{(0)j}=-\lambda^j_{(0)i}$ parameterise infinitesimal transverse $\operatorname{SO}(24)$ rotations, $k^i_{(0)}$ parameterise infinitesimal transverse translations, and the $\lambda^i_{(0)A}$ are the parameters of infinitesimal ``stringy'' boosts, which mix transverse and longitudinal directions. These satisfy the relation
	\begin{equation}
		\lambda^A_{(0)i} = -\delta_{ij}\lambda^j_{(0)B}\eta^{AB}\,.
	\end{equation}
	The infinitesimal parameters $k^A_{(2)}$ and $\lambda^A_{(2)B}$ represent, respectively, subleading longitudinal translations and subleading longitudinal Lorentz transformations as they only act on the subleading embedding scalars $y^A$. As demonstrated in~\cite{Hartong:2022dsx}, the Noether charges associated to the global transformations in~\eqref{eq:string-bargmann-trafos} generate the string Bargmann algebra~\cite{Bergshoeff:2019pij} under the Poisson brackets of the NLO theory.\footnote{Actually, the NLO theory (ignoring the dilaton) has infinitely many symmetries~\cite{Bergshoeff:2019pij}, but these are an artefact of the truncation of the $1/c^2$ expansion. We consider here only those symmetries that follow from the $1/c^2$ expansion of the Poincar\'e symmetries of the relativistic theory or put differently that receive higher-order $1/c$ corrections.}

	So far we have reviewed what the global symmetries are for closed strings. Not all of the transformations~\eqref{eq:string-bargmann-trafos} preserve the boundary conditions that are satisfied by an open string ending on a nrD24-brane for which $x^v$ and $y^v$ both satisfy DD boundary conditions. Requiring that $\dot x^v\big\vert_{\text{ends}} = 0$ is stable under variations and that $x^v$ is equal to the \textit{same} constant at the respective endpoints after variation tells us that 
	\begin{equation}
		0 = \delta \dot x^v(\sigma^0,0) = \lambda^v_{(0)t}\dot x^t(\sigma^0,0)\,,\qquad 0 = k_{(0)}^v\,,
	\end{equation}
	where we picked the left endpoint. Since $\dot x^t(\sigma^0,0)\neq 0$ as it satisfies NN boundary conditions, we conclude that the longitudinal Lorentz transformations $\lambda^A_{(0)B}$ do not preserve the boundary conditions.

	Similarly, requiring that $\dot y^v\big\vert_{\text{ends}} = 0$ is stable under variations gives
	\begin{align}
		\begin{split}
			0 &= \delta \dot y^v(\sigma^0,0) = \lambda^v_{(2)t}\dot x^t(\sigma^0,0) + \lambda^v_{(0)i}\dot x^i(\sigma^0,0)\,,\\
			0 &= \delta \dot y^v(\sigma^0,\pi) = \lambda^v_{(2)t}\dot x^t(\sigma^0,\pi) + \lambda^v_{(0)i}\dot x^i(\sigma^0,\pi)\,,     
		\end{split}
	\end{align}
	which together tells us that $\lambda^v_{(2)t} = \lambda^v_{(0)i} = 0$ (to conclude that the latter is zero we must consider the boundary conditions at \textit{both} ends). Requiring that $y^v$ remains equal to zero at the respective endpoint after variation implies that $a_{(2)}^v = 0$. The transformations that preserve the boundary conditions are thus
	\begin{align}
		\label{eq:Bargmann-trafos}
		\begin{split}
			\delta x^t &= k^t_{(0)}\,,\\
			\delta x^v &= 0\,,\\
			\delta x^i &= \lambda^i{}_{j}x^j+ v^i x^t + k^i_{(0)}\,,\\
			\delta y^t &=   v^i x^i + k^t_{(2)}\,,\\
			\delta y^v &= 0\,,
		\end{split}
	\end{align}
	where we dropped the $(0)$ subscript on the rotation parameter and where we defined
	\begin{equation}
		v^i=\lambda^i_{(0)t}=\lambda^t_{(0)i}\,.
	\end{equation}
	The associated Noether charges generate the Bargmann algebra, as we shall now demonstrate.

	The Noether currents to which the transformations~\eqref{eq:Bargmann-trafos} give rise are
	\begin{align}
		\begin{aligned}
			k_{(0)}^t&:\\
			k_{(2)}^t&:\\
			k_{(0)}^i&:\\
			\lambda_{(0)ij}&:\\
			\lambda_{(0)it}&:
		\end{aligned}
		\qquad    
		\begin{aligned}
			\pi^\alpha_{(0)t}  &= \tilde c\zTeff \D^\alpha y^t\,,\\
			\pi^\alpha_{(-2)t}  &= \tilde c\zTeff \D^\alpha x^t\,,\\
			\pi^\alpha_{(0)t}  &= -\tilde c\zTeff \D^\alpha x^i\,,\\
			j^\alpha_{(0)ij} &= 2x_{[i} \pi^\alpha_{(0)j]}  \,,\\
			j^\alpha_{(0)ti} &= -x^t \pi^\alpha_{(0)i} -x_i \pi^\alpha_{(-2)t}  \,.
		\end{aligned}
	\end{align}
	The notation $\pi_{(n)M}^\alpha$ refers to the coefficient of $c^{-n}$ in the $1/c^2$ expansion of $\pi^\alpha_M=\frac{\partial\mathcal{L}_{\text{P}}}{\partial\partial_\alpha x^M}$ of the relativistic Polyakov string Lagrangian. More explicitly we have
	\begin{eqnarray}
		\pi^\alpha_{(-2)A} & = & \frac{\partial\mathcal{L}_{\text{P-LO}}}{\partial\partial_\alpha x^A}\,,\\
		\pi^\alpha_{(0)A} & = & \frac{\partial\mathcal{L}_{\text{P-NLO}}}{\partial\partial_\alpha y^A}\,,\\
		\pi^\alpha_{(0)i} & = & \frac{\partial\mathcal{L}_{\text{P-NLO}}}{\partial\partial_\alpha x^i}\,.
	\end{eqnarray}
	The equal-$\sigma^0$ Poisson brackets are~\cite{Hartong:2022dsx}
	\begin{align}
		\label{eq:NLO-poisson-brackets}
		\begin{split}
			\{ x^A(\sigma^1),\pi^0_{(0)B}(\tilde\sigma^1) \} &= \delta^A_B\delta(\sigma^1 - \tilde\sigma^1)\,,\\
			\{ y^A(\sigma^1),\pi^0_{(-2)B}(\tilde\sigma^1) \} &= \delta^A_B\delta(\sigma^1 - \tilde\sigma^1)\,,\\
			\{ x^i(\sigma^1),\pi^0_{(0)j}(\tilde\sigma^1) \} &= \delta^i_j\delta(\sigma^1 - \tilde\sigma^1)\,.
		\end{split}
	\end{align}
	The Noether charges are given by
	\begin{align}
		\label{eq:Bargmann-charges}
		\mathcal{P}_{(-2)t} &= \int_0^\pi d\sigma^1\,\pi^0_{(-2)t}   \,,&
		\mathcal{P}_{(0)i} &= \int_0^\pi d\sigma^1\,\pi^0_{(0)i} \,,\nn\\
		\mathcal{P}_{(0)t} &= \int_0^\pi d\sigma^1\,\pi^0_{(0)t}\,,&
		\mathcal{J}_{(0)ij} &=  \int_0^\pi d\sigma^1\,j^0_{(0)ij} \,,\\
		\mathcal{J}_{(0)ti} &=  \int_0^\pi d\sigma^1\,j^0_{(0)ti} \,.&\nn
	\end{align}
	Using the NLO Poisson brackets~\eqref{eq:NLO-poisson-brackets}, these charges generate the Bargmann algebra
	\begin{align}
		\begin{split}
			\{ \mathcal{P}_{(0)i},\mathcal{J}_{(0)tj} \} &= \delta_{ij}\mathcal{P}_{(-2)t}\,,\\
			\{ \mathcal{J}_{(0)ti},\mathcal{P}_{(0)t} \} &= -\mathcal{P}_{(0)i}\,,\\
			\{\mathcal{J}_{(0)ij},\mathcal{P}_{(0)k} \} &= 2\delta_{k[i}\mathcal{P}_{(0)j]}\,,\\
			\{\mathcal{J}_{(0)ij}, \mathcal{J}_{(0)tk} \} &= 2\delta_{k[i}\mathcal{J}_{(0)t\vert j]}\,,\\
			\{\mathcal{J}_{(0)ij}, \mathcal{J}_{(0)kl} \} &= \delta_{ik}\mathcal{J}_{(0)jl} -\delta_{il}\mathcal{J}_{(0)jk} - \delta_{jk}\mathcal{J}_{(0)il} + \delta_{jl}\mathcal{J}_{(0)ik}\,.
		\end{split}
	\end{align}
	In other words, as is well known, the brane spontaneously breaks translation invariance in the $v$-direction, which here breaks the string Bargmann symmetries down to Bargmann. There will be a Goldstone mode associated with this spontaneous breaking of translation symmetry, which corresponds to fluctuations in the position of the brane; we will have more to say about this in the next section.
	
	\section{Non-linear Galilean electrodynamics and nrD\texorpdfstring{$24$}{24}-branes}
	\label{sec:NL-GED}
	In this section, we discuss the inclusion of worldvolume fields on the nrD24-brane. For the open string action we will do this both from an intrinsic non-relativistic perspective as well as from the $1/c^2$ expansion. We will also study the $1/c^2$ expansion of the DBI action of a D24-brane in a flat target spacetime, while postponing the general case to the final section which also includes an intrinsically non-relativistic derivation of the action for a nrD$p$-brane in a string Newton--Cartan target spacetime with a (leading order critical) $B$-field. 
	
	We will derive the properties of the non-relativistic worldvolume gauge fields from the $1/c^2$ expansion of the relativistic world-volume gauge fields and we will also derive their properties directly from the action of a non-relativistic string ending on a nrD24-brane with worldvolume fields turned on.

	We will find that the $1/c^2$ expansion of the DBI action for a D$24$-brane at order ${\alpha'}^{2}_{\text{NR}}$ gives rise to Galilean electrodynamics (GED)~\cite{LeBellac:1973,Santos:2004pq,GEDreview,Bagchi:2014ysa,Festuccia:2016caf} whilst we find that to all orders in $\alpha'_{\text{NR}}$ we obtain a non-linear version of GED that agrees with the results of \cite{Gomis:2020fui} but for which we write down a novel (and more explicit) form of the action.

	\subsection{Target space \texorpdfstring{$B$}{B}-field and worldvolume gauge field}
	\label{sec:critical-B-field}

	Let us now include a $2$-form Kalb-Ramond $B$-field, which in the relativistic parent theory is achieved by adding the Wess--Zumino action 
	\begin{equation}
		\label{eq:WZ-action}
		S_{\text{WZ}} =  \frac{cT}{2}\int d^2\sigma\, \varepsilon^{\alpha\beta}\D_\alpha X^M \D_\beta X^N B_{MN}(X)\,,
	\end{equation}
	to the Polyakov action~\eqref{eq:Polyakov-action-rel}. We remind the reader that in our conventions $\epsilon^{01}=-1$. The $B$-field is assumed to have a $1/c^2$ expansion of the form
	\begin{equation}
		\label{eq:expansion-of-B-field}
		B_{MN} = c^2 B_{(-2)MN} + B_{(0)MN} + \mathcal{O}(c^{-2})\,.
	\end{equation}
	In order to obtain non-relativistic string theory as a strict $c\rightarrow \infty$ limit of relativistic string theory as in~\cite{Gomis:2000bd,Danielsson:2000gi,Bergshoeff:2018yvt}, the divergent LO action in the expansion~\eqref{eq:expansions} must be removed. On a flat target space, this is achieved by introducing a ``critical'' $B$-field
	\begin{equation}\label{eq:criticalB}
		B = -\frac{c^2}{\tilde c}dt\wedge dv \,,
	\end{equation}
	corresponding to
	\begin{equation}
		\label{eq:critical-B-field}
		B_{(-2)MN} = -2 \tilde c^{-1}\delta^t_{[M}\delta^v_{N]}\,,\qquad B_{(0)MN} = 0\,.
	\end{equation}
	In flat target space, this choice trivialises the LO Polyakov action by making the sum of the LO Wess--Zumino action and $S_{\text{P-LO}}$ as defined in~\eqref{eq:expansions} proportional to the LO Virasoro constraints. In particular, this removes the LO contribution to the energy~\eqref{eq:LO-energy}. For more details, we refer to~\cite{Hartong:2021ekg,Hartong:2022dsx}. We note that \eqref{eq:criticalB} has vanishing curvature and so locally can be gauged away. However, this is not possible globally due to the fact that $v$ corresponds to a circle direction.

	Consider now the relativistic theory. It is well known that, in contrast to closed strings, open string boundary conditions imply that the Wess--Zumino action~\eqref{eq:WZ-action} fails to be gauge invariant under $1$-form gauge transformations with parameter $\Omega_M(X)$ of the Kalb--Ramond $2$-form
	\begin{equation}
		\label{eq:1-form-gauge-sym}
		\delta B_{MN}(X) = \pd{\Omega_{N}(X)}{X^M} - \pd{\Omega_M(X)}{X^N}\,,
	\end{equation}
	since the action transforms (up to a total time derivative) into a boundary term
	\begin{equation}
		\begin{split}
			\label{eq:WZ-bdary-term}
			\delta S_{\text{WZ}} &= cT \int d^2\sigma \left( \D_1 (\Omega_M \D_0 X^M) - \D_0(\Omega_M \D_1 X^M ) \right)\\
			&= cT \int d\sigma^0\Big [ \Omega_{a} \D_0 X^{a} \Big]_{\sigma^1 =0}^{\sigma^1=\pi}\,,
		\end{split}
	\end{equation}
	where, as above, we split $M = (a,I)$, where $a$ runs over the NN directions, while $I$ runs over the DD directions. The latter do not contribute to the boundary term. Note also that $\Omega_a$ as defined above does not depend on the DD directions. To counter this non-invariance, we must couple the string to a worldvolume gauge field $A_{a}(X^{b})$ 
	\begin{equation}\label{eq:bdryaction}
		S_A =  \int d\sigma^0\Big [ A_{a} \D_0 X^{a} \Big]_{\sigma^1=0}^{\sigma^1=\pi}\,,
	\end{equation}
	where $A_{a}$ transforms under the $1$-form gauge symmetry so as to cancel the boundary term in~\eqref{eq:WZ-bdary-term} as well as under $U(1)$ gauge transformations 
	\begin{equation}
		\label{eq:U(1)-trafo}
		\delta A_{a} = \D_{a}\Lambda -cT\,\Omega_{a}\,, 
	\end{equation}
	where the parameters $\Lambda$ and $\Omega_{a}$ only depend on the NN directions $X^{a}$.  The dimensions of the boundary term are fixed by requiring that the potential $A_0$ is an energy, which means that $[A_{a}]_{a = 1,\dots, p} = \text{mass}\times\text{length}/\text{time}$. The object $F_{a b} = \D_{a} A_{b} - \D_{b} A_{a}$ is not gauge invariant under $1$-form gauge transformations
	\begin{equation}
		\delta_\Omega F_{a b} = -cT\,\left(\D_{a} \Omega_{b} - \D_{b}\Omega_{a}\right) = - cT\delta_\Omega B_{ab}\,,
	\end{equation}
	implying that we may construct the following gauge invariant field strength
	\begin{equation}
		\label{eq:gauge-inv-combi}
		\mathcal{F}_{ab } = B_{a b}+\frac{1}{cT}F_{a b} \,.
	\end{equation}

	\subsection{Expansion of relativistic open strings with worldvolume fields}
	\label{sec:open-strings-WV-fields}
	
	We now consider the $1/c^2$ expansion of open relativistic strings with a critical $B$-field~\eqref{eq:critical-B-field} in the presence of worldvolume fields. We will consider both the Nambu--Goto and Polyakov formulations, culminating in a rewriting of the Polyakov action to the Gomis--Ooguri form. When the relativistic string ends on a D24-brane, with worldvolume fields turned on, the embedding scalar for the transverse $v$-direction includes a contribution from the transverse fluctuations of the brane described by a worldvolume scalar $\Phi$, i.e.,
	\begin{equation}\label{eq:Xvondynamicalbrane}
		X^v(\sigma^0,\sigma^1) =  X_0^v(\sigma^0,\sigma^1) + \frac{\tilde c}{cT}\Phi(X^\mu(\sigma^0,\sigma^1))\,,
	\end{equation}
	where $\Phi(X^\mu(\sigma^0,\sigma^1))$ is a function of the directions longitudinal to the brane with dimensions of mass, and where $X^v_0$ obeys DD boundary conditions as well as any potential winding conditions.

	The Nambu--Goto action for a relativistic open string ending on a D24-brane with worldvolume fields turned on (where we assume a flat target space metric and a critical $B$-field \eqref{eq:criticalB}) is given by the sum of \eqref{eq:Rel-NG} and \eqref{eq:WZ-action} (evaluated for a critical $B$-field) to which we add \eqref{eq:bdryaction} and in which we take for $X^v$ the expression \eqref{eq:Xvondynamicalbrane}. Explicitly, the action is
	\begin{equation}
		\label{eq:rel-NG-action}
		\begin{split}
			\hspace{-5cm} S_{\text{NG}} &= -cT\int_\Sigma d^2\sigma\,\sqrt{-\det\left(\eta_{\mu\nu}\D_\alpha X^\mu\D_\beta X^\nu + \frac{c^2}{\tilde c^2}\D_\alpha X^v \D_\beta X^v \right)}\\
			&\quad+ \frac{c^3}{\tilde c}T\int_\Sigma d^2\sigma\, \left( \dot X^t X'^v  - X'^t \dot X^v\right)+ \int d\sigma^0\,A_\mu \dot X^\mu \bigg\vert_{\sigma^1=0}^{\sigma^1=\pi}  \,,
		\end{split}
	\end{equation}
	where $X^\mu=(X^t,X^i)$ and where $X^v$ is given by \eqref{eq:Xvondynamicalbrane}. The corresponding Polyakov action is given by 
	\begin{equation}
		\label{eq:Polyakov-action-rel}
		\begin{split}
			S_{\text{P}} &=  - \frac{cT}{2}\int_\Sigma d^2\sigma\,\sqrt{-\gamma}\gamma^{\alpha\beta}\big[\eta_{\mu\nu}\D_\alpha X^\mu\D_\beta X^\nu + \frac{c^2}{\tilde c^2}\D_\alpha  X^v \D_\beta  X^v \big]\\
			&\quad + \frac{c^3}{\tilde c}T\int_\Sigma d^2\sigma\, \left( \dot X^t X'^v  - X'^t \dot X^v\right)+ \int d\sigma^0\,A_\mu \dot X^\mu \bigg\vert_{\sigma^1=0}^{\sigma^1=\pi}\,.
		\end{split}
	\end{equation}
	We will next expand both of these two actions in $1/c^2$. We expand the worldsheet metric and the embedding fields according to
	\begin{equation}
		\begin{split}
			\gamma_{\alpha\beta}(\sigma) &= \gamma_{(0)\alpha\beta}(\sigma) + c^{-2}\gamma_{(2)\alpha\beta}(\sigma) + \mathcal{O}(c^{-2})\,,\\
			X^M(\sigma) &= x^M(\sigma) + c^{-2}y^M(\sigma) + \mathcal{O}(c^{-4})\,,
		\end{split}
	\end{equation}
	where $M = (\mu,v)$ and where $\gamma_{(0)\alpha\beta}(\sigma)$ is assumed to be a Lorentzian worldsheet metric. The worldvolume fields will be expanded as
	\begin{subequations}
		\label{eq:GED-exp-1}
		\begin{align}
			\begin{split}
				\label{eq:expPhi}
				\Phi(X) & =  \phi(X)+c^{-2}\chi(X)+\mathcal{O}(c^{-4})\\
				& =  \phi(x)+c^{-2}\left(\chi(x)+y^\rho\partial_\rho\phi(x)\right)+\mathcal{O}(c^{-4})\,,
			\end{split}
			\\
			A_\mu (X) & =  a_\mu(X)+\mathcal{O}(c^{-2}) = a_\mu(x)+\mathcal{O}(c^{-2})\,, \label{eq:exp-of-A}
		\end{align}
	\end{subequations}
	where the second equality is due to the expansion of the embedding scalars. Using equation \eqref{eq:Xvondynamicalbrane} we then find for the expansion of $X^v$ that
	\begin{equation}
		X^v=x^v+c^{-2}y^v+\mathcal{O}(c^{-4})\,,
	\end{equation}
	with
	\begin{eqnarray}
		x^v & = & x^0_v+\frac{1}{T_{\text{NR}}}\phi(x^\mu)\,,\label{eq:xvplusphi}\\
		y^v & = & \frac{1}{T_{\text{NR}}}\left(\chi(x^\mu)+y^\rho\partial_\rho\phi(x^\mu)\right)\,.
	\end{eqnarray}
	The $1/c^2$ expansion of the relativistic Nambu--Goto action~\eqref{eq:rel-NG-action} is
	\begin{equation}
		S_{\text{NG}} = c^2S_{\text{NG-LO}} + S_{\text{NG-NLO}} + \mathcal{O}(c^{-2})\,,
	\end{equation}
	where, as discussed in Sec.~\ref{sec:critical-B-field} the results of~\cite{Hartong:2022dsx} imply that the LO Nambu--Goto action is identically zero due to the critical $B$-field.\footnote{We assume that $\dot X^t X'^v-X'^t \dot X^v>0$.} This is still true for open strings so that we have
	\begin{equation}
		S_{\text{NG-LO}} = 0\,.
	\end{equation}

	The NLO Nambu--Goto action is
	\begin{equation}\label{eq:NLONG}
		S_{\text{NG-NLO}} =-\frac{\tilde c\zTeff}{2} \int_\Sigma d^2\sigma\, \sqrt{-\tau}\tau^{\alpha\beta}\D_\alpha x^i\D_\beta x^i  + \int d\sigma^0\,a_\mu\dot x^\mu\Big\vert_{\sigma^1=0}^{\sigma^1 = \pi}\,,
	\end{equation}
	where the terms involving $(y^t, y^v)$ cancel out due to the critical $B$-field, and where $\tau^{\alpha\beta}$ is the inverse of
	\begin{equation}
		\tau_{\alpha\beta} = - \D_\alpha x^t \D_\beta x^t + \tilde c^{-2}\D_\alpha x^v \D_\beta x^v\,.
	\end{equation}
	In particular, we notice that this action depends on the LO worldvolume fields $\phi$, and $a_\mu$ but that NLO worldvolume fields such as $\chi$, which feature in the expansion of the transverse-fluctuation field~\eqref{eq:expPhi}, do not appear. The action \eqref{eq:NLONG} can be written as
	\begin{eqnarray}
		S_{\text{NG-NLO}}&  = & \frac{\zTeff}{2} \int_\Sigma d^2\sigma\frac{1}{\dot x^t x'^v-x'^t \dot x^v}\left[\left(x'^v \dot x^i-\dot x^v x'^i\right)^2-\tilde c^2\left(\dot x^t x'^i-x'^t\dot x^i\right)^2\right] \nonumber\\
		&&+ \int d\sigma^0\,a_\mu\dot x^\mu\Big\vert_{\sigma^1=0}^{\sigma^1 = \pi}\,,\label{eq:NLONG2}
	\end{eqnarray}
	which is the form of the action that we will use below when we discuss the action of the Bargmann transformations of section \ref{sec:Bargmann-particles}.

	The Polyakov action~\eqref{eq:Polyakov-action-rel} has an expansion of the form
	\begin{equation}
		S_{\text{P}} = c^2S_{\text{P-LO}} + S_{\text{P-NLO}} + \mathcal{O}(c^{-2})\,.
	\end{equation}
	At leading order we get
	\begin{equation}
		\label{eq:LO-P-not-gauge-fixed}
		S_{\text{P-LO}} = \frac{\tilde c\zTeff}{2}\int_\Sigma d^2\sigma\,\sqrt{-\gamma_{(0)}}\gamma^{\alpha\beta}_{(0)}\big [ \D_\alpha x^t \D_\beta x^t - \tilde c^{-2}\D_\alpha  x^v \D_\beta  x^v \big] + \zTeff \int_\Sigma d^2\sigma\,\left(\dot x^t x'^v-x'^t\dot x^v\right)\,,
	\end{equation}
	where $x^v$ is given by \eqref{eq:xvplusphi}.

	The LO Virasoro constraints are obtained by integrating out $\gamma_{(0)\alpha\beta}$, which we decompose in terms of worldsheet vielbeine as
	\begin{equation}
		\label{eq:WS-vielbeine}
		\gamma_{(0)\alpha\beta} = \eta_{ab}e_\alpha^a e_\beta^b\,,
	\end{equation}
	where $a,b=0,1$ are Lorentzian worldsheet tangent space indices. It is useful to define the lightcone combinations
	\begin{equation}\label{eq:WS-vielbeine2}
		e_\alpha^\pm = e_\alpha^0 \pm e_\alpha^1\,.
	\end{equation}
	The LO Polyakov action~\eqref{eq:LO-P-not-gauge-fixed} can be written as a combination of the Virasoro constraints
	\begin{equation}
		S_{\text{P-LO}} = -2\tilde c\zTeff \int_\Sigma d^2\sigma\,\sqrt{-\gamma_{(0)}} (e^\alpha_+ \D_\alpha x^-)(e^\alpha_- \D_\alpha  x^+)\,,
	\end{equation}
	where as before we defined
	\begin{equation}
		x^\pm =  x^t \pm \tilde c^{-1} x^v\,,
	\end{equation}
	and where we used the identity
	\begin{equation}
		\label{eq:epsilon-identity}
		\varepsilon^{\alpha\beta} = \sqrt{-\gamma_{(0)}}e^\alpha_a e^\beta_b \varepsilon^{ab}\,.
	\end{equation}
	Finally, we consider the NLO Polyakov action which will be equivalent to $S_{\text{NG-NLO}}$ upon integrating out the LO worldsheet metric $\gamma_{(0)\alpha\beta}$ and its NLO correction $\gamma_{(2)\alpha\beta}$. However, rather than writing the NLO Polyakov action in terms of $\gamma_{(2)\alpha\beta}$ we will prefer to write it in ``Gomis--Ooguri'' form. To this end we express the subleading worldsheet metric in vielbein components as
	\begin{equation}
		\gamma_{(2)\alpha\beta} =   e_\alpha ^a e_\beta^b A_{ab}\,,
	\end{equation}
	where $A_{ab} = A_{ba}$. We can then use the  results of~\cite{Hartong:2022dsx}, to express the NLO Polyakov action describing a non-relativistic open string ending on a nrD24-brane as
	\begin{equation}
		\label{eq:NLO-GO-action}
		\begin{split}
			S_{\text{P-NLO}} &= -\frac{\tilde c\zTeff}{2}\int_\Sigma d^2\sigma\,  \left( \sqrt{-\gamma_{(0)}}\gamma_{(0)}^{\alpha\beta} \D_\alpha x^i \D_\beta x^i +\lambda_{++}\varepsilon^{\alpha\beta}e_\alpha^+ \D_\beta  x^+ + \lambda_{--} \varepsilon^{\alpha\beta} e_\alpha^- \D_\beta  x^- \right) \\
			&\quad+ \int d\sigma^0\,a_\mu \dot x^\mu \Big\vert_{\sigma^1=0}^{\sigma^1 = \pi}\,,
		\end{split}
	\end{equation}
	where $\lambda_{++}$ and $\lambda_{--}$ are Lagrange multipliers given by
	\begin{equation}
		\begin{split}
			\lambda_{++} &= -4e^\alpha_-\D_\alpha  x^- A_{++} - 2 e^\alpha_+ \D_\alpha  y^-\,,\\
			\lambda_{--} &= 4 e^\alpha_+ \D_\alpha  x^+ A_{--} + 2  e^\alpha_- \D_\alpha  y^+\,.
		\end{split}
	\end{equation}
	In here we defined 
	\begin{equation}
		y^\pm = y^t \pm  \tilde c^{-1}y^v\,,
	\end{equation}
	and we assumed that $e^\alpha_-\D_\alpha  x^-$ and $e^\alpha_+\D_\alpha  x^+$ are nonzero. This form of the NLO Polyakov action is what we refer to as the Gomis--Ooguri form~\cite{Gomis:2000bd,Hartong:2022dsx}.

	The form of the action \eqref{eq:NLONG2} is useful to study the fate of the global symmetries \eqref{eq:Bargmann-trafos} once we turn on the worldvolume fields $\phi, a_\mu$. When we act on the embedding scalars $x^\mu$ the fields $\phi, a_\mu$ which depend on $x^\mu$, transform as 
	\begin{equation}
		\delta_x\phi=\delta x^\rho\partial_\rho\phi\,,\qquad \delta_x a_\mu=\delta x^\rho\partial_\rho a_\mu\,.
	\end{equation}
	Under the variation $\delta x^\mu$, as given in \eqref{eq:Bargmann-trafos}, the action \eqref{eq:NLONG2} transforms as
	\begin{equation}\label{eq:trafoNLONG}
		\delta_x S_{\text{NG-NLO}}=\delta_\phi S_{\text{NG-NLO}}+\int d\sigma^0\,\dot x^\mu\left(\mathcal{L}_\xi a_\mu+\delta_\mu^i v^i\phi\right)\Big\vert_{\sigma^1=0}^{\sigma^1 = \pi}\,,
	\end{equation}
	up to a total time derivative, where $\xi^\mu=\delta x^\mu$ and where the first term on the right-hand side is the variation of $S_{\text{NG-NLO}}$ with respect to a diffeomorphism acting on $\phi$, i.e., where $\delta\phi=\mathcal{L}_\xi \phi$. 
	The last term in \eqref{eq:trafoNLONG} is due to the fact that the kinetic term in \eqref{eq:NLONG2} is not Galilean boost invariant but rather transforms into a boundary term. We thus see that the variation of the embedding scalars is equivalent to not varying the $x^\mu$, but to instead transform the worldvolume fields as follows
	\begin{equation}
		\begin{split}
			\delta\phi & =  \mathcal{L}_\xi \phi\,,\\
			\delta a_\mu & =  \mathcal{L}_\xi a_\mu+\delta_\mu^i v^i\phi\,,
		\end{split}
	\end{equation}where for convenience we repeat here that
	\begin{equation}
		\xi^t= k^t_{(0)}\,,\qquad
		\xi^i = \lambda^i{}_{j}x^j+ v^i x^t + k^i_{(0)}\,.
	\end{equation}
	The pair $(\phi, a_\mu)$ transform exactly as the fields of Galilean electrodynamics (GED) \cite{LeBellac:1973,Santos:2004pq,GEDreview,Bagchi:2014ysa,Festuccia:2016caf}. We thus expect that the action of the nrD24-brane is given by some non-linear version of GED. In the next subsection we will see that this is indeed the case. We mention that the GED fields only transform under the Galilei algebra and not under the Bargmann algebra as the mass generator (with parameter $k^t_{(2)}$) has a trivial action on $x^\mu$.

	\subsection{Worldvolume theory of a nrD\texorpdfstring{$24$}{24}-brane}
	\label{sec:expansion-of-DBI}
	
	In this section we will derive the worldvolume theory of a nrD24-brane by $1/c^2$ expanding the relativistic DBI action for a D24-brane in a flat target space with a critical $B$-field. In Section~\ref{sec:DBI-from-T-duality-covariance} we will derive the same action from an intrinsic non-relativistic perspective using invariance under certain target space gauge symmetries and by demanding covariance under transverse T-duality.

	The DBI action for a relativistic D$24$-brane on a flat target space described by the metric~\eqref{eq:flat-metric} and with a critical $B$-field~\eqref{eq:critical-B-field} is given by
	\begin{equation}
		\label{eq:DBI-flat}
		S_{\text{D}24} = - cT_{\text{D24}}\int d^{25}\sigma\, \sqrt{-\det M_{\hat\alpha\hat\beta}}\,,
	\end{equation}
	where $T_{\text{D24}}$ is the tension of the D$24$-brane which has dimensions of mass density, i.e., $ML^{-24}$, and where we defined
	\begin{equation}
		\label{eq:def-of-M}
		M_{\hat\alpha\hat\beta} :=  \frac{\partial X^M}{\partial\sigma^{\hat\alpha}}\frac{\partial X^N}{\partial\sigma^{\hat\beta}} \eta_{MN} + B_{\hat\alpha\hat\beta} + \frac{1}{cT}F_{\hat\alpha\hat\beta} \,,
	\end{equation}
	where $T$ is the fundamental string tension and where $\hat\alpha,\hat\beta=0,1,\ldots,24$. At this point, we emphasise that the coordinates $\sigma^{\hat\alpha}$ on the worldvolume are \textit{dimensionful}, in contrast to the coordinates on the string worldsheet. In static gauge, which we will use in this section, we have
	\begin{equation}
		\label{eq:brane-monge-gauge}
		\sigma^0 = X^t\,,\qquad \sigma^i = X^i\,,
	\end{equation}
	with $i=1,\ldots,24$. In static gauge the distinction between $\sigma^{\hat\alpha}$ and $X^a$ is gone and for ease of notation we will continue to denote worldvolume indices by $a,b$.
	
	Using \eqref{eq:Xvondynamicalbrane} and static gauge the pullback of $\eta_{MN}$ becomes
	\begin{equation}
		\label{eq:pullback-mink}
		\frac{\partial X^M}{\partial\sigma^{a}}\frac{\partial X^N}{\partial\sigma^{b}} \eta_{MN} = -c^2\delta^t_{a}\delta^t_{b}+\delta^i_{a}\delta^i_{b} +  \frac{c^2}{\tilde c^2\zTeff^2}\D_{a} \Phi \D_{b} \Phi \,,
	\end{equation}
	where $\Phi$ describes fluctuations in the position of the D24-brane. The pullback of the critical $B$-field is given by
	\begin{equation}
		B_{ab}=\frac{\partial X^M}{\partial\sigma^{a}}\frac{\partial X^N}{\partial\sigma^{b}} B_{MN}=\frac{c^2}{\tilde c T_{\text{NR}}}\left(\delta^t_{b}\partial_{a}\Phi-\delta^t_{a}\partial_{b}\Phi\right)\,.
	\end{equation}
	Combining all our results we find that
	\begin{equation}
		M_{ab}=-c^2 T^-_{a}T^+_{b}+\delta^i_{a}\delta^i_{b}+\frac{1}{\tilde c T_{\text{NR}}}F_{ab}\,,
	\end{equation}
	where we defined
	\begin{equation}
		T^\pm_{a}=\delta^t_{a}\pm \frac{1}{\tilde c T_{\text{NR}}}\partial_{a}\Phi\,.
	\end{equation}
	In order to expand the action \eqref{eq:DBI-flat} the following identity is useful
	\begin{equation}
		\label{eq:useful-identity-T-duality}
		\text{det}\left(-c^2 T^-_{a} T^+_{b} + X_{ab}\right) = c^2\text{det}\left(\begin{array}{cc}
			c^{-2} ~&~ T^+_{b}\\
			T^-_{a} ~&~ X_{ab}
		\end{array}\right)\,,
	\end{equation}
	where $X_{ab}$ is any matrix. Using the expansion of the worldvolume fields \eqref{eq:expPhi} and \eqref{eq:exp-of-A} we find that at LO the action \eqref{eq:DBI-flat} is given by
	\begin{equation}
		\label{eq:rel-D24}
		S_{\text{D}24} = - c^2T_{\text{D24}}\int d^{25}\sigma\, \left(\sqrt{-\det \mathcal{M}}+\mathcal{O}(c^{-2})\right)\,,
	\end{equation}
	where we defined
	\begin{equation}
		\mathcal{M}=\left(\begin{array}{cc}
			0 ~&~ \tau^+_{b}\\
			\tau^-_{a} ~&~ \delta^i_{a}\delta^i_{b}+\frac{1}{\tilde c T_{\text{NR}}}f_{ab}
		\end{array}\right)\,,
	\end{equation}
	with
	\begin{equation}
		\tau^\pm_{a}=\delta^t_{a}\pm \frac{1}{\tilde c T_{\text{NR}}}\partial_{a}\phi\,,\qquad f_{ab}=\partial_{a}a_{b}-\partial_{b}a_{a}\,.
	\end{equation}
	The LO action in $S_{\text{D}24}$ in Eq.~\eqref{eq:rel-D24} is, neglecting the dilaton, the action in Eq.~(4.6) in~\cite{Gomis:2020fui}.
	
	In order to compute the determinant of $\mathcal{M}$ we will write it as
	\begin{equation}
		\mathcal{M}=\left(\begin{array}{cc}
			A ~&~ B\\
			C ~&~ D
		\end{array}\right)\,,
	\end{equation}
	where $A$ is a $2\times 2$ matrix, $B$ a $2\times 24$ matrix, $C$ a $24\times 2$ matrix and $D$ an invertible $24\times 24$ matrix that are given by
	\begin{subequations}
		\begin{eqnarray}
			A & = & \left(\begin{array}{cc}
				0 ~&~ \tau^+_t\\
				\tau^-_t ~&~ 0
			\end{array}\right)\,,\\
			B & = & \left(\begin{array}{c}
				\frac{1}{\tilde c T_{\text{NR}}}\partial_j\phi \\
				\frac{1}{\tilde c T_{\text{NR}}}f_{tj}
			\end{array}\right)\,,\\
			C & = & \left(\begin{array}{cc}
				-\frac{1}{\tilde c T_{\text{NR}}}\partial_i\phi ~&~ -\frac{1}{\tilde c T_{\text{NR}}}f_{ti}
			\end{array}\right)\,,\\
			D & = & \left(\delta_{ij}+\frac{1}{\tilde c T_{\text{NR}}}f_{ij}\right)\,.\label{eq:D}
		\end{eqnarray}  
	\end{subequations}
	We then use the identity
	\begin{equation}\label{eq:detid}
		\det\left(\begin{array}{cc}
			A ~&~ B\\
			C ~&~ D
		\end{array}\right)=\text{det}\, D\,\text{det}\left(A-B D^{-1}C\right)\,,
	\end{equation}
	to find
	\begin{eqnarray}
		\det \mathcal{M} & = & \det\left(\delta_{ij}+\frac{1}{\tilde c T_{\text{NR}}}f_{ij}\right)\left(-\tau^+_t\tau^-_t+\frac{1}{\tilde c^2 T^2_{\text{NR}}}\tau^+_tD^{-1}_{kl}f_{tk}\partial_l\phi+\frac{1}{\tilde c^2 T^2_{\text{NR}}}\tau^-_tD^{-1}_{kl}f_{tl}\partial_k\phi\right.\nonumber\\
		&&\left.+\frac{1}{\tilde c^4 T^4_{\text{NR}}}\left(D^{-1}_{kl}D^{-1}_{mn}-D^{-1}_{km}D^{-1}_{ln}\right)f_{tk}f_{tl}\partial_m\phi\partial_n\phi\right)\,,
	\end{eqnarray}
	where $D^{-1}$ is the inverse of the matrix \eqref{eq:D}. For the D24-brane action~\eqref{eq:rel-D24} at LO we then obtain
	\begin{equation}\label{eq:NLGED}
		\begin{split}
			S_{\text{nrD24}} &= -T_{\text{D}24}c^2\int d^{25}\sigma \sqrt{\det\left( \delta_{ij} + \frac{1}{\tilde c\zTeff} f_{ij} \right)}\times\left[1-\frac{1}{\tilde c^2 T^2_{\text{NR}}}\left(\partial_t\phi\right)^2\right.\\
			&\left.\hspace{-1cm}-\frac{1}{\tilde c^2 T^2_{\text{NR}}}\left(D^{-1}_{kl}+D^{-1}_{lk}\right)f_{tk}\partial_l\phi-\frac{1}{\tilde c^3 T^3_{\text{NR}}}\partial_t\phi\left(D^{-1}_{kl}-D^{-1}_{lk}\right)f_{tk}\partial_l\phi\right.\\
			&\left.+\frac{1}{\tilde c^4 T^4_{\text{NR}}}\left(D^{-1}_{km}D^{-1}_{ln}-D^{-1}_{kl}D^{-1}_{mn}\right)f_{tk}f_{tl}\partial_m\phi\partial_n\phi\right]^{1/2}\,.
		\end{split}
	\end{equation}
	We remark that the DBI action for a D24-brane also contains a dilaton-dependent factor. In this work we do not consider dilatons but the relation between the relativistic and the non-relativistic dilaton may contain a factor of $c$ which is why we refrain from redefining the D24-brane tension.

	The action \eqref{eq:NLGED} is the action for non-linear GED that describes the dynamics of a nrD$24$-brane. The linear GED theory is obtained by expanding \eqref{eq:NLGED} in $T^{-1}_{\text{NR}}$. Expanding to order $T^{-2}_{\text{NR}}$ we find
	\begin{equation}
		\begin{split}
			S_{\text{nrD24}} &= T_{\text{D}24}c^2\int d^{25}\sigma \left[-1+\frac{1}{\tilde c^2 T^2_{\text{NR}}}\left(\frac{1}{2}\left(\partial_t\phi\right)^2+f_{tk}\partial_k\phi-\frac{1}{4}f_{ij}f_{ij}\right)+\mathcal{O}(T^{-4}_{\text{NR}})\right]\,.
		\end{split}
	\end{equation}
	The term in parenthesis at order $T^{-2}_{\text{NR}}$ is the GED Lagrangian.
	
	It is not surprising that we end up with GED at this order. For the relativistic DBI action of a D24-brane, at order $T^{-2}$, we obtain the action of Maxwell plus a free real scalar field and it is known that GED can be obtained as the large $c$ limit of this action \cite{Bergshoeff:2015sic,Festuccia:2016caf}. In these works the LO behaviour of the fields in their $1/c$ expansion differs from what we have here and that difference is precisely accounted for by the critical $B$-field which is absent in \cite{Bergshoeff:2015sic,Festuccia:2016caf}.

	\section{Transverse T-duality}
	\label{sec:transverse-T-duality}
	In this section, we explore non-relativistic T-duality in the transverse directions. The resulting Buscher rules will be used in the next section when we discuss T-duality covariance of the non-relativistic counterpart of the DBI action for nrD$p$-branes for general backgrounds. Our results for transverse T-duality are in agreement with~\cite{Bergshoeff:2018yvt,Gomis:2020izd} where non-relativistic strings are viewed in the limit $c\rightarrow\infty$, whereas in this work we adopt the perspective of the $1/c^2$ expansion.
	
	One can also consider T-duality along the compact longitudinal direction. However, in that case the theory is not invariant because longitudinal T-duality invariance is broken by the non-relativistic expansion. To put it differently, it was shown in~\cite{Hartong:2022dsx} that longitudinal T-duality in the compact $v$-direction does not commute with the $1/c^2$ expansion. If we perform a longitudinal T-duality before we expand in $1/c^2$ it amounts to switching from an expansion in the dimensionless parameter $\epsilon = \frac{\alpha' \hbar}{cR^2}$ (introduced in equation \eqref{eq:dimless-param}), to one where we expand instead in the parameter $\tilde\epsilon = \frac{\alpha' \hbar}{c \tilde R^2}$, where $\tilde R = \frac{\hbar \alpha'}{c R}$. If we perform a longitudinal T-duality after the $1/c^2$ expansion we either end up with the DLCQ of a relativistic open string or with non-commutative open string theory depending on the details of the T-duality (see \cite{Gomis:2020izd}).

	\subsection{Transverse T-duality for closed non-relativistic strings}
	\label{sec:T-duality-closed-strings}
	We begin with a discussion of transverse T-duality for closed non-relativistic strings obtained from a $1/c^2$ expansion of relativistic strings with a critical $B$-field~\eqref{eq:critical-B-field}. We consider a setup where, in addition to the longitudinal $v$-direction, one of the transverse directions is compact. We call this compact transverse direction the $\theta$-direction, and we call its radius $R_\perp$. Therefore, the topology of the target space of the string we consider here is $\RR^{1,23}\times S^1_R\times S^1_{R_\perp}$, where the subscripts on the circles indicate their radii. This leads to a split of the spacetime index of the form $M = (\bar\mu,\theta,v)$, where $\bar \mu = (t,\bar i)$ with $\bar i$ ranging over the $23$ non-compact transverse directions.

	\subsubsection{Flat target space}
	\label{sec:closed-T-duality-flat}
	We begin by considering T-duality at the level of the mode expansions of the closed string embedding scalars. The embedding field in the compact $\theta$-direction is $x^{\theta}$. Using results from \cite{Hartong:2021ekg,Hartong:2022dsx} we obtain the following gauge-fixed mode expansion
	\begin{align}
		\label{eq:mode-expansions-closed-string}
		\begin{split}
			x^\pm &= x_0^\pm + w\tilde c^{-1}\zReff \sigma^\pm\,,\\
			y^+ &= \tilde c^{-1}y^t_0 - \frac{\zaeff}{\tilde c} p_{(0)+}(\sigma^++\sigma^-)  +  \frac{i}{\sqrt{4\pi \tilde c\zTeff}}\sum_{k\neq 0} \frac{1}{k}\beta_k^+ e^{-ik\sigma^-}\,,\\
			y^- &= \tilde c^{-1}y^t_0 - \frac{\zaeff}{\tilde c} p_{(0)-}(\sigma^++\sigma^-)  +  \frac{i}{\sqrt{4\pi \tilde c\zTeff}}\sum_{k\neq 0} \frac{1}{k}\tilde\beta_k^- e^{-ik\sigma^+}\,,\\
			x^{\bar i} &= x_0^{\bar i} + \frac{\zaeff}{\tilde c} p_{(0){\bar i}} \sigma^0+ \frac{i}{\sqrt{4\pi \tilde c \zTeff}}\sum_{k\neq 0} \frac{1}{k}\left(\alpha_k^{\bar i} e^{-ik\sigma^-}+\tilde{\alpha}^{\bar i}_k e^{-ik\sigma^+} \right)\,,\\
			x^{\theta} &= x_0^{\theta} + \frac{\zaeff}{\tilde c}\frac{\hbar  n_\perp}{R_\perp}\sigma^0+ w_\perp R_\perp\sigma^1 + \frac{i}{\sqrt{4\pi \tilde c \zTeff}}\sum_{k\neq 0} \frac{1}{k}\left(\alpha_k^{\theta} e^{-ik\sigma^-}+\tilde{\alpha}^{\theta}_k e^{-ik\sigma^+} \right)\,,
		\end{split}
	\end{align}
	where 
	\begin{equation}
		p_{(0)\pm} = \frac{1}{2}(p_{(0)t} \pm \tilde c p_{(0)v} )\,,
	\end{equation}
	has dimensions of energy, and where $p_{(0)v}$ is quantised according to
	\begin{equation}
		\label{eq:quantisation}
		p_{(0)v} = \frac{\hbar n}{\zReff}\,,
	\end{equation}
	where $w$ and $n$ are, respectively, the winding number and momentum number in the compact longitudinal $v$-direction. These mode expansions are derived in the same way as the open string mode expansions in Section~\ref{sec:string-expansions-without-WV-fields}: solving the equations of motion, which is most conveniently done in lightcone coordinates, imposing the Virasoro constraints and then fixing the residual gauge invariance. For more details, we refer to~\cite{Hartong:2021ekg,Hartong:2022dsx}. The mode expansion for $x^{\theta}$ can also be written as
	\begin{equation}
		\begin{split}
			x^{\theta} &= x_0^{\theta}+ \frac{\zaeff}{2 \tilde c}p^{\theta}_+\sigma^+ + \frac{\zaeff}{2 \tilde c}p^{\theta}_-\sigma^- + \frac{i}{\sqrt{4\pi \tilde c\zTeff}}\sum_{k\neq 0} \frac{1}{k}\left(\alpha_k^{\theta} e^{-ik\sigma^-}+\tilde{\alpha}^{\theta}_k e^{-ik\sigma^+} \right)\\
			&=: x^{\theta}_+(\sigma^+) + x^{\theta}_-(\sigma^-)  \,,
		\end{split}
	\end{equation}
	where we split the embedding scalar $x^\theta$ into left and right movers, and where we defined
	\begin{equation}
		\label{eq:24-momenta}
		p_\pm^{\theta} := \frac{\hbar n_\perp}{R_\perp} \pm \tilde c\frac{w_\perp R_\perp}{\zaeff}\,.
	\end{equation}
	The mode expansions~\eqref{eq:mode-expansions-closed-string} imply that the lightcone derivatives have mode expansions of the form
	\begin{align}
		\begin{aligned}
			\D_- x^{\bar i} &= \frac{1}{\sqrt{4\pi \tilde c \zTeff}}\sum_{k\in\mathbb{Z}} \alpha_k^{\bar i} e^{-ik\sigma^-}\,,& \D_+ x^{\bar i} &= \frac{1}{\sqrt{4\pi \tilde c\zTeff}}\sum_{k\in\mathbb{Z}} \tilde\alpha_k^{\bar i} e^{-ik\sigma^+}\,,\\
			\D_- x^{\theta} &= \frac{1}{\sqrt{4\pi \tilde c\zTeff}}\sum_{k\in\mathbb{Z}} \alpha_k^{\theta} e^{-ik\sigma^-}\,, & \D_+ x^{\theta} &= \frac{1}{\sqrt{4\pi \tilde c\zTeff}}\sum_{k\in\mathbb{Z}} \tilde\alpha_k^{\theta} e^{-ik\sigma^+}\,,  
		\end{aligned}
	\end{align}
	where the zero modes are given by
	\begin{align}
		\tilde\alpha_0^{\bar i} = \alpha_0^{\bar i} = \sqrt{\frac{\zaeff}{2 \tilde c}}p_{(0)\bar i}\,,\qquad \alpha^{\theta}_0 = \sqrt{\frac{\zaeff}{2\tilde c}}p^{\theta}_-\,,\qquad \tilde\alpha^{\theta}_0 = \sqrt{\frac{\zaeff}{2\tilde c}}p^{\theta}_+\,.
	\end{align}
	The Virasoro constraints for the closed string are (cf.~their open string counterparts in~\eqref{eq:NLO-Virasoro})
	\begin{equation}
		\D_+ y^- = \frac{\tilde c}{w \zReff}\D_+ x^i\D_+ x^i\,,\qquad\D_- y^+ = \frac{\tilde c}{w \zReff}\D_- x^i\D_- x^i\,,
	\end{equation}
	where we remind the reader that $i = (\bar i,\theta)$, and they imply that
	\begin{equation}
		\label{eq:NLO-energy-closed}
		E_{\text{NLO}}:=-p_{(0)t}  = \frac{\tilde c}{\zReff}\left(N_{(0)} + \tilde N_{(0)}\right) + \frac{\zaeff}{2w\zReff}\left(  p_{(0)\bar i} p_{(0)\bar i} + \frac{\hbar^2 n_\perp^2}{R_\perp^2} + \frac{\hbar^2 w_\perp^2 R_\perp^2}{l^4_{s,\text{NR}}} \right)\,,
	\end{equation}
	where the NR string length $l_{s,\text{NR}}$ is defined in \eqref{eq:lsNR}.
	Similarly, since $p_{(0)v}$ is quantised as in~\eqref{eq:quantisation}, we get the following level matching constraint
	\begin{equation}
		\label{eq:level-matching-closed}
		N_{(0)} - \tilde N_{(0)} = \hbar n w + \hbar n_\perp w_\perp\,.
	\end{equation}
	The energy~\eqref{eq:NLO-energy-closed} and the level matching constraint~\eqref{eq:level-matching-closed} are both invariant under the following transverse T-duality transformation
	\begin{equation}
		\label{eq:tilde-L}
		n_\perp\leftrightarrow w_\perp\,,\quad\text{and}\quad R_\perp \leftrightarrow \frac{l^2_{s,\text{NR}}}{R_\perp}=:\tilde R_\perp\,,
	\end{equation}
	where the T-dual circle has radius $\tilde R_\perp$. For the momenta defined in~\eqref{eq:24-momenta}, this T-duality transformation amounts to 
	\begin{equation}
		p_+^{\theta} \leftrightarrow p_+^{\theta}\,,\qquad p_-^{\theta} \leftrightarrow -p_-^{\theta}\,.
	\end{equation}
	The T-dual direction will be denoted by $\tilde \theta$ which is compact with radius $\tilde R_\perp$. The embedding field in the $\tilde \theta$-direction will be denoted by $x^{\tilde \theta}$.

	\subsubsection{Ro\v{c}ek--Verlinde procedure}
	We will study the Ro\v{c}ek--Verlinde method \cite{Rocek:1991ps} (see, e.g.,~\cite{Alvarez:1994dn} for a review) for the non-relativistic closed string to NLO with a compact direction that is also an isometry. The LO Polyakov action in flat gauge is~\cite{Hartong:2021ekg,Hartong:2022dsx}
	\begin{equation}
		\label{eq:LO-Lagrangian-with-WZ-term}
		\mathcal{L}_{\text{P-LO}} = -\frac{\tilde c\zTeff}{2}\tau_{MN}(x)\D^\alpha x^M\D_\alpha x^N + \frac{\tilde c\zTeff}{2}\epsilon^{\alpha\beta}B_{(-2)MN}(x)\D_\alpha x^M\D_\beta x^N\,,
	\end{equation}
	where we assume the LO Kalb--Ramond field $B_{(-2)MN}$ to be longitudinal, i.e., 
	\begin{equation}
		\label{eq:longitudinal-LO-B-field}
		B_{(-2)MN} = \tilde c^{-1}B_{(-2)}\varepsilon_{AB}\tau_M{^A}\tau_N{^B}\,,
	\end{equation}
	where $B_{(-2)}$ is a constant. Unlike in the previous subsubsection we do not require the $B$ field to be critical, cf.~\eqref{eq:critical-B-field}. A critical $B$-field corresponds to $B_{(-2)}=-1$.

	Denoting the Killing vector associated with the compact isometry by $k^M$, the statement that this direction is transverse translates to the requirement $k^M\tau_M{^A} = 0$. In adapted coordinates in which $k^M=\delta^M_\theta$ we get 
	\begin{equation}
		\tau_\theta^A = 0\,.
	\end{equation}
	This means that the first term in the LO action~\eqref{eq:LO-Lagrangian-with-WZ-term} is independent of $x^\theta$. Since $B_{(-2)MN}$ is longitudinal, it satisfies $B_{(-2)\theta N} = 0$. Thus, the LO action is independent of $x^\theta$. For the same reasons the same is true for the LO action on the T-dual background where we have $\tau_{\tilde\theta}^A = 0=B_{(-2)\tilde\theta N}$. We thus have the following trivial Buscher rules for the background fields appearing in the LO action 
	\begin{equation}
		\label{eq:LO-Buscher}
		\tilde \tau_{\bar M\bar N} = \tau_{\bar M\bar N}\,,\qquad \tilde B_{(-2)\bar M\bar N} = B_{(-2)\bar M\bar N}\,, 
	\end{equation}
	where $\bar M, \bar N=(\bar\mu,v)$. 
	Turning our attention to the NLO theory, the NLO Polyakov Lagrangian for a closed string moving in a general SNC background is~\cite{Hartong:2021ekg,Hartong:2022dsx}
	\begin{eqnarray}
		\mathcal{L}_{\text{P-NLO}} & = & -\frac{\tilde c\zTeff}{2}\sqrt{-\gamma_{(0)}}\gamma_{(0)}^{\alpha\beta}H_{MN}(x)\D_\alpha x^M\D_\beta x^N + \frac{\tilde c\zTeff}{2}\varepsilon^{\alpha\beta}B_{(0)MN}(x)\D_\alpha x^M\D_\beta x^N \nonumber\\
		&&-\frac{\tilde c T_{\text{NR}}}{2}\left(\lambda_{++}\varepsilon^{\alpha\beta}e^+_\alpha\tau^+_\beta+\lambda_{--}\varepsilon^{\alpha\beta}e^-_\alpha\tau^-_\beta\right)+ y^M\frac{\delta \mathcal{L}_{\text{NG-LO}}}{\delta x^M}\,,\label{eq:P-NLOclosed}
	\end{eqnarray}
	where 
	\begin{equation}
		\label{eq:tau-pm}
		\tau^\pm_\alpha=\tau^0_\alpha\pm\tilde c^{-1}\tau^1_\alpha\,,
	\end{equation}
	and where $B_{(0)MN}$ is the subleading $B$-field as defined in~\eqref{eq:expansion-of-B-field} and $H_{MN}$ is defined in \eqref{eq:metric-exp}. We note that the $y^M$-dependent term involves the LO Nambu--Goto Lagrangian. For a LO critical $B$ field this term vanishes. We remind the reader that $e^\pm_\alpha$ are worldsheet zweibeine related to $\gamma_{(0)\alpha\beta}$ via \eqref{eq:WS-vielbeine} and \eqref{eq:WS-vielbeine2}.
	
	For the purpose of deriving the T-duality transformation rules for the SNC geometry (the non-relativistic analogue of the Buscher rules) we do not need to consider the second line of \eqref{eq:P-NLOclosed} since this only depends on $\tau_{MN}$ and $B_{(-2)MN}$ which, as we argued above, do not transform under T-duality. We will therefore restrict our attention to the first line of \eqref{eq:P-NLOclosed} and fix a flat worldsheet gauge since this also has no bearing on the derivation. We will denote this part of the (partially gauge-fixed) NLO Polyakov Lagrangian by 
	\begin{equation}
		\mathcal{L}_{\text{P-NLO}}[H,B] = -\frac{\tilde c\zTeff}{2}H_{MN}(x)\D_\alpha x^M\D^\alpha x^N + \frac{\tilde c\zTeff}{2}\varepsilon^{\alpha\beta}B_{(0)MN}(x)\D_\alpha x^M\D_\beta x^N\,,\label{eq:P-NLOclosed2}
	\end{equation}
	where $[H,B]$ is supposed to indicate the part of $\mathcal{L}_{\text{P-NLO}}$ that depends on $H_{MN}$ and $B_{(0)MN}$.

	In adapted coordinates, the NLO Lagrangian $\mathcal{L}_{\text{P-NLO}}[H,B]$ above can be written as
	\begin{equation}
		\label{eq:split-polyakov-NR}
		\begin{split}
			\mathcal{L}_{\text{P-NLO}}[H,B] &= - \frac{\tilde c\zTeff}{2}H_{\bar M\bar N}(x)\D_\alpha x^{\bar M}\D^\alpha x^{\bar N} - \tilde c\zTeff H_{\bar M \theta}(x)\D_\alpha x^{\bar M}\D^\alpha x^\theta - \frac{\tilde c\zTeff}{2}H_{\theta\theta}\D_\alpha x^\theta\D^\alpha x^\theta \\
			&\quad + \frac{\tilde c\zTeff}{2}\epsilon^{\alpha\beta}B_{(0)\bar M\bar N}(x)\D_\alpha x^{\bar M}\D_\beta x^{\bar N} + \tilde c\zTeff \varepsilon^{\alpha\beta}B_{(0)\bar M \theta} \D_\alpha x^{\bar M} \D_\beta x^\theta   \,.
		\end{split}
	\end{equation}
	Since the $\theta$-direction is an isometry, the NLO Lagrangian has a global shift symmetry of the form
	\begin{equation}
		\label{eq:global-sym-NR}
		x^\theta \rightarrow x^\theta + \Lambda\,,
	\end{equation}
	where $\Lambda$ is a constant. The first step of the Ro\v cek--Verlinde procedure is to gauge the global symmetry~\eqref{eq:global-sym-NR} by promoting $\Lambda $ to an arbitrary function $\Lambda(\sigma^\alpha)$ on the worldsheet by introducing a gauge field $A_\alpha$
	transforming as
	\begin{equation}
		\delta A_\alpha = \D_\alpha \Lambda\,.
	\end{equation}
	To ensure gauge invariance, we couple the string to the worldsheet gauge field $A_\alpha$ by making the following replacement
	\begin{equation}
		\label{eq:replacement}
		\D_\alpha x^\theta \rightarrow \D_\alpha x^\theta - A_\alpha\,,
	\end{equation}
	which is gauge invariant. The second step is to demand that the connection $A_\alpha$ is flat on shell, $F_{\alpha\beta} := \D_\alpha A_\beta - \D_\beta A_\alpha = 0$, which we enforce at the level of the Lagrangian by introducing a Lagrange multiplier $\lambda$. These two steps give rise to a theory that is given by
	\begin{equation}
		\begin{split}
			\mathcal{L}'_{\text{P-NLO}}[H,B] &= -\frac{\tilde c\zTeff}{2}H_{\bar M\bar N}\D_\alpha x^{\bar M}\D^\alpha x^{\bar N} + \tilde c\zTeff H_{\bar M \theta}\D_\alpha  x^{\bar M}\tilde A^\alpha\\
			&\quad  - \frac{\tilde c\zTeff}{2}H_{\theta\theta}\tilde A_\alpha\tilde A^\alpha+ \frac{\tilde c\zTeff}{2}\varepsilon^{\alpha\beta}B_{(0)\bar M\bar N}\D_\alpha x^{\bar M}\D_\beta x^{\bar N}\\
			&\quad - \tilde c\zTeff \varepsilon^{\alpha\beta}B_{(0)\bar M \theta} \D_\alpha x^{\bar M}\tilde A_\beta - \frac{\tilde c\zTeff}{2}\lambda \varepsilon^{\alpha\beta}\left( \D_\alpha \tilde A_\beta - \D_\beta\tilde A_\alpha \right) \,,\label{eq:parentaction}
		\end{split}
	\end{equation}
	where we defined the gauge invariant field
	\begin{equation}
		\tilde A_\alpha = A_\alpha - \D_\alpha x^\theta\,.
	\end{equation}
	Integrating out $\lambda$ imposes $\D_\alpha \tilde A_\beta - \D_\beta \tilde A_\alpha = 0$, which means that we may globally write 
	\begin{equation}
		\tilde A_\alpha = \D_\alpha f\,,
	\end{equation}
	where the scalar $f$ is multi-valued if $\tilde A_\alpha$ has non-trivial holonomy along the string. Substituting this back into $\mathcal{L}'_{\text{NLO}}$, we recover the Lagrangian~\eqref{eq:split-polyakov-NR} upon identifying
	\begin{equation}
		f=-x^\theta \,,
	\end{equation}
	where the multi-valuedness of $f$ corresponds to the winding of $x^\theta$. Rather than integrating out $\lambda$ we now integrate out the non-dynamical field $\tilde A_\alpha$, which produces the equation of motion
	\begin{equation}
		\label{eq:relation-between-T-dual-fields}
		\tilde A_\alpha = \frac{1}{H_{\theta\theta}}\left( H_{\bar M \theta}\D_\alpha x^{\bar M} - \eta_{\alpha\beta}\varepsilon^{\gamma\beta}B_{(0)\bar M \theta} \D_\gamma x^{\bar M} + \eta_{\alpha\beta} \varepsilon^{\gamma\beta}\D_\gamma\lambda  \right)\,,
	\end{equation}
	which we can substitute back into~\eqref{eq:parentaction} to get
	\begin{equation}
		\label{eq:Buscher-action}
		\begin{split}
			\tilde{\mathcal{L}}_{\text{P-NLO}}[\tilde H, \tilde B] &= - \frac{\tilde c\zTeff}{2}\left( H_{\bar M\bar N} - \frac{H_{\bar M \theta}H_{\bar N \theta} - B_{(0)\bar M \theta}B_{(0)\bar N \theta} }{H_{\theta\theta}} \right)\D_\alpha x^{\bar M}\D^\alpha x^{\bar N}\\
			&\quad - \frac{\tilde c\zTeff}{2}\frac{1}{H_{\theta\theta}}\D_\alpha \lambda \D^\alpha \lambda  +\tilde c\zTeff \frac{B_{(0)\bar M \theta}}{H_{\theta\theta}}  \D_\alpha x^{\bar M} \D^\alpha \lambda - \tilde c\zTeff\varepsilon^{\alpha\beta}\frac{H_{\bar M \theta}}{H_{\theta\theta}}\D_\alpha x^{\bar M}\D_\beta\lambda\nonumber\\
			&\quad + \frac{\tilde c\zTeff}{2}\varepsilon^{\alpha\beta} \left( B_{(0)\bar M\bar N} + \frac{H_{\bar M \theta}B_{(0)\bar N \theta} - H_{\bar N \theta} B_{(0)\bar M \theta} }{H_{\theta\theta}} \right)\D_\alpha x^{\bar M} \D_\beta x^{\bar N}\,.
		\end{split}
	\end{equation}
	We then identify the embedding field in the T-dual $\tilde \theta$-direction with the Lagrange multiplier $\lambda$, i.e.
	\begin{equation}
		\label{eq:identification}
		x^{\tilde \theta} := \lambda\,.
	\end{equation}
	Since the actions $\tilde{\mathcal{L}}_{\text{P-NLO}}[\tilde H, \tilde B]$ and $\mathcal{L}_{\text{P-NLO}}[H, B]$ follow from the same parent action they are equivalent. 
	Demanding that \eqref{eq:Buscher-action} takes the same form as \eqref{eq:split-polyakov-NR} we obtain the non-relativistic Buscher rules
	\begin{align}
		\label{eq:NLO-closed-Buscher-rules}
		\begin{aligned}
			\tilde H_{\bar M\bar N} &=  H_{\bar M\bar N} - \frac{H_{\bar M \theta}H_{\bar N \theta} - B_{(0)\bar M \theta}B_{(0)\bar N \theta} }{H_{\theta\theta}}\,, \\
			\tilde B_{(0)\bar M\bar N} &= B_{(0)\bar M\bar N} + \frac{H_{\bar M \theta}B_{(0)\bar N \theta} - H_{\bar N \theta} B_{(0)\bar M \theta} }{H_{\theta\theta}} \,,\\
			\tilde H_{\tilde \theta\tilde \theta} &=  \frac{1}{H_{\theta\theta}}\,,\\
			\tilde B_{(0)\bar M\tilde \theta  } &= -\frac{H_{\bar M \theta}}{H_{\theta\theta}}\,,\\
			\tilde H_{\bar M \tilde \theta} &= -\frac{B_{(0)\bar M\theta}}{H_{\theta\theta}}\,.
		\end{aligned}
	\end{align}
	Together with the (trivial) LO Buscher rules~\eqref{eq:LO-Buscher} these form the T-duality transformations of SNC geometry (plus a $B$-field). This result agrees with \cite{Bergshoeff:2018yvt} where they used a different sign for the $B$-field.
	
	Going back to the case of a flat target spacetime (as we discussed in Sec.~\ref{sec:closed-T-duality-flat}) we observe that
	\begin{equation}
		H_{MN}\D_\alpha x^M \D_\beta x^N = \D_\alpha x^{\bar i} \D_\beta x^{\bar i} + R_\perp^2 \D_\alpha x^\theta\D_\beta x^\theta\,,
	\end{equation}
	where in contrast to what was done above we now take $x^\theta$ to be dimensionless.
	The parent Lagrangian~\eqref{eq:parentaction} now takes the form 
	\begin{equation}
		\label{eq:flat-L-prime}
		\begin{split}
			\mathcal{L}'_{\text{P-NLO}} &= -\frac{\tilde c\zTeff}{2}\D_\alpha x^{\bar i}\D^\alpha x^{\bar i}  - \frac{\tilde c\zTeff}{2}R_\perp^2\tilde A^\alpha\tilde A_\alpha- \tilde c\zTeff \varepsilon^{\alpha\beta} \tilde A_\alpha \D_\beta \lambda \,,
		\end{split}
	\end{equation}
	where $\tilde A_\alpha$ is dimensionless, while $\lambda$ has dimensions of length$^2$. Integrating out $\tilde A_\alpha$ gives
	\begin{equation}
		\label{eq:rel-for-A}
		\tilde A_\alpha = \frac{1}{R_\perp^2}\eta_{\alpha\beta} \varepsilon^{\gamma\beta}\D_\gamma\lambda\,,
	\end{equation}
	and, writing
	\begin{equation}
		\label{eq:rel-for-lambda}
		\lambda = l_{s,\text{NR}}^2 x^{\tilde\theta}\,,
	\end{equation}
	where $x^{\tilde\theta}$ is the dimensionless T-dual embedding field in the $\tilde\theta$-direction parametrising the dual circle, the Lagrangian~\eqref{eq:flat-L-prime} becomes
	\begin{equation}
		\begin{split}
			\tilde{\mathcal{L}}_{\text{P-NLO}} &= -\frac{\tilde c\zTeff}{2}\D_\alpha x^{\bar i}\D^\alpha x^{\bar i}  - \frac{\zTeff}{2}\frac{l_{s,\text{NR}}^4}{R_\perp^2} \D_\alpha x^{\tilde\theta}\D^\alpha  x^{\tilde\theta} \,.
		\end{split}
	\end{equation}
	Combining the relations~\eqref{eq:rel-for-A} and~\eqref{eq:rel-for-lambda} gives
	\begin{equation}
		\D_\alpha x^\theta = \frac{l_{s,\text{NR}}^2}{ R_\perp^2}\eta_{\alpha\beta} \varepsilon^{\beta\gamma} \D_\gamma x^{\tilde \theta}\,, 
	\end{equation}
	which upon redefining
	\begin{align}
		x^\theta \rightarrow R_\perp x^\theta\qquad\text{and}\qquad   x^{\tilde \theta} \rightarrow \tilde R_\perp  x^{\tilde\theta} =\frac{l_{s,\text{NR}}^2}{R_\perp} x^{\tilde\theta}\,,
	\end{align}
	becomes 
	\begin{equation}\label{eq:transverse-T-duality}
		\D_\alpha x^\theta = \eta_{\alpha\beta}\varepsilon^{\beta\gamma}\D_\gamma x^{\tilde \theta}\,,
	\end{equation}
	which is a well-known relation between the embedding scalars for the two T-dual circles.
	
	\subsection{Transverse T-duality for the open non-relativistic string}
	\label{sec:T-duality-open-strings}
	In this subsection, we discuss T-duality for open non-relativistic strings as discussed in Section~\ref{sec:NR-open-strings}. First we do this at the level of the mode expansions, just as we did for closed non-relativistic strings in Section~\ref{sec:closed-T-duality-flat}. Then we discuss the Ro\v cek--Verlinde method for the non-relativistic open string in a flat target space ending on a nrD24-brane and demonstrate how this turns into a non-relativistic open string ending on a nrD23-brane. 
	
	\subsubsection{Mode expansions and T-duality}
	When the transverse $\theta$-direction is compact with radius $R_\perp$, the gauge-fixed mode expansions we found in Section~\ref{sec:NR-open-strings} for the NLO string take the form
	\begin{align}
		\begin{split}
			x^\pm &= x^\pm_0 + 2 w\tilde c^{-1} \zReff \sigma^\pm \,,\\
			y^\pm &= \tilde c^{-1}y^t_0 - \frac{1}{2\pi\tilde c \zTeff}p_{(0)t}(\sigma^+ + \sigma^-) +  \frac{i}{\sqrt{4\pi\tilde c \zTeff}}\sum_{k\neq 0}\frac{1}{k}\beta_k^+ e^{-ik\sigma^\mp}\,,\\
			x^{\bar i} &= x^{\bar i}_0 + \frac{1}{\tilde c\pi \zTeff}p_{(0)\bar i}\sigma^0 + \frac{i}{\sqrt{\pi\tilde c\zTeff}}\sum_{k\neq 0}\frac{1}{k}\alpha_k^{\bar i} e^{-ik\sigma^0}\cos(k\sigma^1)\,,\\
			x^{\theta} &= x^{\theta}_0 + \frac{1}{\tilde c\pi \zTeff}\frac{\hbar n_\perp}{R_\perp}\sigma^0 + 2 w_\perp R_\perp\sigma^1 + \text{oscillations}\,,\quad  \begin{cases}
				w_\perp = 0\,,& \text{if } x^\theta\text{ NN}\\
				n_\perp = 0\,,& \text{if } x^\theta\text{ DD}
			\end{cases}\,,
		\end{split}
	\end{align}
	where $\bar i = 1,\dots,23$ labels the NN directions, and where $n_\perp$ is the momentum mode and $w_\perp$ the winding number in the compact $\theta$-direction. We emphasise that at this stage, we remain agnostic about whether $x^{\theta}$ satisfies DD or NN boundary conditions, since we kept both momentum and winding. However, if $x^{\theta}$ is a NN direction, it cannot have winding, while if it is a DD direction, it cannot have momentum. In what follows, we will keep both $n_\perp$ and $w_\perp$ in the expressions so as to be able to treat NN and DD boundary conditions simultaneously with the understanding that one of them will always be zero.
	
	We can also write the mode expansion for $x^{\theta}$ as
	\begin{equation}
		x^{\theta} = x^{\theta}_0 + \frac{\zaeff}{\tilde c} p^{\theta}_-\sigma^- + \frac{\zaeff}{\tilde c} p^{\theta}_+\sigma^+ + \text{oscillations}\,,
	\end{equation}
	where
	\begin{equation}
		p^{\theta}_\pm = \frac{\hbar n_\perp}{ R_\perp} \pm  \tilde c\frac{w_\perp R_\perp}{\zaeff}\,.
	\end{equation}
	The Virasoro constraints~\eqref{eq:NLO-Virasoro} tell us that
	\begin{align}
		\begin{split}
			-p_{(0)t} &= \frac{\zaeff}{2w \zReff}((p^{\theta}_+)^2 + p_{(0)\bar i}p_{(0)\bar i}) + \frac{\tilde c N_{(0)}}{2w\zReff}\,,\\
			-p_{(0)t} &= \frac{\zaeff}{2w\zReff}((p^{\theta}_-)^2 + p_{(0)\bar i}p_{(0)\bar i}) + \frac{\tilde c N_{(0)}}{2w\zReff}\,,
		\end{split}
	\end{align}
	and hence we find that
	\begin{equation}
		{(p^{\theta}_-)}^2 = {(p^{\theta}_+)}^2\,,
	\end{equation}
	which implies that
	\begin{equation}
		w_\perp n_\perp = 0\,,
	\end{equation}
	that is to say, we cannot simultaneously have momentum and winding: this is just the statement that $x^{\theta}$ cannot simultaneously satisfy NN and DD boundary conditions. Hence, the NLO energy is 
	\begin{equation}
		\begin{split}
			E^{\text{NN}}_{\text{NLO}} &= \frac{\zaeff}{2w\zReff}\left(\frac{ \hbar^2 n_\perp^2}{R_\perp^2}  + p_{(0)\bar i}p_{(0)\bar i}\right) + \frac{\tilde c N_{(0)}}{2w\zReff}\,,\\
			E^{\text{DD}}_{\text{NLO}} &= \frac{\zaeff}{2w\zReff}\left( \frac{\tilde c^2 w_\perp^2 R_\perp^2}{\zaeff^2} + p_{(0)\bar i}p_{(0)\bar i}\right) + \frac{\tilde c N_{(0)}}{2w\zReff}\,,
		\end{split}
	\end{equation}
	depending on whether $x^\theta$ satisfies NN or DD boundary conditions. Note that these expressions transform into each other under T-duality
	\begin{equation}
		n_\perp\leftrightarrow w_\perp\,,\qquad R_\perp \leftrightarrow \frac{\hbar \zaeff}{\tilde c R_\perp}\,.
	\end{equation}
	This T-duality transformation also changes between NN and DD boundary conditions as dictated by~\eqref{eq:transverse-T-duality}. Explicitly, starting with $x^{\theta}$ an NN direction with radius $R_\perp$, the associated mode expansion is
	\begin{equation}
		x^{\theta} = x^{\theta}_0 + \frac{1}{\pi \tilde c \zTeff}\frac{\hbar n_\perp}{R_\perp}\sigma^0 + \frac{i}{\sqrt{\pi\tilde c\zTeff}}\sum_{k\neq 0}\frac{1}{k}\alpha_k^{\theta} e^{-ik\sigma^0}\cos(k\sigma^1)\,.
	\end{equation}
	The T-dual embedding scalar $x^{\tilde \theta}$, which satisfies~\eqref{eq:transverse-T-duality}, is now a DD direction and acquires the mode expansion
	\begin{equation}
		x^{\tilde \theta} = x^{\tilde \theta}_0 + 2 w_\perp \tilde R_\perp\sigma^1 - \frac{1}{\sqrt{\pi\tilde c\zTeff}}\sum_{k\neq 0}\frac{1}{k}\alpha_k^{\tilde \theta} e^{-ik\sigma^0}\sin(k\sigma^1)\,,
	\end{equation}
	where the radius of the T-dual DD $\tilde \theta$-direction is $\tilde R_\perp = \frac{\hbar \zaeff}{\tilde c R_\perp}$.

	\subsubsection{Ro\v{c}ek--Verlinde procedure for the NLO open string}
	\label{sec:RV-open}
	Here we consider the Ro\v cek--Verlinde procedure for the NLO open string with a critical $B$-field (cf.~Eq.~\eqref{eq:critical-B-field}) and with GED fields on the nrD$24$-brane. We take the background to be flat space with an isometry in the transverse and compact $\theta$-direction as above. In flat gauge, $\gamma_{(0)\alpha\beta}=\eta_{\alpha\beta}$, the NLO Polyakov action for an open string ending on a nrD$24$-brane~\eqref{eq:NLO-GO-action} takes the form
	\begin{equation}
		\label{eq:Open-string-action-for-T-duality}
		\begin{split}
			S_{\text{P-NLO}} &= -\frac{\tilde c\zTeff}{2}\int d^2\sigma\,  \left(  \D_\alpha x^i \D^\alpha x^i + \lambda_{++}\varepsilon^{\alpha\beta}e_\alpha^+ \D_\beta  x^+ + \lambda_{--} \varepsilon^{\alpha\beta} e_\alpha^- \D_\beta  x^- \right) \\
			&\quad+ \int d\sigma^0\,a_\mu \dot x^\mu \bigg\vert_{\sigma^1=0}^{\sigma^1 = \pi}\\
			&= -\frac{\tilde c\zTeff}{2}\int d^2\sigma\,  \left(  \D_\alpha x^i \D^\alpha x^i + \lambda_{++}\varepsilon^{\alpha\beta}e_\alpha^+ \D_\beta  x^+ + \lambda_{--} \varepsilon^{\alpha\beta} e_\alpha^- \D_\beta  x^- \right.\\
			&\quad \left.- \frac{1}{\tilde c\zTeff}\varepsilon^{\alpha\beta}f_{\alpha\beta} \right) \,,
		\end{split}
	\end{equation}
	where $e^\pm_\alpha$ are such that $\eta_{\alpha\beta}=-\left(e^+_\alpha e^-_\beta+e^+_\beta e^-_\alpha\right)/2$ and 
	where we used that 
	\begin{equation}
		\int d\sigma^0\,a_\mu \dot x^\mu\bigg\vert_{\sigma^1=0}^{\sigma^1 = \pi} = \frac{1}{2}\int d^2\sigma\, \varepsilon^{\alpha\beta} \D_\alpha x^\mu \D_\beta x^\nu f_{\mu\nu}\,.
	\end{equation}
	Furthermore, we defined $f_{\alpha\beta} := \D_\alpha x^\mu \D_\beta x^\nu f_{\mu\nu}$ in~\eqref{eq:Open-string-action-for-T-duality}. 
	
	Starting with \eqref{eq:NLO-GO-action} we split the index $\mu=(t,i)=(t,\bar i, \theta)=(\bar\mu,\theta)$ and we apply the Ro\v cek--Verlinde procedure by gauging the $\theta$-direction with a flat connection and defining $\tilde A_\alpha=A_\alpha-\partial_\alpha x^\theta$. The steps are very similar to what we did above for closed strings. The only difference is that we must now impose suitable boundary conditions in order that the variational problem be well defined. The parent action is now given by
	\begin{equation}
		\label{eq:OS-T-dual-action}
		\begin{split}
			S'_{\text{P-NLO}} &= -\frac{\tilde c\zTeff}{2}\int d^2\sigma\,  \bigg(  \D_\alpha x^{\bar i} \D^\alpha x^{\bar i} + \tilde A_\alpha \tilde A^\alpha + 2\varepsilon^{\alpha\beta}\lambda \D_\alpha \tilde A_\beta + \tilde c^{-1}\varepsilon^{\alpha\beta}\left(\lambda_{++}e_\alpha^+ \D_\beta x^+ + \lambda_{--}  e_\alpha^- \D_\beta x^- \right)\\
			&- \frac{1}{\tilde c\zTeff}\varepsilon^{\alpha\beta}f_{\bar\mu\bar\nu}\D_\alpha x^{\bar\mu}\D_\beta x^{\bar\nu} -\frac{2}{\tilde c\zTeff}\varepsilon^{\alpha\beta}f_{\theta \bar\mu} \D_\alpha x^{\bar\mu} \tilde A_\beta \bigg) + \tilde c\zTeff \int d\sigma^0\, \lambda \tilde A_0 \bigg\vert_{\sigma^1=0}^{\sigma^1 = \pi}  \,,
		\end{split}
	\end{equation}
	where the gauge field $\tilde A_\alpha$ and the Lagrange multiplier $\lambda$ are supplemented with the following boundary conditions\footnote{We remind the reader that the embedding fields $x^i,x^t,y^t$ satisfy NN boundary conditions, while $x^v,y^v$ satisfy DD boundary conditions.}
	\begin{equation}
		\label{eq:open-T-duality-BCs}
		\tilde A_1\big\vert_{\text{ends}} = 0\,,\qquad \delta \lambda\big\vert_{\text{ends}} = 0\,.
	\end{equation}
	The first condition ensures that $x^\theta$ becomes an NN direction upon integrating out $\lambda$ and identifying $\tilde A_\alpha = -\D_\alpha x^\theta$, while the second condition turns the T-dual direction, which is identified with $\lambda$ as in~\eqref{eq:identification}, into a DD direction. The boundary term in~\eqref{eq:OS-T-dual-action} is there to ensure invariance (up to a total time derivative) under the constant shift 
	\begin{equation}
		\label{eq:lambda-sym}
		\lambda \rightarrow \lambda + k\,,
	\end{equation}
	which will correspond to translation invariance along the T-dual circle. To obtain the T-dual Lagrangian by integrating out $\tilde A_\alpha$, we first perform an integration by parts which exactly cancels the boundary term introduced in~\eqref{eq:OS-T-dual-action}, leading to
	\begin{equation}
		\label{eq:ready-to-integrate-out}
		\begin{split}
			S'_{\text{P-NLO}} &= -\frac{\tilde c \zTeff}{2}\int d^2\sigma\,  \bigg(  \D_\alpha x^{\bar i} \D^\alpha x^{\bar i} + \tilde A_\alpha \tilde A^\alpha - 2\varepsilon^{\alpha\beta} \tilde A_\beta \D_\alpha \lambda - \frac{1}{\tilde c \zTeff}\varepsilon^{\alpha\beta}f_{\bar\mu\bar\nu}\D_\alpha x^{\bar\mu}\D_\beta x^{\bar\nu} \\
			& -\frac{2}{\tilde c \zTeff}\varepsilon^{\alpha\beta} f_{\theta \bar\mu} \D_\alpha x^{\bar\mu} \tilde A_\beta + \varepsilon^{\alpha\beta}\left(\lambda_{++}e_\alpha^+ \D_\beta  x^+ + \lambda_{--}  e_\alpha^- \D_\beta x^-\right)\bigg)  \,.
		\end{split}
	\end{equation}
	The equation of motion for $\tilde A_\alpha$ takes the form
	\begin{equation}
		\tilde A_\alpha = - \eta_{\alpha\beta} \varepsilon^{\beta\gamma}\D_\gamma \lambda - \frac{1}{\tilde c \zTeff} \eta_{\alpha\beta} \varepsilon^{\beta \gamma} f_{\theta \bar\mu} \D_\gamma x^{\bar \mu}\,,
	\end{equation}
	and therefore integrating out $\tilde A_\alpha$ in~\eqref{eq:ready-to-integrate-out} leads to 
	\begin{equation}
		\begin{split}
			\tilde S_{\text{P-NLO}} &= -\frac{\tilde c \zTeff}{2}\int d^2\sigma\,  \bigg(  \D_\alpha x^{\bar i} \D^\alpha x^{\bar i} + \D_\alpha(x^{\tilde \theta} - \tilde c^{-1}\zTeff^{-1}a_\theta)\D^\alpha(x^{\tilde \theta} - \tilde c^{-1}\zTeff^{-1}a_\theta) \\
			&\quad - \frac{1}{\tilde c\zTeff}\varepsilon^{\alpha\beta} \D_\alpha x^{\bar\mu} \D_\beta x^{\bar\nu} f_{\bar\mu\bar\nu}+ \varepsilon^{\alpha\beta}\left(\lambda_{++}e_\alpha^+ \D_\beta  x^+ +\lambda_{--}  e_\alpha^- \D_\beta  x^-\right)\bigg) \,,
		\end{split}
	\end{equation}
	where we identified $\lambda = x^{\tilde \theta}$ with the T-dual $\tilde \theta$-direction as above. In particular, this reveals that the component $a_\theta$ of the gauge field in the $\theta$-direction describes the fluctuations of the dual nrD$23$-brane in the transverse $\tilde \theta$-direction (just like in relativistic string theory).

	\section{Non-relativistic D\texorpdfstring{$p$}{p}-branes}
	\label{sec:nrDp-branes}
	
	We now turn our attention to the description of nrD$p$-branes for arbitrary $p\leq 24$ and with arbitrary background fields. We first derive the form of the non-relativistic DBI action by expanding the relativistic DBI action in powers of $1/c^2$ which, just like the nrD$24$-brane we considered in Section~\ref{sec:NL-GED}, is done with a LO critical $B$-field. We then demonstrate that this action may also be derived ``intrinsically'' without recourse to the relativistic parent theory by demanding that it has the right gauge symmetries and is T-duality covariant under the transverse T-duality transformations derived in Section~\ref{sec:transverse-T-duality}. The action for non-relativistic D$p$-branes we obtain in this section agrees with the one proposed earlier in the literature~\cite{Gomis:2020fui,Ebert:2021mfu}.

	\subsection{Action for nrD\texorpdfstring{$p$}{p}-branes from \texorpdfstring{$1/c^2$}{large-c} expansion}
	\label{sec:nrDBI-from-exp}
	We will denote the embedding fields for a D$p$-brane with $p\le 24$ by the same symbol as for the open string, namely $X^M(\sigma) = (X^{a}(\sigma), X^{I}(\sigma))$ where now $\sigma$ is shorthand notation for the worldvolume coordinates $\sigma^{\hat\alpha}$ with $\hat\alpha = 0,\dots,p$ (cf.~Table~\ref{tab:indices}). The embedding scalars split into NN directions $X^{a}(\sigma)$ and DD directions $X^{I}(\sigma)$, which includes the compact $v$-direction.

	Associated to each DD direction is a Goldstone boson due to the spontaneous breaking of translation symmetry in that direction, leading to the following generalisation of~\eqref{eq:Xvondynamicalbrane}
	\begin{equation}
		\label{eq:DD-goldstone}
		X^{I}(\sigma) = X^{I}_0 + \frac{\tilde c}{cT}\Phi^{I}(X^{a}(\sigma))\,, 
	\end{equation}
	where $X^{I}_0$ describes the location of the brane.
	In curved space, a LO critical $B$-field corresponds to (cf.~\eqref{eq:critical-B-field})
	\begin{equation}
		\label{eq:curved-critical-B-field}
		B_{MN} = -2\tilde c^{-1} c^2\tau_{[M}^0 \tau_{N]}^1 + B_{(0)MN} + \mathcal{O}(c^{-2})\,.
	\end{equation}
	For the string $1/c^2$ expansion of the target space metric we can use \eqref{eq:metric-exp} which we repeat here for convenience in the following form
	\begin{equation}
		g_{MN}=c^2\left(-\tau^0_M\tau^0_N+\frac{1}{\tilde c^2}\tau^1_M\tau^1_N\right)+H_{MN}+\mathcal{O}(c^{-2})\,.
	\end{equation}
	It will be convenient to
	define the combination
	\begin{equation}
		\label{eq:M-combo}
		E_{MN} := g_{MN} + B_{MN}\,,
	\end{equation}
	whose pullback appears in the DBI action. For a LO critical $B$-field this admits the following expansion
	\begin{equation}\label{eq:EMN}
		E_{MN}=-c^2 \tau^-_M \tau^+_N+H_{MN}+B_{(0)MN}+\mathcal{O}(c^{-2})\,,
	\end{equation}
	where we defined 
	\begin{equation}
		\tau^\pm_M =\tau^0_M\pm \tilde c^{-1}\tau^1_M\,.
	\end{equation}
	The action for a D$p$-brane with $p\le 24$ is given by (cf.~\eqref{eq:DBI-flat})
	\begin{equation}
		\label{eq:Dp}
		S_{\text{D}p}=-cT_{\text{D}p}\int d^{p+1}\sigma\sqrt{-\det M_{\hat\alpha\hat\beta}}\,,
	\end{equation}
	where 
	\begin{equation}
		M_{\hat\alpha\hat\beta}= E_{\hat\alpha\hat\beta} +\frac{1}{cT}F_{\hat\alpha\hat\beta}\,,
	\end{equation}
	where 
	\begin{equation}
		E_{\hat\alpha\hat\beta}= \D_{\hat \alpha}X^M\D_{\hat \beta} X^N E_{MN}\,,
	\end{equation}
	and where $F_{\hat\alpha\hat\beta}$ is the field strength of the wordvolume gauge field $A_{\hat\alpha}$. We remind the reader that we do not consider the dilaton in this work. Using \eqref{eq:EMN} and the expansion 
	\begin{equation}
		A_{\hat\alpha}=a_{\hat\alpha}+\mathcal{O}(c^{-2})
	\end{equation}
	we find
	\begin{equation}
		M_{\hat\alpha\hat\beta}= -c^2  \tau^-_{\hat\alpha} \tau^+_{\hat\beta} + H_{\hat\alpha\hat\beta}+B_{(0)\hat\alpha\hat\beta} +\frac{1}{\tilde c T_{\text{NR}}}f_{\hat\alpha\hat\beta}+\mathcal{O}(c^{-2})\,,
	\end{equation}
	where $f_{\hat\alpha\hat\beta}=(da)_{\hat\alpha\hat\beta}$. 
	The pullbacks at this point are computed with respect to the unexpanded embedding scalars $X^M$ with $X^I$ given by \eqref{eq:DD-goldstone}. The scalars $\Phi^I$ expand as
	\begin{equation}
		\Phi^I=\phi^I+\mathcal{O}(c^{-2})\,.
	\end{equation}
	Using the identity~\eqref{eq:useful-identity-T-duality}, we find that the $1/c^2$-expansion of the DBI action~\eqref{eq:Dp} takes the form
	\begin{equation}
		\label{eq:nrDp}
		S_{\text{D}p} = -T_{\text{D}p}c^2 \int d^{p+1}\sigma\,\sqrt{-\det\begin{pmatrix}
				0 ~ & ~ \tau^+_{\hat \beta}\\
				\tau^-_{\hat \alpha} ~&~  H_{\hat \alpha \hat \beta} +B_{(0)\hat \alpha \hat \beta} + \frac{1}{\tilde c T_{\text{NR}}}f_{\hat\alpha\hat\beta}
		\end{pmatrix}}+\cdots\,,
	\end{equation}
	where the dots denote subleading corrections and where now the pullbacks are computed with respect to the LO embedding scalars $X^M=x^M+\mathcal{O}(c^{-2})$ with
	\begin{equation}
		x^I=x_0^I+\frac{1}{T_{\text{NR}}}\phi^I\,.
	\end{equation}
	For $p=24$ with a flat target space and using static gauge, the action~\eqref{eq:nrDp} reduces to the action for non-linear GED~\eqref{eq:NLGED}.

	\subsection{Target space gauge symmetries}
	\label{sec:syms}
	
	In this subsection, we discuss the target space gauge symmetries that arise in the string $1/c^2$ expansion of Section~\ref{sec:string-expansion} of open and closed strings. These symmetries should also leave invariant the nrD$p$-brane action as we will discuss in the next subsection. 
	
	The following statements are true regardless of whether we work with open or closed strings and they hold both in the Nambu--Goto and Polyakov formulation. For definiteness let us consider the Nambu--Goto formulation of a closed non-relativistic string. If we $1/c^2$ expand the Nambu--Goto action of a relativistic string in the background of a metric and a LO critical $B$-field we end up with the following action
	\begin{equation}
		\label{eq:NG-NLO}
		S_{\text{NG-NLO}} = -\frac{\zTeff}{2}\int d^2\sigma\,\left( \sqrt{-\tau} \tau^{\alpha\beta}H_{\alpha\beta} + \varepsilon^{\alpha\beta}B_{(0)\alpha\beta}  \right)\,,
	\end{equation}
	where we remind the reader (see \eqref{eq:combinations}) that $H_{\alpha\beta}$ is the pullback of 
	\begin{equation}
		H_{MN}=H^\perp_{MN} + 2\eta_{AB}\tau_{(M}^A m_{N)}^B\,,
	\end{equation}
	where $H^\perp_{MN}$ has signature $(0,0,1,\ldots,1)$.

	This theory is known to possess a number of target space gauge symmetries which we summarise here (see, e.g.,~\cite{Bergshoeff:2018yvt,Harmark:2019upf,Hartong:2022dsx,Bidussi:2023rfs}). First of all we have the Stueckelberg symmetry
	\begin{equation}\label{eq:Stsym}
		\delta H_{MN} = 2C_{(M}^A\tau_{N)}^B\eta_{AB}\,,\qquad \delta B_{(0)MN} = - 2C_{[M}^A\tau_{N]}^B\varepsilon_{AB}\,.
	\end{equation}
	The reason this is called a Stueckelberg symmetry is because it shifts $m_M^A$ by $C^A_M$ and so we can fix this symmetry entirely by setting $m_M^A=0$. We briefly list the other symmetries. The string Galilei boost symmetry leaves $H_{MN}$ and $B_{(0)MN}$ invariant but acts on $H_{\perp MN}$ and $m_M^A$. When we gauge fix $m_M^A=0$ this symmetry acts on $H^\perp_{MN}$ and $B_{(0) MN}$. For explicit transformations we refer to \cite{Bergshoeff:2018yvt,Harmark:2019upf,Hartong:2022dsx,Bidussi:2023rfs}. Then there is a target space Weyl symmetry that rescales $\tau_{MN}$ with an arbitrary conformal factor. The Weyl rescaling also acts on the dilaton and since the latter is not included here we will not consider it further.
	Finally, there is the usual gauge transformation acting on the 2-form $B_{(0)MN}$, cf.~Eq.~\eqref{eq:1-form-gauge-sym}, which for the case of the open string leads to the requirement of having gauge fields at the endpoints of the string.

	We are thus looking for a nrD$p$-brane action that is an action for the embedding scalars and that is invariant under all these gauge transformations as well as under worldvolume diffeomorphisms. This will be the subject of the next and final subsection.

	\subsection{Non-relativistic DBI action from symmetries \textit{\&} T-duality covariance}
	\label{sec:DBI-from-T-duality-covariance}

	The geometrical setting is a 26-dimensional string Newton--Cartan target space and a $(p+1)$-dimensional submanifold. For simplicity we will assume $\tau^0_M=\delta^t_M$ and $\tau^1_M=\delta^v_M$ but the final result holds for any $\tau_{MN}$.
	The submanifold is a Newton--Cartan space with fields $\phi^I$ and $a_{\hat a}$ defined on it. When we turn off the worldvolume fields, so that $\partial_{\hat\alpha}X^I=0$, the induced objects $\tau_{\alpha\beta}$ and $h^\perp_{\hat\alpha\hat\beta}$ form a Newton--Cartan structure of a $(p+1)$-dimensional Newton--Cartan submanifold, i.e., they have signature $(-1,0,\ldots,0)$ and $(0,1,\ldots,1)$, respectively. 
	
	The starting point for building a nrD$p$-brane action is an integration measure built from pullbacks of metric data for the case of a brane that has no worldvolume fields turned on. Such an integration measure is provided by $\sqrt{-\text{det}\,\mathcal{M}}$, where 
	\begin{equation}\label{eq:buildingblock}
		\text{det}\,\mathcal{M} = \frac{1}{p!}\varepsilon^{\hat\alpha_1\hat\alpha_2\cdots\hat\alpha_{p+1}}\varepsilon^{\hat\beta_1\hat\beta_2\cdots\hat\beta_{p+1}}\tau_{\hat\alpha_1\hat\beta_1}h^\perp_{\hat\alpha_2\hat\beta_2}\cdots h^\perp_{\hat\alpha_{p+1}\hat\beta_{p+1}}\,,
	\end{equation}
	in which $\tau_{\hat\alpha\hat\beta}=-\tau^0_{\hat\alpha}\tau^0_{\hat\beta}$. This is invariant under the usual Galilean boost symmetry acting on $h^\perp_{\hat\alpha\hat\beta}$. We will take \eqref{eq:buildingblock} as our building block from which we will construct an action for the embedding scalars and worldvolume fields that respects all the gauge symmetries mentioned in the previous subsection.  The next step is to turn on the $\phi^I$ fields, thus allowing the brane to fluctuate in all transverse directions. In particular this means that $\tau_{\hat\alpha\hat\beta}=-\tau^0_{\hat\alpha}\tau^0_{\hat\beta}+\tau^1_{\hat\alpha}\tau^1_{\hat\beta}$ with $\tau^1_{\hat\alpha}\neq 0$. This leads to fields that are defined on the Newton--Cartan geometry induced on the brane. If we take again \eqref{eq:buildingblock}, but this time with $\partial_{\hat\alpha} x^I\neq 0$, the expression \eqref{eq:buildingblock} fails to be invariant under stringy Galilei boosts. To remedy this we can replace $h^\perp_{\hat\alpha\hat\beta}$ by $H_{\hat\alpha\hat\beta}$ leading to
	\begin{equation}
		\text{det}\,\mathcal{M} = \frac{1}{p!}\varepsilon^{\hat\alpha_1\hat\alpha_2\cdots\hat\alpha_{p+1}}\varepsilon^{\hat\beta_1\hat\beta_2\cdots\hat\beta_{p+1}}\tau_{\hat\alpha_1\hat\beta_1}H_{\hat\alpha_2\hat\beta_2}\cdots H_{\hat\alpha_{p+1}\hat\beta_{p+1}}\,.
	\end{equation}
	However this is not invariant under the Stueckelberg transformation \eqref{eq:Stsym}. To ensure that this is the case we need to replace $H_{\hat\alpha\hat\beta}$ by $H_{\hat\alpha\hat\beta}+B_{(0)\hat\alpha\hat\beta}$, leading to
	\begin{equation}\label{eq:buildingblock2}
		\text{det}\,\mathcal{M} = \frac{1}{p!}\varepsilon^{\hat\alpha_1\hat\alpha_2\cdots\hat\alpha_{p+1}}\varepsilon^{\hat\beta_1\hat\beta_2\cdots\hat\beta_{p+1}}\tau_{\hat\alpha_1\hat\beta_1}\left(H+B_{(0)}\right)_{\hat\alpha_2\hat\beta_2}\cdots \left(H+B_{(0)}\right)_{\hat\alpha_{p+2}\hat\beta_{p+2}}\,.
	\end{equation}
	To show that this is indeed invariant under \eqref{eq:Stsym} we first of all observe that
	\begin{equation}
		\delta\left(H_{MN}+B_{(0)MN}\right) = 2C_{(M}^A\tau_{N)}^B\eta_{AB}- 2C_{[M}^A\tau_{N]}^B\varepsilon_{AB}=-C^-_M\tau^+_N-C^+_N\tau^-_M\,,    
	\end{equation}
	so that
	\begin{equation}
		\delta\left(H+B_{(0)}\right)_{\hat\alpha\hat\beta} = -C^-_{\hat\alpha}\tau^+_{\hat\beta}-C^+_{\hat\beta}\tau^-_{\hat\alpha}\,.
	\end{equation}
	Secondly, we use that
	\begin{eqnarray}
		\text{det}\,\mathcal{M} & = &  \frac{1}{p!}\varepsilon^{\hat\alpha_1\hat\alpha_2\cdots\hat\alpha_{p+1}}\varepsilon^{\hat\beta_1\hat\beta_2\cdots\hat\beta_{p+1}}\tau_{\hat\alpha_1\hat\beta_1}X_{\hat\alpha_2\hat\beta_2}\cdots X_{\hat\alpha_{p+1}\hat\beta_{p+1}}\nonumber\\
		& = &  -\frac{1}{p!}\varepsilon^{\hat\alpha_1\hat\alpha_2\cdots\hat\alpha_{p+1}}\varepsilon^{\hat\beta_1\hat\beta_2\cdots\hat\beta_{p+1}}\tau^-_{\hat\alpha_1}\tau^+_{\hat\beta_1}X_{\hat\alpha_2\hat\beta_2}\cdots X_{\hat\alpha_{p+1}\hat\beta_{p+1}}\,,
	\end{eqnarray}
	where $X_{\hat\alpha\hat\beta}=\left(H+B_{(0)}\right)_{\hat\alpha\hat\beta}$ and where the second line follows from the fact that the antisymmetric part of $\tau^-_{\hat\alpha_1}\tau^+_{\hat\beta_1}$ does not contribute. Hence, we have arrived at the point where $\sqrt{-\text{det}\,\mathcal{M}}$ is a scalar density under $(p+1)$-dimensional worldvolume diffeomorphisms that is furthermore invariant under target space string Galilei boosts and Stueckelberg transformations (similar observations were made in \cite{Gomis:2020fui}). To make it also invariant under the 2-form gauge transformation we add $\tilde c^{-1}T^{-1}_{\text{NR}}f_{\hat\alpha\hat\beta}$ to the $B_{(0)}$ term. 
	
	If we define
	\begin{equation}\label{eq:calM}
		\mathcal{M}=\left(\begin{array}{cc}
			0 & \tau^+_{\hat\beta}\\
			\tau^-_{\hat\alpha} & X_{\hat\alpha\hat\beta}
		\end{array}\right)
	\end{equation}
	then we can write 
	\begin{eqnarray}
		\text{det}\,\mathcal{M} & = & -\frac{1}{p!}\varepsilon^{\hat\alpha_1\hat\alpha_2\cdots\hat\alpha_{p+1}}\varepsilon^{\hat\beta_1\hat\beta_2\cdots\hat\beta_{p+1}}\tau^-_{\hat\alpha_1}\tau^+_{\hat\beta_1}X_{\hat\alpha_2\hat\beta_2}\cdots X_{\hat\alpha_{p+1}\hat\beta_{p+1}}\,,
	\end{eqnarray}
	where $X_{\hat\alpha\hat\beta}$ is any tensor. So if we take 
	\begin{equation}\label{eq:X}
		X_{\hat\alpha\hat\beta}=H_{\hat\alpha\hat\beta}+B_{(0)\hat\alpha\hat\beta}+\tilde c^{-1}T^{-1}_{\text{NR}}f_{\hat\alpha\hat\beta}\,,
	\end{equation}
	we obtain an object that has all the local string Newton--Cartan symmetries mentioned in the previous section so that the nrD$p$-brane action is given by
	\begin{equation}\label{eq:NRDBI}
		S_{\text{nrDp}}=-T_{\text{nrDp}}\int d^{p+1}\sigma \sqrt{-\text{det}\,\mathcal{M}}\,,
	\end{equation}
	where $T_{\text{nrDp}}$ is a constant that is required on dimensional grounds and where $\mathcal{M}$ is given by \eqref{eq:calM} with 
	$X_{\hat\alpha\hat\beta}$ given by \eqref{eq:X}.

	We did not consider invariance under the target space Weyl transformations. To make this possible we need to introduce the dilaton $\Phi$. The dilaton field in non-relativistic string theory is subject to a Stueckelberg gauge transformation \cite{Bergshoeff:2018yvt}. This is a shift symmetry whose parameter is precisely the one that corresponds to the target space Weyl transformation acting on $\tau_{MN}$ that was mentioned in the previous section. The combination $e^{-\Phi}\sqrt{-\tau}$ is invariant under this target space Weyl transformation. So to include the dilaton in \eqref{eq:NRDBI} we just multiply the integrand with $e^{-\Phi}$. 
	
	We have shown that the action \eqref{eq:NRDBI} has all the symmetries that we want a nrD$p$-brane action to have. It would be nice to see if this is the unique action with such symmetries. In the remainder of this subsection we will check that \eqref{eq:NRDBI} is also transverse T-duality covariant.

	To demonstrate T-duality covariance we start with the action for a nrD$24$-brane, and show that after a T-duality transformation this becomes the action for a nrD23-brane. 
	To this end we define 
	\begin{equation}
		e_{MN}=H_{MN}+B_{(0)MN}\,,
	\end{equation}
	which transforms nicely under the transverse Buscher rules \eqref{eq:NLO-closed-Buscher-rules} which can be written as 
	\begin{equation}
		\label{eq:Buscher-rules-E}
		\begin{split}
			\tilde e_{\tilde\theta\tilde\theta} &= e_{\theta\theta}^{-1}\,,\\
			\tilde e_{\bar M\bar N} &= e_{\bar M \bar N} - e_{\bar M\theta}e^{-1}_{\theta\theta}e_{\theta\bar N}\,,\\
			\tilde e_{\bar M\tilde \theta} &= e_{\theta\theta}^{-1}e_{\bar M \theta}\,,\\
			\tilde e_{\tilde\theta\bar M}  &= -e_{\theta\theta}^{-1}e_{\theta \bar M}\,,
		\end{split}    
	\end{equation}
	where we remind the reader that $\bar M, \bar N$ run over all directions except $\theta$ or its T-dual direction $\tilde\theta$.

	We now consider a nrD$24$-brane in static gauge $x^t = \sigma^0$, and $x^i=\sigma^i$. Since we are performing a T-duality along the worldvolume $\theta$-direction, we will end up with a nrD$23$-brane (cf.~Section~\ref{sec:RV-open}) with the T-dual $\tilde\theta$-direction in the transverse space.

	We will use $\bar\mu = (t,\bar i)$ to denote the NN directions after the T-duality, i.e., $\mu = (t,i)=(\bar \mu, \theta)$. 
	Using \eqref{eq:NRDBI} the action for the nrD$24$-brane in static gauge is 
	\begin{equation}
		\label{eq:T-duality-starting-point}
		S_{\text{nrD}24} = -T_{\text{nrD}24}\int d^{25}\sigma\,\sqrt{-\det\begin{pmatrix}
				0 ~&~ \tau^+_{\nu}+T^{-1}_{\text{NR}}\partial_\nu\phi\tau^+_v\\
				\tau^-_{ \mu}+T^{-1}_{\text{NR}}\partial_\mu\phi\tau^-_v ~&~  \mathcal{E}_{\mu\nu}+ \tilde c^{-1}\zTeff^{-1}f_{\mu\nu}
		\end{pmatrix}}\,,
	\end{equation}
	where we defined 
	\begin{equation}
		\mathcal{E}_{\mu\nu}=e_{\mu \nu} +T^{-1}_{\text{NR}}\partial_\mu\phi e_{v\nu}+T^{-1}_{\text{NR}}\partial_\nu\phi e_{\mu v}+T^{-2}_{\text{NR}}\partial_\mu\phi \partial_\nu\phi e_{vv}\,.
	\end{equation}
	Since $\theta$ is an isometry, we have that
	\begin{equation}
		\begin{split}
			\tau^\pm_{\theta} &= 0\,,\qquad  f_{\bar\mu \theta} = \D_{\bar\mu} a_{\theta}\,,\qquad\partial_\theta\phi=0\,.
		\end{split}
	\end{equation}
	We can write the $25\times 25$ matrix appearing in \eqref{eq:T-duality-starting-point} as
	\begin{equation}
		\begin{pmatrix}
			A ~&~ B\\
			C ~&~ D
		\end{pmatrix}\,,
	\end{equation}
	where
	\begin{subequations}
		\begin{eqnarray}
			A & = & \left(\begin{array}{cc}
				0 & \tau^+_{\bar\nu}+T^{-1}_{\text{NR}}\partial_{\bar\nu}\phi\tau^+_v\\
				\tau^-_{\bar\mu}+T^{-1}_{\text{NR}}\partial_{\bar\mu}\phi\tau^-_v &  \mathcal{E}_{\bar\mu\bar\nu}+ \tilde c^{-1}\zTeff^{-1}f_{\bar\mu\bar\nu}
			\end{array}
			\right)
			\,,\\
			B & = & \left(\begin{array}{c}
				0 \\
				\mathcal{E}_{\bar\mu\theta}+ \tilde c^{-1}\zTeff^{-1}\partial_{\bar\mu}a_\theta
			\end{array}\right)\,,\\
			C & = & \left(\begin{array}{cc}
				0 & \mathcal{E}_{\theta\bar\nu}- \tilde c^{-1}\zTeff^{-1}\partial_{\bar\nu}a_\theta
			\end{array}\right)\,,\\
			D & = & e_{\theta\theta}\,.
		\end{eqnarray}    
	\end{subequations}
	In here $A$ is a $24\times 24$ matrix, $B$ is a $24\times 1$ matrix and $C$ is a $1\times 24$ matrix. Using the determinant identity \eqref{eq:detid} we have
	\begin{eqnarray}\label{eq:D23intermediate}
		&&\det\begin{pmatrix}
			0 ~&~ \tau^+_{\nu}+T^{-1}_{\text{NR}}\partial_\nu\phi\tau^+_v\\
			\tau^-_{ \mu}+T^{-1}_{\text{NR}}\partial_\mu\phi\tau^-_v ~&~  \mathcal{E}_{\mu\nu}+ \tilde c^{-1}\zTeff^{-1}f_{\mu\nu}
		\end{pmatrix}=\nonumber\\
		&&e_{\theta\theta}\det\begin{pmatrix}
			0 ~&~ \tau^+_{\bar\nu}+T^{-1}_{\text{NR}}\partial_{\bar\nu}\phi\tau^+_v\\
			\tau^-_{\bar\mu}+T^{-1}_{\text{NR}}\partial_{\bar\mu}\phi\tau^-_v ~&~  \tilde{\mathcal{E}}_{\bar\mu\bar\nu}+ \tilde c^{-1}\zTeff^{-1}f_{\bar\mu\bar\nu}
		\end{pmatrix}\,,
	\end{eqnarray}
	where
	\begin{eqnarray}
		\tilde{\mathcal{E}}_{\bar\mu\bar\nu} & = & \mathcal{E}_{\bar\mu\bar\nu}-e_{\theta\theta}^{-1}\left(\mathcal{E}_{\bar\mu\theta}+ \tilde c^{-1}\zTeff^{-1}\partial_{\bar\mu}a_\theta\right)\left(\mathcal{E}_{\theta\bar\nu}- \tilde c^{-1}\zTeff^{-1}\partial_{\bar\nu}a_\theta\right)\nonumber\\
		& = & \tilde e_{\bar\mu\bar\nu} +T^{-1}_{\text{NR}}\partial_{\bar\mu}\phi^I \tilde e_{I\bar\nu}+T^{-1}_{\text{NR}}\partial_{\bar\nu}\phi^I \tilde e_{\bar\mu I}+T^{-2}_{\text{NR}}\partial_{\bar\mu}\phi^I \partial_{\bar\nu}\phi^J \tilde e_{IJ}\,,\nonumber
	\end{eqnarray}
	where we used \eqref{eq:Buscher-rules-E} and where we defined $I=(v,\tilde\theta)$ with 
	\begin{equation}
		\phi^I=(\phi^v,\phi^{\tilde\theta}):=(\phi,\tilde c^{-1}a_\theta)\,.
	\end{equation}
	The above expression for $\tilde{\mathcal{E}}_{\bar\mu\bar\nu}$ is precisely the pullback of $\tilde e_{MN}$ for a nrD23-brane using static gauge. Furthermore, we have
	\begin{equation}
		\tau^\pm_{\bar\nu}+T^{-1}_{\text{NR}}\partial_{\bar\nu}\phi\tau^\pm_v=\tilde   \tau^\pm_{\bar\nu}  +T^{-1}_{\text{NR}}\partial_{\bar\nu}\phi^I\tilde\tau^\pm_I\,.
	\end{equation}
	Hence, the upper right and lower left elements in \eqref{eq:D23intermediate}
	are precisely the pullback of $\tilde \tau^\pm_M$ in static gauge where we used the Buscher rule $\tilde \tau^\pm_M=\tau^\pm_M$ (see eq. \eqref{eq:LO-Buscher}) and where we used that $\tau^\pm_\theta=\tau^\pm_{\tilde\theta}=0$.  We thus conclude that
	\begin{eqnarray}
		S_{\text{nrD}24} & = &  -T_{\text{nrD}24}\int d^{25}\sigma\,\sqrt{-\det\begin{pmatrix}
				0 ~&~ \tau^+_{\nu}+T^{-1}_{\text{NR}}\partial_\nu\phi\tau^+_v\\
				\tau^-_{ \mu}+T^{-1}_{\text{NR}}\partial_\mu\phi\tau^-_v ~&~  \mathcal{E}_{\mu\nu}+ \tilde c^{-1}\zTeff^{-1}f_{\mu\nu}
		\end{pmatrix}}\nonumber\\
		& = & -T_{\text{nrD}23}\int d^{24}\sigma\,e^{1/2}_{\theta\theta}\sqrt{-\det\begin{pmatrix}
				0 ~&~ \tilde   \tau^+_{\bar\nu}  +T^{-1}_{\text{NR}}\partial_{\bar\nu}\phi^I\tilde\tau^+_I\\
				\tilde   \tau^-_{\bar\mu}  +T^{-1}_{\text{NR}}\partial_{\bar\mu}\phi^I\tilde\tau^-_I ~&~  \tilde{\mathcal{E}}_{\bar\mu\bar\nu}+ \tilde c^{-1}\zTeff^{-1}f_{\bar\mu\bar\nu}
		\end{pmatrix}}\,,\nonumber\\
		&  & \label{eq:D24=D23}
	\end{eqnarray}
	where we defined the nrD23-brane tension via
	\begin{equation}
		T_{\text{nrD}23} = 2\pi R_\perp T_{\text{nrD}24}\,,    
	\end{equation}
	with $2\pi R_\perp$ the length of the compact $\theta$ direction, i.e.,
	$2\pi R_\perp=\int d\theta$. Ignoring the overall factor of $e^{1/2}_{\theta\theta}$ in~\eqref{eq:D24=D23}, which gets absorbed by the T-duality transformation of the dilaton which we ignored, the second line in \eqref{eq:D24=D23} is the action for a nrD23-brane in static gauge.
	
	This concludes the demonstration of T-duality covariance: by applying T-duality, we find that the action of a nrD$24$-brane becomes the action of a nrD$23$-brane, whose action takes the same general form and which agrees with the result~\eqref{eq:nrDp} we obtained by expanding the relativistic DBI action in powers of $1/c^2$. 
	
	\section{Discussion}
	\label{sec:discussion}
	
	In this article, we have considered the $1/c^2$ expansion of relativistic open strings and D-branes up to NLO. This work opens up several worthwhile directions for future study, which we describe below.
	
	First of all, our expansion of the DBI action assumed that the $B$-field is critical at LO. It would be interesting to study what happens for a general LO $B$-field. Are there expansions of the DBI action that lead to nonlinear versions of electric and magnetic large $c$ (Galilean) limits of Maxwell's theory? 
	
	We focused on single branes, which in the case of a nrD$24$-brane led to a non-linear theory of GED. It would be interesting to study coincident branes with $U(n)$ Chan--Paton factors, which would lead to formulations of non-linear Galilean Yang--Mills theory rather than GED (see, e.g., \cite{Bagchi:2022twx}). This was studied in the context of NRST in~\cite{Gomis:2020fui}, and it would be interesting to revisit this from the perspective of $1/c^2$ expansions. In addition, we only expanded in the parameter $c$, not in $\tilde c$. It would be very interesting to investigate links to spin-matrix theory~\cite{Harmark:2014mpa,Harmark:2020vll,Bidussi:2023rfs} by either taking a strict limit $\tilde c\to \infty$ or considering an expansion in powers of $1/\tilde c^2$. This might provide an interesting avenue to study open strings and D-branes in the context of spin matrix theory. Relatedly, one could also consider Carrollian, i.e., $\tilde c\to 0$, limits and expansions in $\tilde c$, which was done in~\cite{Bidussi:2023rfs} for the closed string. 
	
	Just as for closed strings in~\cite{Hartong:2021ekg,Hartong:2022dsx}, we only consider an expansion in even inverse powers of the speed of light. It would be interesting to see what novel effects the inclusion of odd powers would bring about. In the particle $1/c^2$ expansion, odd powers were considered in~\cite{Ergen:2020yop,Hartong:2023ckn}.
	
	In the present manuscript, we considered transverse T-duality. It would be interesting to further study longitudinal T-duality along the $v$-direction from the perspective of $1/c^2$ expansions and to compare with the results of \cite{Gomis:2020izd}. More generally, placing the non-relativistic open string theories and D-branes obtained by expanding in powers of $1/c^2$ within the duality web uncovered in~\cite{Blair:2023noj,Gomis:2023eav} would be very interesting.

	\section*{Acknowledgements} 
	We thank Dibakar Roychowdhury for collaboration in initial stages of this project. We are grateful to Niels Obers and Ziqi Yan for useful discussions. The work of JH was supported by the Royal Society University Research Fellowship Renewal "Non-Lorentzian String Theory" (grant number URF\textbackslash R\textbackslash 221038).
	The work of EH was supported by the Villum Foundation Experiment project 00050317, ``Exploring the wonderland of Carrollian physics''.

	\addcontentsline{toc}{section}{\refname}

	\providecommand{\href}[2]{#2}\begingroup\raggedright\endgroup


\begin{thebibliography}{10}
		
		\bibitem{Gomis:2020fui}
		J.~Gomis, Z.~Yan, and M.~Yu, ``{Nonrelativistic Open String and Yang-Mills Theory},'' \href{http://dx.doi.org/10.1007/JHEP03(2021)269}{{\em JHEP} {\bfseries 03} (2021) 269}, \href{http://arxiv.org/abs/2007.01886}{{\ttfamily arXiv:2007.01886 [hep-th]}}.
		
		\bibitem{Gomis:2000bd}
		J.~Gomis and H.~Ooguri, ``{Nonrelativistic closed string theory},'' \href{http://dx.doi.org/10.1063/1.1372697}{{\em J. Math. Phys.} {\bfseries 42} (2001) 3127--3151},
		\href{http://arxiv.org/abs/hep-th/0009181}{{\ttfamily arXiv:hep-th/0009181 [hep-th]}}.
		
		\bibitem{Danielsson:2000gi}
		U.~H. Danielsson, A.~Guijosa, and M.~Kruczenski, ``{IIA/B, wound and wrapped},'' \href{http://dx.doi.org/10.1088/1126-6708/2000/10/020}{{\em JHEP} {\bfseries 10} (2000) 020},
		\href{http://arxiv.org/abs/hep-th/0009182}{{\ttfamily arXiv:hep-th/0009182 [hep-th]}}.
		
		\bibitem{Danielsson:2000mu}
		U.~H. Danielsson, A.~Guijosa, and M.~Kruczenski, ``{Newtonian gravitons and d-brane collective coordinates in wound string theory},'' \href{http://dx.doi.org/10.1088/1126-6708/2001/03/041}{{\em JHEP} {\bfseries 03} (2001) 041}, \href{http://arxiv.org/abs/hep-th/0012183}{{\ttfamily arXiv:hep-th/0012183}}.
		
		\bibitem{Harmark:2017rpg}
		T.~Harmark, J.~Hartong, and N.~A. Obers, ``{Nonrelativistic strings and limits of the AdS/CFT correspondence},'' \href{http://dx.doi.org/10.1103/PhysRevD.96.086019}{{\em Phys. Rev.} {\bfseries D96} no.~8, (2017) 086019},
		\href{http://arxiv.org/abs/1705.03535}{{\ttfamily arXiv:1705.03535 [hep-th]}}.
		
		\bibitem{Kluson:2018egd}
		J.~Kluson, ``{Remark About Non-Relativistic String in Newton-Cartan Background and Null Reduction},'' \href{http://dx.doi.org/10.1007/JHEP05(2018)041}{{\em JHEP} {\bfseries 05} (2018) 041},
		\href{http://arxiv.org/abs/1803.07336}{{\ttfamily arXiv:1803.07336 [hep-th]}}.
		
		\bibitem{Bergshoeff:2018yvt}
		E.~Bergshoeff, J.~Gomis, and Z.~Yan, ``{Nonrelativistic String Theory and T-Duality},'' \href{http://dx.doi.org/10.1007/JHEP11(2018)133}{{\em JHEP} {\bfseries 11} (2018) 133},
		\href{http://arxiv.org/abs/1806.06071}{{\ttfamily arXiv:1806.06071 [hep-th]}}.
		
		\bibitem{Harmark:2018cdl}
		T.~Harmark, J.~Hartong, L.~Menculini, N.~A. Obers, and Z.~Yan, ``{Strings with Non-Relativistic Conformal Symmetry and Limits of the AdS/CFT Correspondence},'' \href{http://dx.doi.org/10.1007/JHEP11(2018)190}{{\em JHEP} {\bfseries 11} (2018) 190},
		\href{http://arxiv.org/abs/1810.05560}{{\ttfamily arXiv:1810.05560 [hep-th]}}.
		
		\bibitem{Bergshoeff:2018vfn}
		E.~A. Bergshoeff, K.~T. Grosvenor, C.~Simsek, and Z.~Yan, ``{An Action for Extended String Newton-Cartan Gravity},'' \href{http://dx.doi.org/10.1007/JHEP01(2019)178}{{\em JHEP} {\bfseries 01} (2019) 178},
		\href{http://arxiv.org/abs/1810.09387}{{\ttfamily arXiv:1810.09387 [hep-th]}}.
		
		\bibitem{Harmark:2019upf}
		T.~Harmark, J.~Hartong, L.~Menculini, N.~A. Obers, and G.~Oling, ``{Relating non-relativistic string theories},'' \href{http://dx.doi.org/10.1007/JHEP11(2019)071}{{\em JHEP} {\bfseries 11} (2019) 071}, \href{http://arxiv.org/abs/1907.01663}{{\ttfamily arXiv:1907.01663 [hep-th]}}.
		
		\bibitem{Bergshoeff:2021bmc}
		E.~A. Bergshoeff, J.~Lahnsteiner, L.~Romano, J.~Rosseel, and C.~\c{S}im\c{s}ek, ``{A non-relativistic limit of NS-NS gravity},'' \href{http://dx.doi.org/10.1007/JHEP06(2021)021}{{\em JHEP} {\bfseries 06} (2021) 021}, \href{http://arxiv.org/abs/2102.06974}{{\ttfamily arXiv:2102.06974 [hep-th]}}.
		
		\bibitem{Yan:2021lbe}
		Z.~Yan, ``{Torsional deformation of nonrelativistic string theory},'' \href{http://dx.doi.org/10.1007/JHEP09(2021)035}{{\em JHEP} {\bfseries 09} (2021) 035}, \href{http://arxiv.org/abs/2106.10021}{{\ttfamily arXiv:2106.10021 [hep-th]}}.
		
		\bibitem{Hartong:2021ekg}
		J.~Hartong and E.~Have, ``{Nonrelativistic Expansion of Closed Bosonic Strings},'' \href{http://dx.doi.org/10.1103/PhysRevLett.128.021602}{{\em Phys. Rev. Lett.} {\bfseries 128} no.~2, (2022) 021602}, \href{http://arxiv.org/abs/2107.00023}{{\ttfamily arXiv:2107.00023 [hep-th]}}.
		
		\bibitem{Bidussi:2021ujm}
		L.~Bidussi, T.~Harmark, J.~Hartong, N.~A. Obers, and G.~Oling, ``{Torsional string Newton-Cartan geometry for non-relativistic strings},'' \href{http://dx.doi.org/10.1007/JHEP02(2022)116}{{\em JHEP} {\bfseries 02} (2022) 116}, \href{http://arxiv.org/abs/2107.00642}{{\ttfamily arXiv:2107.00642 [hep-th]}}.
		
		\bibitem{Hartong:2022dsx}
		J.~Hartong and E.~Have, ``{Nonrelativistic approximations of closed bosonic string theory},'' \href{http://dx.doi.org/10.1007/JHEP02(2023)153}{{\em JHEP} {\bfseries 02} (2023) 153}, \href{http://arxiv.org/abs/2211.01795}{{\ttfamily arXiv:2211.01795 [hep-th]}}.
		
		\bibitem{Bidussi:2023rfs}
		L.~Bidussi, T.~Harmark, J.~Hartong, N.~A. Obers, and G.~Oling, ``{Longitudinal Galilean and Carrollian limits of non-relativistic strings},'' \href{http://dx.doi.org/10.1007/JHEP12(2023)141}{{\em JHEP} {\bfseries 12} (2023) 141}, \href{http://arxiv.org/abs/2309.14467}{{\ttfamily arXiv:2309.14467 [hep-th]}}.
		
		\bibitem{Isberg:1993av}
		J.~Isberg, U.~Lindstrom, B.~Sundborg, and G.~Theodoridis, ``{Classical and quantized tensionless strings},'' \href{http://dx.doi.org/10.1016/0550-3213(94)90056-6}{{\em Nucl. Phys.} {\bfseries B411} (1994) 122--156},
		\href{http://arxiv.org/abs/hep-th/9307108}{{\ttfamily arXiv:hep-th/9307108 [hep-th]}}.
		
		\bibitem{Bagchi:2020fpr}
		A.~Bagchi, A.~Banerjee, S.~Chakrabortty, S.~Dutta, and P.~Parekh, ``{A tale of three \textemdash{} tensionless strings and vacuum structure},'' \href{http://dx.doi.org/10.1007/JHEP04(2020)061}{{\em JHEP} {\bfseries 04} (2020) 061}, \href{http://arxiv.org/abs/2001.00354}{{\ttfamily arXiv:2001.00354 [hep-th]}}.
		
		\bibitem{Harksen:2024bnh}
		M.~Harksen, D.~Hidalgo, W.~Sybesma, and L.~Thorlacius, ``{Carroll strings with an extended symmetry algebra},'' \href{http://dx.doi.org/10.1007/JHEP05(2024)206}{{\em JHEP} {\bfseries 05} (2024) 206}, \href{http://arxiv.org/abs/2403.01984}{{\ttfamily arXiv:2403.01984 [hep-th]}}.
		
		\bibitem{Bagchi:2023cfp}
		A.~Bagchi, A.~Banerjee, J.~Hartong, E.~Have, K.~S. Kolekar, and M.~Mandlik, ``{Strings near black holes are Carrollian},'' \href{http://arxiv.org/abs/2312.14240}{{\ttfamily arXiv:2312.14240 [hep-th]}}.
		
		\bibitem{Blair:2023noj}
		C.~D.~A. Blair, J.~Lahnsteiner, N.~A.~J. Obers, and Z.~Yan, ``{Unification of Decoupling Limits in String and M Theory},'' \href{http://dx.doi.org/10.1103/PhysRevLett.132.161603}{{\em Phys. Rev. Lett.} {\bfseries 132} no.~16, (2024) 161603}, \href{http://arxiv.org/abs/2311.10564}{{\ttfamily arXiv:2311.10564 [hep-th]}}.
		
		\bibitem{Gomis:2023eav}
		J.~Gomis and Z.~Yan, ``{Worldsheet Formalism for Decoupling Limits in String Theory},'' \href{http://arxiv.org/abs/2311.10565}{{\ttfamily arXiv:2311.10565 [hep-th]}}.
		
		\bibitem{Harmark:2014mpa}
		T.~Harmark and M.~Orselli, ``{Spin Matrix Theory: A quantum mechanical model of the AdS/CFT correspondence},'' \href{http://dx.doi.org/10.1007/JHEP11(2014)134}{{\em JHEP} {\bfseries 11} (2014) 134},
		\href{http://arxiv.org/abs/1409.4417}{{\ttfamily arXiv:1409.4417 [hep-th]}}.
		
		\bibitem{Harmark:2020vll}
		T.~Harmark, J.~Hartong, N.~A. Obers, and G.~Oling, ``{Spin Matrix Theory String Backgrounds and Penrose Limits of AdS/CFT},'' \href{http://dx.doi.org/10.1007/JHEP03(2021)129}{{\em JHEP} {\bfseries 03} (2021) 129}, \href{http://arxiv.org/abs/2011.02539}{{\ttfamily arXiv:2011.02539 [hep-th]}}.
		
		\bibitem{Gomis:2005pg}
		J.~Gomis, J.~Gomis, and K.~Kamimura, ``{Non-relativistic superstrings: A New soluble sector of AdS(5) x S\textasteriskcentered\textasteriskcentered5},'' \href{http://dx.doi.org/10.1088/1126-6708/2005/12/024}{{\em JHEP} {\bfseries 12} (2005) 024},
		\href{http://arxiv.org/abs/hep-th/0507036}{{\ttfamily arXiv:hep-th/0507036 [hep-th]}}.
		
		\bibitem{Kim:2007pc}
		B.~S. Kim, ``{Non-relativistic superstring theories},'' \href{http://dx.doi.org/10.1103/PhysRevD.76.126013}{{\em Phys. Rev. D} {\bfseries 76} (2007) 126013}, \href{http://arxiv.org/abs/0710.3203}{{\ttfamily arXiv:0710.3203 [hep-th]}}.
		
		\bibitem{Blair:2019qwi}
		C.~D.~A. Blair, ``{A worldsheet supersymmetric Newton-Cartan string},'' \href{http://dx.doi.org/10.1007/JHEP10(2019)266}{{\em JHEP} {\bfseries 10} (2019) 266},
		\href{http://arxiv.org/abs/1908.00074}{{\ttfamily arXiv:1908.00074 [hep-th]}}.
		
		\bibitem{Bergshoeff:2022pzk}
		E.~Bergshoeff, J.~Lahnsteiner, L.~Romano, and J.~Rosseel, ``{The supersymmetric Neveu-Schwarz branes of non-relativistic string theory},'' \href{http://dx.doi.org/10.1007/JHEP08(2022)218}{{\em JHEP} {\bfseries 08} (2022) 218}, \href{http://arxiv.org/abs/2204.04089}{{\ttfamily arXiv:2204.04089 [hep-th]}}.
		
		\bibitem{Ko:2015rha}
		S.~M. Ko, C.~Melby-Thompson, R.~Meyer, and J.-H. Park, ``{Dynamics of Perturbations in Double Field Theory \& Non-Relativistic String Theory},'' \href{http://dx.doi.org/10.1007/JHEP12(2015)144}{{\em JHEP} {\bfseries 12} (2015) 144},
		\href{http://arxiv.org/abs/1508.01121}{{\ttfamily arXiv:1508.01121 [hep-th]}}.
		
		\bibitem{Berman:2019izh}
		D.~S. Berman, C.~D.~A. Blair, and R.~Otsuki, ``{Non-Riemannian geometry of M-theory},'' \href{http://dx.doi.org/10.1007/JHEP07(2019)175}{{\em JHEP} {\bfseries 07} (2019) 175},
		\href{http://arxiv.org/abs/1902.01867}{{\ttfamily arXiv:1902.01867 [hep-th]}}.
		
		\bibitem{Cho:2019ofr}
		K.~Cho and J.-H. Park, ``{Remarks on the non-Riemannian sector in Double Field Theory},'' \href{http://dx.doi.org/10.1140/epjc/s10052-020-7648-9}{{\em Eur. Phys. J. C} {\bfseries 80} no.~2, (2020) 101}, \href{http://arxiv.org/abs/1909.10711}{{\ttfamily arXiv:1909.10711 [hep-th]}}.
		
		\bibitem{Blair:2020ops}
		C.~D.~A. Blair, ``{Non-relativistic duality and $T \bar T$ deformations},'' \href{http://dx.doi.org/10.1007/JHEP07(2020)069}{{\em JHEP} {\bfseries 07} (2020) 069}, \href{http://arxiv.org/abs/2002.12413}{{\ttfamily arXiv:2002.12413 [hep-th]}}.
		
		\bibitem{Park:2020ixf}
		J.-H. Park and S.~Sugimoto, ``{String Theory and non-Riemannian Geometry},'' \href{http://dx.doi.org/10.1103/PhysRevLett.125.211601}{{\em Phys. Rev. Lett.} {\bfseries 125} no.~21, (2020) 211601}, \href{http://arxiv.org/abs/2008.03084}{{\ttfamily arXiv:2008.03084 [hep-th]}}.
		
		\bibitem{Gallegos:2020egk}
		A.~D. Gallegos, U.~G\"ursoy, S.~Verma, and N.~Zinnato, ``{Non-Riemannian gravity actions from double field theory},'' \href{http://dx.doi.org/10.1007/JHEP06(2021)173}{{\em JHEP} {\bfseries 06} (12, 2020) 173}, \href{http://arxiv.org/abs/2012.07765}{{\ttfamily arXiv:2012.07765 [hep-th]}}.
		
		\bibitem{Blair:2020gng}
		C.~D.~A. Blair, G.~Oling, and J.-H. Park, ``{Non-Riemannian isometries from double field theory},'' \href{http://dx.doi.org/10.1007/JHEP04(2021)072}{{\em JHEP} {\bfseries 04} (2021) 072}, \href{http://arxiv.org/abs/2012.07766}{{\ttfamily arXiv:2012.07766 [hep-th]}}.
		
		\bibitem{Roychowdhury:2019sfo}
		D.~Roychowdhury, ``{Semiclassical dynamics for torsional Newton-Cartan strings},'' \href{http://dx.doi.org/10.1016/j.nuclphysb.2020.115132}{{\em Nucl. Phys. B} {\bfseries 958} (2020) 115132}, \href{http://arxiv.org/abs/1911.10473}{{\ttfamily arXiv:1911.10473 [hep-th]}}.
		
		\bibitem{Fontanella:2021hcb}
		A.~Fontanella and J.~M. Nieto~Garc\'\i{}a, ``{Light-cone gauge in non-relativistic AdS$_{5}$\texttimes{}S$^{5}$ string theory},'' \href{http://dx.doi.org/10.1007/JHEP11(2023)053}{{\em JHEP} {\bfseries 11} (2023) 053}, \href{http://arxiv.org/abs/2102.00008}{{\ttfamily arXiv:2102.00008 [hep-th]}}.
		
		\bibitem{Fontanella:2021btt}
		A.~Fontanella and J.~M.~N. Garc\'\i{}a, ``{Classical string solutions in non-relativistic AdS$_{5}\times S^{5}$: closed and twisted sectors},'' \href{http://dx.doi.org/10.1088/1751-8121/ac4abd}{{\em J. Phys. A} {\bfseries 55} no.~8, (2022) 085401}, \href{http://arxiv.org/abs/2109.13240}{{\ttfamily arXiv:2109.13240 [hep-th]}}.
		
		\bibitem{Roychowdhury:2021wte}
		D.~Roychowdhury, ``{Decoding the Spin-Matrix limit of strings on AdS5\texttimes{}S5},'' \href{http://dx.doi.org/10.1016/j.physletb.2021.136499}{{\em Phys. Lett. B} {\bfseries 820} (2021) 136499}, \href{http://arxiv.org/abs/2101.06513}{{\ttfamily arXiv:2101.06513 [hep-th]}}.
		
		\bibitem{Fontanella:2022fjd}
		A.~Fontanella and S.~J. van Tongeren, ``{Coset space actions for nonrelativistic strings},'' \href{http://dx.doi.org/10.1007/JHEP06(2022)080}{{\em JHEP} {\bfseries 06} (2022) 080}, \href{http://arxiv.org/abs/2203.07386}{{\ttfamily arXiv:2203.07386 [hep-th]}}.
		
		\bibitem{Fontanella:2022pbm}
		A.~Fontanella and J.~M. Nieto~Garc\'\i{}a, ``{Extending the nonrelativistic string AdS coset},'' \href{http://dx.doi.org/10.1103/PhysRevD.106.L121901}{{\em Phys. Rev. D} {\bfseries 106} no.~12, (2022) L121901}, \href{http://arxiv.org/abs/2208.02295}{{\ttfamily arXiv:2208.02295 [hep-th]}}.
		
		\bibitem{Fontanella:2024rvn}
		A.~Fontanella and J.~M. Nieto~Garc\'\i{}a, ``{Constructing Non-Relativistic AdS$_5$/CFT$_4$ Holography},'' \href{http://arxiv.org/abs/2403.02379}{{\ttfamily arXiv:2403.02379 [hep-th]}}.
		
		\bibitem{deLeeuw:2024uaq}
		M.~de~Leeuw, A.~Fontanella, and J.~M. Nieto~Garc\'\i{}a, ``{A perturbative approach to the non-relativistic string spectrum},'' \href{http://arxiv.org/abs/2403.09563}{{\ttfamily arXiv:2403.09563 [hep-th]}}.
		
		\bibitem{Oling:2022fft}
		G.~Oling and Z.~Yan, ``{Aspects of Nonrelativistic Strings},'' \href{http://dx.doi.org/10.3389/fphy.2022.832271}{{\em Front. in Phys.} {\bfseries 10} (2022) 832271}, \href{http://arxiv.org/abs/2202.12698}{{\ttfamily arXiv:2202.12698 [hep-th]}}.
		
		\bibitem{Gomis:2004pw}
		J.~Gomis, K.~Kamimura, and P.~K. Townsend, ``{Non-relativistic superbranes},'' \href{http://dx.doi.org/10.1088/1126-6708/2004/11/051}{{\em JHEP} {\bfseries 11} (2004) 051}, \href{http://arxiv.org/abs/hep-th/0409219}{{\ttfamily arXiv:hep-th/0409219}}.
		
		\bibitem{Brugues:2004an}
		J.~Brugues, T.~Curtright, J.~Gomis, and L.~Mezincescu, ``{Non-relativistic strings and branes as non-linear realizations of Galilei groups},'' \href{http://dx.doi.org/10.1016/j.physletb.2004.05.024}{{\em Phys. Lett. B} {\bfseries 594} (2004) 227--233}, \href{http://arxiv.org/abs/hep-th/0404175}{{\ttfamily arXiv:hep-th/0404175}}.
		
		\bibitem{Gomis:2005bj}
		J.~Gomis, F.~Passerini, T.~Ramirez, and A.~Van~Proeyen, ``{Non relativistic Dp branes},'' \href{http://dx.doi.org/10.1088/1126-6708/2005/10/007}{{\em JHEP} {\bfseries 10} (2005) 007}, \href{http://arxiv.org/abs/hep-th/0507135}{{\ttfamily arXiv:hep-th/0507135}}.
		
		\bibitem{Brugues:2006yd}
		J.~Brugues, J.~Gomis, and K.~Kamimura, ``{Newton-Hooke algebras, non-relativistic branes and generalized pp-wave metrics},'' \href{http://dx.doi.org/10.1103/PhysRevD.73.085011}{{\em Phys. Rev. D} {\bfseries 73} (2006) 085011}, \href{http://arxiv.org/abs/hep-th/0603023}{{\ttfamily arXiv:hep-th/0603023}}.
		
		\bibitem{Mazzucato:2008tr}
		L.~Mazzucato, Y.~Oz, and S.~Theisen, ``{Non-relativistic Branes},'' \href{http://dx.doi.org/10.1088/1126-6708/2009/04/073}{{\em JHEP} {\bfseries 04} (2009) 073}, \href{http://arxiv.org/abs/0810.3673}{{\ttfamily arXiv:0810.3673 [hep-th]}}.
		
		\bibitem{Kluson:2019avy}
		J.~Kluso\v{n}, ``{Non-Relativistic D-brane from T-duality Along Null Direction},'' \href{http://dx.doi.org/10.1007/JHEP10(2019)153}{{\em JHEP} {\bfseries 10} (2019) 153}, \href{http://arxiv.org/abs/1907.05662}{{\ttfamily arXiv:1907.05662 [hep-th]}}.
		
		\bibitem{Roychowdhury:2019qmp}
		D.~Roychowdhury, ``{Probing tachyon kinks in Newton-Cartan background},'' \href{http://dx.doi.org/10.1016/j.physletb.2019.06.031}{{\em Phys. Lett. B} {\bfseries 795} (2019) 225--229}, \href{http://arxiv.org/abs/1903.05890}{{\ttfamily arXiv:1903.05890 [hep-th]}}.
		
		\bibitem{Kluson:2020kyp}
		J.~Kluso\v{n}, ``{D-brane actions in nonrelativistic string theory and T duality},'' \href{http://dx.doi.org/10.1103/PhysRevD.104.086009}{{\em Phys. Rev. D} {\bfseries 104} no.~8, (2021) 086009}, \href{http://arxiv.org/abs/2011.14323}{{\ttfamily arXiv:2011.14323 [hep-th]}}.
		
		\bibitem{Gomis:2020izd}
		J.~Gomis, Z.~Yan, and M.~Yu, ``{T-Duality in Nonrelativistic Open String Theory},'' \href{http://dx.doi.org/10.1007/JHEP02(2021)087}{{\em JHEP} {\bfseries 02} (2021) 087}, \href{http://arxiv.org/abs/2008.05493}{{\ttfamily arXiv:2008.05493 [hep-th]}}.
		
		\bibitem{Kluson:2020aoq}
		J.~Kluso\v{n}, ``{Unstable D-brane in Torsional Newton-Cartan Background},'' \href{http://dx.doi.org/10.1007/JHEP09(2020)191}{{\em JHEP} {\bfseries 09} (2020) 191}, \href{http://arxiv.org/abs/2001.11543}{{\ttfamily arXiv:2001.11543 [hep-th]}}.
		
		\bibitem{Ebert:2021mfu}
		S.~Ebert, H.-Y. Sun, and Z.~Yan, ``{Dual D-brane actions in nonrelativistic string theory},'' \href{http://dx.doi.org/10.1007/JHEP04(2022)161}{{\em JHEP} {\bfseries 04} (2022) 161}, \href{http://arxiv.org/abs/2112.09316}{{\ttfamily arXiv:2112.09316 [hep-th]}}.
		
		\bibitem{Guijosa:2023qym}
		A.~Guijosa and I.~C. Rosas-L\'opez, ``{Geometry from D-branes in Nonrelativistic String Theory},'' \href{http://arxiv.org/abs/2312.03332}{{\ttfamily arXiv:2312.03332 [hep-th]}}.
		
		\bibitem{Lambert:2024yjk}
		N.~Lambert and J.~Smith, ``{Non-Relativistic Intersecting Branes, Newton-Cartan Geometry and AdS/CFT},'' \href{http://arxiv.org/abs/2405.06552}{{\ttfamily arXiv:2405.06552 [hep-th]}}.
		
		\bibitem{Lambert:2024uue}
		N.~Lambert and J.~Smith, ``{Non-relativistic M2-branes and the AdS/CFT correspondence},'' \href{http://dx.doi.org/10.1007/JHEP06(2024)009}{{\em JHEP} {\bfseries 06} (2024) 009}, \href{http://arxiv.org/abs/2401.14955}{{\ttfamily arXiv:2401.14955 [hep-th]}}.
		
		\bibitem{VandenBleeken:2017rij}
		D.~Van~den Bleeken, ``Torsional Newton-Cartan gravity from the large c expansion of general relativity,'' \href{http://dx.doi.org/10.1088/1361-6382/aa83d4}{{\em Class. Quant. Grav.} {\bfseries 34} no.~18, (2017) 185004},
		\href{http://arxiv.org/abs/1703.03459}{{\ttfamily arXiv:1703.03459 [gr-qc]}}.
		
		\bibitem{Hansen:2018ofj}
		D.~Hansen, J.~Hartong, and N.~A. Obers, ``{Action Principle for Newtonian Gravity},'' \href{http://dx.doi.org/10.1103/PhysRevLett.122.061106}{{\em Phys. Rev. Lett.} {\bfseries 122} no.~6, (2019) 061106},
		\href{http://arxiv.org/abs/1807.04765}{{\ttfamily arXiv:1807.04765 [hep-th]}}.
		
		\bibitem{VandenBleeken:2019gqa}
		D.~Van~den Bleeken, ``{Torsional Newton-Cartan gravity and strong gravitational fields},'' in {\em {15th Marcel Grossmann Meeting on Recent Developments in Theoretical and Experimental General Relativity, Astrophysics, and Relativistic Field Theories (MG15) Rome, Italy, July 1-7, 2018}}.
		\newblock 2019.
		\newblock
		\href{http://arxiv.org/abs/1903.10682}{{\ttfamily arXiv:1903.10682 [gr-qc]}}.
		\newblock
		
		\bibitem{Hansen:2019vqf}
		D.~Hansen, J.~Hartong, and N.~A. Obers, ``{Gravity between Newton and Einstein},'' \href{http://dx.doi.org/10.1142/S0218271819440103}{{\em Int. J. Mod. Phys. D} {\bfseries 28} no.~14, (2019) 1944010},
		\href{http://arxiv.org/abs/1904.05706}{{\ttfamily arXiv:1904.05706 [gr-qc]}}.
		
		\bibitem{Hansen:2020pqs}
		D.~Hansen, J.~Hartong, and N.~A. Obers, ``{Non-Relativistic Gravity and its Coupling to Matter},'' \href{http://dx.doi.org/10.1007/JHEP06(2020)145}{{\em JHEP} {\bfseries 06} (2020) 145}, \href{http://arxiv.org/abs/2001.10277}{{\ttfamily arXiv:2001.10277 [gr-qc]}}.
		
		\bibitem{Ergen:2020yop}
		M.~Ergen, E.~Hamamci, and D.~Van~den Bleeken, ``{Oddity in nonrelativistic, strong gravity},'' \href{http://dx.doi.org/10.1140/epjc/s10052-020-8112-6}{{\em Eur. Phys. J. C} {\bfseries 80} no.~6, (2020) 563}, \href{http://arxiv.org/abs/2002.02688}{{\ttfamily arXiv:2002.02688 [gr-qc]}}. [Erratum: Eur.Phys.J.C 80, 657 (2020)].
		
		\bibitem{Seiberg:2000ms}
		N.~Seiberg, L.~Susskind, and N.~Toumbas, ``{Strings in background electric field, space / time noncommutativity and a new noncritical string theory},'' \href{http://dx.doi.org/10.1088/1126-6708/2000/06/021}{{\em JHEP} {\bfseries 06} (2000) 021}, \href{http://arxiv.org/abs/hep-th/0005040}{{\ttfamily arXiv:hep-th/0005040}}.
		
		\bibitem{Gopakumar:2000na}
		R.~Gopakumar, J.~M. Maldacena, S.~Minwalla, and A.~Strominger, ``{S duality and noncommutative gauge theory},'' \href{http://dx.doi.org/10.1088/1126-6708/2000/06/036}{{\em JHEP} {\bfseries 06} (2000) 036}, \href{http://arxiv.org/abs/hep-th/0005048}{{\ttfamily arXiv:hep-th/0005048}}.
		
		\bibitem{LeBellac:1973}
		M.~L. Bellac and J.~M. L\'evy-Leblond, ``{Galilean Electromagnetism},'' {\em Nuovo Cim.} {\bfseries B14} (1973) 217--33.
		
		\bibitem{Santos:2004pq}
		E.~S. Santos, M.~de~Montigny, F.~C. Khanna, and A.~E. Santana, ``{Galilean covariant Lagrangian models},''
		\href{http://dx.doi.org/10.1088/0305-4470/37/41/011}{{\em J. Phys.} {\bfseries A37} (2004) 9771--9789}.
		
		\bibitem{GEDreview}
		G.~Rousseaux, ``Forty years of Galilean Electromagnetism (1973--2013),'' {\em The European Physical Journal Plus} {\bfseries 128} no.~8, (2013) 81.
		
		\bibitem{Bagchi:2014ysa}
		A.~Bagchi, R.~Basu, and A.~Mehra, ``{Galilean Conformal Electrodynamics},'' \href{http://dx.doi.org/10.1007/JHEP11(2014)061}{{\em JHEP} {\bfseries 11} (2014) 061},
		\href{http://arxiv.org/abs/1408.0810}{{\ttfamily arXiv:1408.0810 [hep-th]}}.
		
		\bibitem{Festuccia:2016caf}
		G.~Festuccia, D.~Hansen, J.~Hartong, and N.~A. Obers, ``{Symmetries and Couplings of Non-Relativistic Electrodynamics},'' \href{http://dx.doi.org/10.1007/JHEP11(2016)037}{{\em JHEP} {\bfseries 11} (2016) 037},
		\href{http://arxiv.org/abs/1607.01753}{{\ttfamily arXiv:1607.01753 [hep-th]}}.
		
		\bibitem{Bergshoeff:2019ctr}
		E.~Bergshoeff, J.~M. Izquierdo, T.~Ortin, and L.~Romano, ``{Lie Algebra Expansions and Actions for Non-Relativistic Gravity},'' \href{http://dx.doi.org/10.1007/JHEP08(2019)048}{{\em JHEP} {\bfseries 08} (2019) 048}, \href{http://arxiv.org/abs/1904.08304}{{\ttfamily arXiv:1904.08304 [hep-th]}}.
		
		\bibitem{Bergshoeff:2022fzb}
		E.~Bergshoeff, K.~van Helden, J.~Lahnsteiner, L.~Romano, and J.~Rosseel, ``{Generalized Newton\textendash{}Cartan geometries for particles and strings},'' \href{http://dx.doi.org/10.1088/1361-6382/acbe8c}{{\em Class. Quant. Grav.} {\bfseries 40} no.~7, (2023) 075010}, \href{http://arxiv.org/abs/2207.00363}{{\ttfamily arXiv:2207.00363 [hep-th]}}.
		
		\bibitem{Bergshoeff:2019pij}
		E.~A. Bergshoeff, J.~Gomis, J.~Rosseel, C.~\c{S}im\c{s}ek, and Z.~Yan, ``{String Theory and String Newton-Cartan Geometry},'' \href{http://dx.doi.org/10.1088/1751-8121/ab56e9}{{\em J. Phys. A} {\bfseries 53} no.~1, (2020) 014001}, \href{http://arxiv.org/abs/1907.10668}{{\ttfamily arXiv:1907.10668 [hep-th]}}.
		
		\bibitem{Bergshoeff:2015sic}
		E.~Bergshoeff, J.~Rosseel, and T.~Zojer, ``{Non-relativistic fields from arbitrary contracting backgrounds},'' \href{http://dx.doi.org/10.1088/0264-9381/33/17/175010}{{\em Class. Quant. Grav.} {\bfseries 33} no.~17, (2016) 175010}, \href{http://arxiv.org/abs/1512.06064}{{\ttfamily arXiv:1512.06064 [hep-th]}}.
		
		\bibitem{Rocek:1991ps}
		M.~Rocek and E.~P. Verlinde, ``{Duality, quotients, and currents},'' \href{http://dx.doi.org/10.1016/0550-3213(92)90269-H}{{\em Nucl. Phys.} {\bfseries B373} (1992) 630--646},
		\href{http://arxiv.org/abs/hep-th/9110053}{{\ttfamily arXiv:hep-th/9110053 [hep-th]}}.
		
		\bibitem{Alvarez:1994dn}
		E.~Alvarez, L.~Alvarez-Gaume, and Y.~Lozano, ``{An Introduction to T duality in string theory},'' \href{http://dx.doi.org/10.1016/0920-5632(95)00429-D}{{\em Nucl. Phys. B Proc. Suppl.} {\bfseries 41} (1995) 1--20}, \href{http://arxiv.org/abs/hep-th/9410237}{{\ttfamily arXiv:hep-th/9410237}}.
		
		\bibitem{Bagchi:2022twx}
		A.~Bagchi, R.~Basu, M.~Islam, K.~S. Kolekar, and A.~Mehra, ``{Galilean gauge theories from null reductions},'' \href{http://dx.doi.org/10.1007/JHEP04(2022)176}{{\em JHEP} {\bfseries 04} (2022) 176}, \href{http://arxiv.org/abs/2201.12629}{{\ttfamily arXiv:2201.12629 [hep-th]}}.
		
		\bibitem{Hartong:2023ckn}
		J.~Hartong and J.~Musaeus, ``{Toward a covariant framework for post-Newtonian expansions for radiative sources},'' \href{http://dx.doi.org/10.1103/PhysRevD.109.124058}{{\em Phys. Rev. D} {\bfseries 109} no.~12, (2024) 124058}, \href{http://arxiv.org/abs/2311.07546}{{\ttfamily arXiv:2311.07546 [gr-qc]}}.
		
	\end{thebibliography}
\end{document}